\documentclass[oneside, a4paper, onecolumn, 11pt]{article}
\usepackage[left=2.5cm,top=2.5cm,bottom=2.5cm,right=2.5cm]{geometry}
\usepackage{amsmath}
\usepackage{graphicx}
\usepackage{fancyhdr}
\usepackage{calc}
\usepackage{amssymb}
\usepackage{setspace}
\usepackage{amsfonts}
\usepackage{commath}
\usepackage[innercaption]{sidecap}
\usepackage[framemethod=TikZ]{mdframed}
\usepackage{wrapfig}
\usepackage{hyperref}
\usepackage{parskip}
\usepackage{mathrsfs}
\usepackage{verbatim}
\usepackage[english]{babel}
\usepackage{amsthm}
\usepackage[normalem]{ulem}
\usepackage{amssymb}
\usepackage{todonotes}
\usepackage[backend=biber,style=nature,natbib=true]{biblatex}

\newcommand{\Rom}[1]{\expandafter\@slowromancap\romannumeral #1@}
\makeatother
{ \normalfont} 

\expandafter\def\expandafter\normalsize\expandafter{%
	\normalsize
	\setlength\abovedisplayskip{0pt}
	\setlength\belowdisplayskip{5pt}
	\setlength\abovedisplayshortskip{0pt}
	\setlength\belowdisplayshortskip{5pt}
}

\newcommand{\av}[1]{\langle #1 \rangle}
\newcommand{\M}{\mathcal M}
\providecommand{\A}{}
\renewcommand{\A}{\mathcal A}
\newcommand{\X}{\mathcal X}

\newcommand{\edited}[1]{{\color{blue}#1}}

\definecolor{Gray}{gray}{0.75}

\newmdenv[backgroundcolor=Gray, leftmargin = 0pt, rightmargin = 0pt, linewidth = 0pt, roundcorner = 2 pt, innerleftmargin=5pt, innerrightmargin=5pt, innertopmargin=5pt, innerbottommargin=5pt]{Frame}

\graphicspath{{Figures/}}

\begin{document}

\begin{center}
{\color{blue} \LARGE \textbf{Cost-effective network robustness}}

\vspace{3mm}
Rotem Brand$^{1,2}$, Simcha Haber$^{1}$,  Reuven Cohen$^{1}$, Avi Leon$^{3}$, Tomer Ron$^{3}$, Emanuel Zuckerberger$^{3}$, Sorin Avram$^{3}$, Shachar Ganem Thabit$^{3}$ \& Baruch Barzel$^{1,2}$ 
\end{center}

\begin{enumerate}
\small{
\item
Department of Mathematics, Bar-Ilan University, Ramat-Gan, Israel 
\item
Gonda Multidisciplinary Brain Research Center, Bar-Ilan University, Ramat-Gan, Israel
\item 
Israel Electric Company, Haifa, Israel
}
\end{enumerate}

\vspace{2mm}

\hspace{-2.27mm} 
\textbf{
Modern society relies heavily on infrastructure networks, from communication to power systems, making their reliability under failure and disruption critical. Robustness typically requires redundancy, providing alternative paths that ensure continuous service, yet redundancy is costly and often limited. Real-world networks therefore operate under a fundamental trade-off between resource investment and reliability. Here we derive theoretical bounds governing this trade-off and identify the network architectures that minimize expected downtime under constraints of cost, component durability, and spatial embedding. We show that, in the relevant limit, optimal configurations consist of a $3$-connected structural core linked by intermediate chains of uniform weight. This reveals a previously unrecognized family of spatial networks that achieves near-optimal robustness with orders-of-magnitude fewer resources. Distinct from commonly studied network models, these architectures expose general structural principles underlying the efficient organization of reliable infrastructure networks.
}

\begin{refsection}

Critical infrastructure — power lines, communication channels, transportation routes, and water and sewage networks — is often pushed to its limits~\supercite{PowerResilience,prudhvi2025vulnerability}. Scarce resources and growing demand frequently produce sparse, cost-minimized networks that are highly vulnerable to local, sporadic failures~\supercite{cohen2000resilience,motter2002cascade,universal_gao_2016,robustness_artime_2024,albert2000error}.

To balance cost and reliability, we seek optimized networks that incorporate a minimal number of redundant links while preserving robustness. In practice, this is often achieved by introducing a limited set of carefully placed connections that guarantee at least two disjoint paths between each pair of nodes~\supercite{JainRaviSingh03,TarjanConencted,kerivin2005design,Frederickson1981ApproximationAF,SeboeVygen14,monma1989methods}. Such redundancy produces $2$-connected graph structures, ensuring that no single link failure can isolate nodes, thereby reducing the probability of service loss~\supercite{moore1956reliable,Boesch1986OnUP,wang1997structure,On_the_validity,romero2021uniformly}

The challenge, we show here, is that the reliability of such $2$-connected networks decreases asymptotically with network size $N$, and therefore they become inevitably unstable as $N$ grows. To address this, we analytically derive the reliability phase-space, uncovering how the interplay among network size $N$, link redundancy $R$, and component failure probability $p$ jointly determine the network’s reliability $F$. This allows us to identify the optimal graph structures that maximize reliability under the given resources.

Ultimately, our theory leads to a universal algorithm that ingests the spatial~\supercite{spatial_barthelemy_2010,shape_gastner_2020,Shekhtman2014RobustnessOA,tomassini2025enhancing,catastrophic_buldyrev_2010,complex_vespignani_2010} configuration of nodes and the upper bound on allowable redundancy, and outputs the optimal network that maximizes reliability. When tested on both simulated and real-world infrastructure networks, we find that reliability can be improved — sometimes by orders of magnitude — not through additional resources, but by optimally deploying the existing network capacity.

\vspace{3mm}
\textbf{\Large \color{blue} Modeling framework}

Consider an infrastructure network, such as a power or communication system, whose connectivity is described by the graph $G(S,V,E)$. Here $V = \{1,\dots,N\}$ denotes the set of system nodes, of which $S = \{1,\dots,K\} \subseteq V$ represents the subset of source nodes, or generators, while the remaining nodes act as consumers. The set $E = \{1,\dots,L\}$ enumerates the system’s $L$ undirected links. The network is fully functional as long as every consumer $v \in V$ is connected to at least one source $s \in S$ via a finite path, \textit{i.e.}, $v \leftrightarrow S$. If, however, a node $v$ cannot reach any source through the existing network paths, namely $v \nleftrightarrow S$, it is denied service until connectivity to $S$ is restored.

In power systems, we typically consider a small number of sources, with $K \sim O(1)$, reflecting the fact that each generator serves multiple consumers. In other systems, such as communication~\supercite{neumayer2011assessing} or transportation, this distinction is absent:\ nodes can both send and receive flows of information or passengers, and hence $S$ covers most or all of $V$ and $K \sim O(N)$. Our analysis is designed to accommodate both types of networks. However, the differing nature of the $S$–$V$ relationship requires dedicated treatment in each case. Accordingly, we focus here primarily on power networks, and leave the generalization to multi-source networks to Supplementary Section~1.

\vspace{3mm}
\textbf{\color{blue} Cost}.\
In $G$, the links require investment of material resources, such as wires, cables or pipelines, and hence the infrastructure wiring cost amounts to

\begin{equation}
C = \sum_{l \in E} c \lambda_l = c \lambda_{\text{Net}},
\label{eq:Cost}
\end{equation}

where $c$ is the link cost per unit length and $\lambda_l$ is the length of the $l$th link; $\lambda_{\text{Net}} = L \av \lambda$ denotes the total length of all links in $G$. To reduce construction expenses, we seek to connect all consumers to $S$, while minimizing $C$. As a result the network graph is typically sparse (small $L$), and the links are mostly local (small $\lambda_l$), extended between geographically proximate nodes.  

\vspace{3mm}
\textbf{\color{blue} Service reliability}.\ 
The challenge is that an overly parsimonious $G$, while cost-effective, incurs the risk of service loss due to link failures. Hence, we denote the link failure probability by $p_l = p \lambda_l$, where $p$, the average failure probability per unit length, represents the intrinsic component risk~\supercite{allan2013reliability,neumayer2011assessing}. We assume the limit where $p_l \ll 1$, \textit{i.e}.\ individual links are generally stable. For a given network $G$ and component risk $p$, we seek the System Average Interruption Duration Index (SAIDI)~\supercite{allan2013reliability,IEEE_guide,trifunovic2006water} 

\begin{equation}
F = \frac{1}{W}\sum_{v \in V} w_v P(S \nleftrightarrow v \mid G,p),
\label{SAIDI}
\end{equation}

where $P(S \nleftrightarrow v | G,p)$ is the probability that a consumer $v$ experiences service denial over time, and $w_v$ denotes the consumer weight, prioritizing, \textit{e.g}., critical infrastructure over residential users, who may tolerate occasional service disruption. The index in \eqref{SAIDI} is normalized by $W = \sum_{v \in V} w_v$, ensuring $0 \le F \le 1$. 

The SAIDI index $F$ is frequently used to quantify the impact of sporadic link failures on network reliability in power, communication, and transport systems~\supercite{colbourn1987combinatorics,trifunovic2006water,perez2018sixty,bolch2006queueing,monte_carlo} (Supplementary Section~1.2). It is designed to capture service continuity by quantifying the average fraction of consumers denied service at any given time, or equivalently, the average duration of service denial per consumer per unit time. Hence, $F \to 0$ indicates a highly reliable network, whereas $F \sim 1$ corresponds to a network in a constant state of failure. 

Our goal is to design optimal networks $G$ that, for a given component risk $p$, minimize both the SAIDI index $F$ in \eqref{SAIDI} and the construction cost $C$ in \eqref{eq:Cost}. This optimization is subject to constraints imposed by the spatial layout of sources $S$ and consumers in $V$. In particular, the network must ensure that when all links are operational, every consumer remains connected to at least one source, \textit{i.e.}\ $v \leftrightarrow S$ for all $v \in V$. This requirement sets a lower bound of $L = N$ links, guaranteeing at least one incoming link per consumer. 

To enhance reliability we consider adding a fraction $\rho$ of redundant links, amounting to a total of $R = \rho N$ excess links, thus setting the total number of links to

\begin{equation}
L = N - 1 + R \approx (1 + \rho)N,
\label{LNR}
\end{equation}

and the subsequent cost in \eqref{eq:Cost} to $C \approx c (1 + \rho) N \av \lambda$. With this, minimizing the overall cost is primarily achieved by maintaining a small $R$ (or small $\rho$), and, as much as possible, also by avoiding excessively long links (small $\lambda_l$).

\vspace{3mm}
\textbf{\color{blue} \Large Results - the New York City network}

\begin{figure}
\begin{centering}
\includegraphics[width=0.87\linewidth]{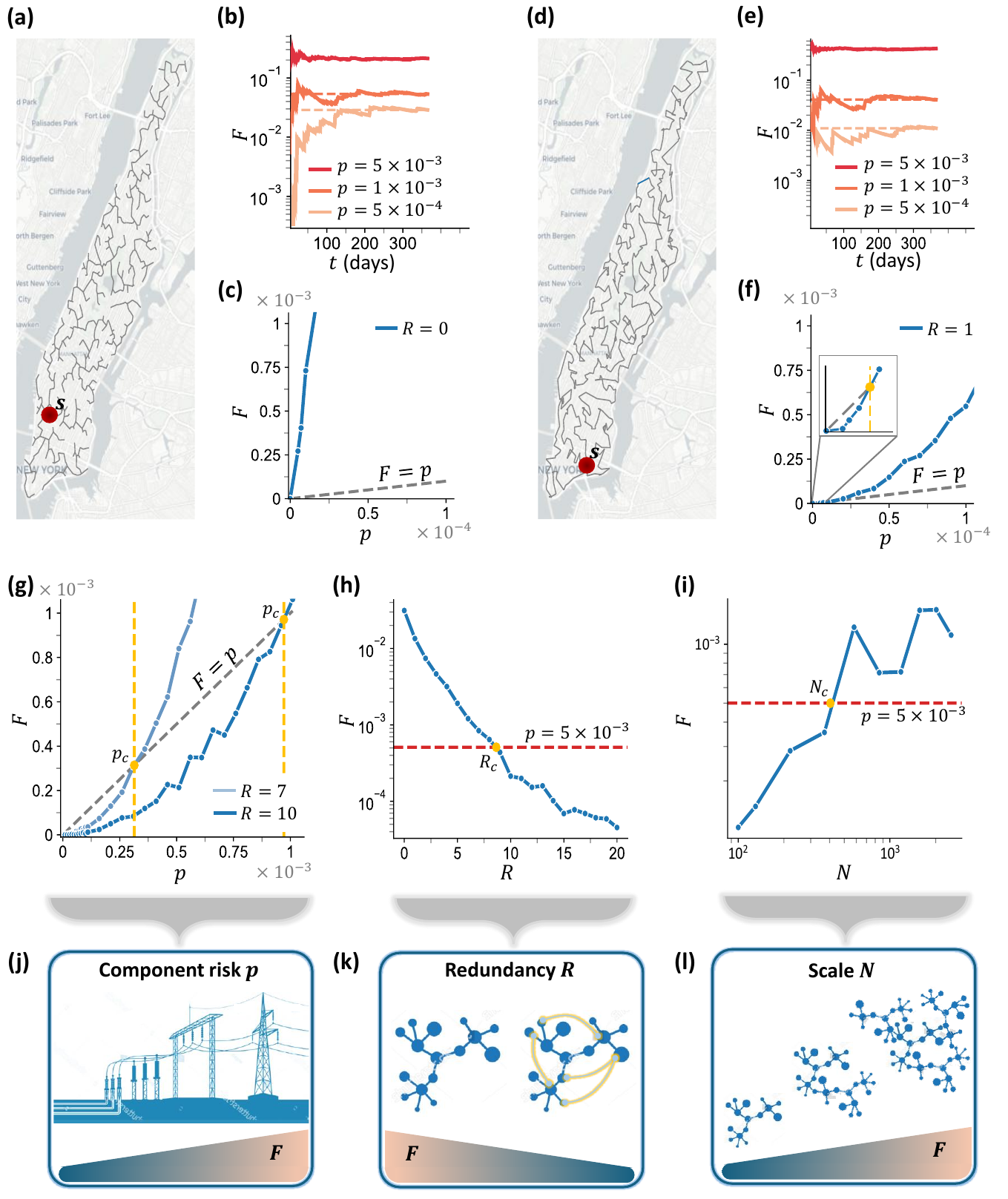}
\par\end{centering}
\vspace{-4mm}
\caption{\footnotesize \textbf{\color{blue} Power network layout in New-York City (NYC)}.\ 
\textbf{a} We used a minimum spanning tree to layout an optimally efficient network, supplying power to $N = 600$ nodes from a single source $s$ (red dot). Normalizing all link lengths to $\av \lambda  = 1$, and setting the cost per unit length to $c = 1$, the total network cost is $C = 600$.\ 
\textbf{b} The fraction of disconnected nodes $F$ vs.\ time, as obtained from numerical simulations under three values of component risk $p$ (solid lines). The average value of $F$ is also shown (dashed lines). For example, setting $p = 10^{-4}$, a low risk scenario (light red) we have $F \to 2.5 \times 10^{-2}$.
\textbf{c} $F$ vs.\ $p$ for the NYC network in \textbf{a} (blue solid line). The network risk rises sharply as $p$ increases. The $F = p$ line (dashed) represents the case where the \textit{network} is as risky as its individual \textit{components}. The minimal-cost network in \textbf{a} is found to be consistently above this line.
\textbf{d} We redesigned the NYC network to be $2$-connected, such than no single link failure can lead to service loss. This was achieved by adding just $R = 1$ redundant links.
\textbf{e} $F$ vs.\ $t$ (solid lines) and the average SAIDI (dashed) for the $2$-connected configuration. 
\textbf{f} $F$ vs.\ $p$ is now of quadratic form (blue solid line). For small $p$, $F$ resides below the $F = p$ line (inset). However, as $p$ increases, we observe a critical $p_c$ (yellow dot) above which the network enters the $F > p$ regime. The $p \le p_c$ region captures the network's \textit{reliability window}. 
\textbf{g} To enhance reliability we added a set of $R > 1$ redundant links, showing $F$ vs.\ $p$ for $R = 7$ and $10$ (blue solid lines). The increased redundancy expands the reliability window, but still has $F > p$ for $p > p_c$ (yellow dashed lines). 
\textbf{h} $F$ vs.\ $R$ (blue) under $p = 5 \times 10^{-3}$. We observe a critical redundancy $R_c$ above which $F$ dips below $p$ (red dashed line). For $R < R_c$ the network remains in the $F > p$ regime.
\textbf{i} We gradually increased the number of nodes $N$, adding sources accordingly, and measured $F$ vs.\ $N$. We find that network size plays a destabilizing role. 
The three factors controlling $F$:\ 
\textbf{j} Component risk $p$ describes the reliability of the links themselves, affected \textit{e.g}., by underground (left) vs. above ground (right) layout; $F$ increased with $p$. 
\textbf{k} Redundancy $R$ describes the available budget for link enrichment (yellow); $F$ decreases with $R$. 
\textbf{l} Scale $N$ denotes the number of consumers; $F$ increases with $N$.
}
\label{FigNYC}
\end{figure}

To illustrate the challenge, we begin with a simulation of power distribution in New York City (NYC). We first construct the network $G$ for a sample of nodes (Fig.\ \ref{FigNYC}a), comprising $N = 600$ consumers in Manhattan, served by a single source, $s = 1$ (red). To construct the most cost-effective network we build the minimal spanning tree, which connects all vertices with exactly $N$ local links at average length $\av \lambda = 1$. This construction strictly avoids any link redundancies - therefore $R = 0$. We set the component risk to $p$ and the node weights uniformly to $w_v = 1$, then numerically evaluate $F$ via \eqref{SAIDI}. The results for selected values of $p$ are presented in Fig.\ \ref{FigNYC}b (solid lines). The average value of $F$ is also shown (dashed lines). 

In Fig.\ \ref{FigNYC}c we show $F$ vs.\ $p$ as obtained from our numerical simulations. We observe that $F$ rises sharply with increasing $p$, indicating that the zero-redundancy network is potentially unreliable. For comparison, we also show the $F = p$ line (grey), which captures a state in which the \textit{network} is as reliable as its \textit{components}. To interpret this, consider a set of $N$ isolated nodes, each independently linked to the source $s$ via links of unit length. At any given time, on average, a fraction $p$ of these nodes will be denied service, and hence for this decomposed network we have $F = p$. The fact that our network features $F > p$, above this line, indicates that, in this case, the network has a destabilizing effect:\ consumers are denied service more frequently as part of the network than they would be if independently connected to $s$. In simple terms, in this regime the \textit{network} risk ($F$) is greater than the \textit{component} risk ($p$). 

This observation is expected. Under $R = 0$, most nodes can reach the source $s$ only through a series of intermediate links. Hence failing any of those links along the $v \leftrightarrow s$ path is sufficient for $v$ to lose service. Another way to interpret this is to note that, in such a sparse $G$, the failure of a single link often disconnects many nodes - hence, quite naturally, we observe $F > p$.

To enhance robustness, we reconfigure the system into a $2$-connected topology by forming a single closed loop that traverses all $N$ nodes (Fig.\ \ref{FigNYC}d). In this configuration, each node can reach $s$ along two disjoint directions, therefore single link failures no longer result in service loss. This wiring requires $L = N$ links, a marginal redundancy of $R = 1$. Still, even with this minor addition, we observe an improved robustness:\ the $F$–$p$ curve now falls below the grey reference line (Fig.\ \ref{FigNYC}f, see inset), revealing a window of $p$ values for which $F < p$. Within this region, the network, indeed, performs better than its individual components.

The observed improvement, however, is limited. As $p$ increases beyond a threshold value $p_c$, $F$ rapidly crosses the reference line, and the network, once again, becomes unreliable. Increasing $R$ expands the reliability window (Fig.\ \ref{FigNYC}g), pushing $p_c$ upwards, but nonetheless, it remains bounded (vertical dashed lines). In all cases, once $p > p_c$, the system enters the $F > p$ \textit{runaway} zone - a state where the network injects instability and $F$ rapidly increases.

This illustrates the trade-off between $p$ and $R$:\ to ensure network reliability, we must avoid the $F > p$ regime. We can achieve this either by enhancing the \textit{component} reliability, pushing $p$ below $p_c$, or by bolstering the \textit{network} reliability, adding redundancy $R$ and thus driving $p_c$ above $p$. The former approach requires investing resources in the network components, such as maintenance to decrease the failure risk $p$. The latter incurs additional construction costs due to the inclusion of redundant links. 

To demonstrate this trade-off in our NYC-based network, we fix the component risk at $p = 5 \times 10^{-4}$ and search for the minimal redundancy $R$ that ensures $F < p$ (Fig.\ \ref{FigNYC}h, blue). At $R = 0$, the network yields $F = 3 \times 10^{-2}$, significantly exceeding $p$. As we gradually add redundant links, $F$ begins to decline, as expected. We find that with a relatively modest redundancy, around $R \approx 10$, or $\rho \approx 2 \times 10^{-2}$, $F$ drops below the component failure rate, entering the $F < p$ region. This identifies a critical redundancy threshold $R_c$, beyond which the network shifts from playing a destabilizing role to providing stability.        

The difficulty emerges as the network is densified to serve additional consumers. In Fig.\ \ref{FigNYC}i, we progressively expand network coverage from $N = 10^2$, spanning only a small segment of Manhattan, to $N = 3 \times 10^3$, covering the entire region. As the network grows, additional sources (red) are introduced to meet the rising demand. Despite this, we observe that $F$ increases with $N$, revealing a critical network size $N_c$ beyond which the network becomes unreliable. Hence, network size itself plays a destabilizing role.

Taken together, our analysis of the NYC power network illustrates the inherent challenge of achieving cost-effective robust infrastructure design. Specifically, it reveals that
$\bullet$
For a given component risk $p$, network reliability can be improved by introducing a modest fraction $R$ of redundant links
$\bullet$ 
This redundancy suppresses $F$, but only within a finite reliability window $p \le p_c$. Beyond this window the system enters the $F > p$ runaway regime - an unstable state in which reliability rapidly deteriorates
$\bullet$
Most critically, network size $N$ itself acts as a destabilizing factor: designs that perform well at small scale can become unreliable as the network expands. This casts severe limitations on our ability to construct reliable networks at scale. 

To address these challenges, we next turn to a systematic analysis of the interplay between the three principal factors that govern network reliability:\ component durability ($p$, Fig.\ \ref{FigNYC}j), link redundancy ($R,\rho$, Fig.\ \ref{FigNYC}k), and network size ($N$, Fig.\ \ref{FigNYC}l). We seek optimal construction strategies that support large-scale ($N \gg 1$), low-cost ($\rho \ll 1$), and highly reliable ($F \ll 1$) networks.

\vspace{3mm}
\textbf{\color{blue} \Large Analysis}

\begin{figure}
\begin{centering}
\includegraphics[width=0.7\linewidth]{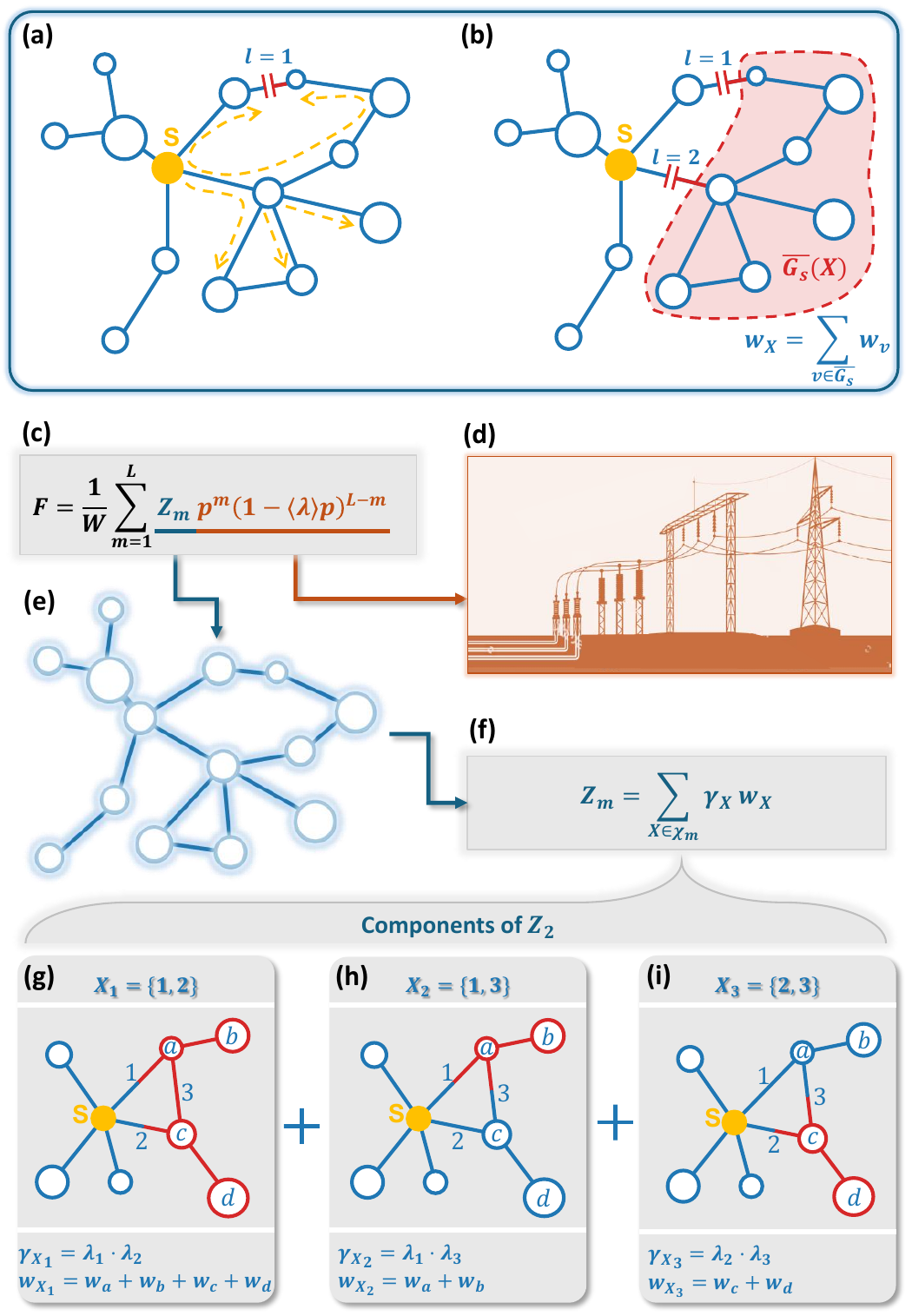}
\par\end{centering}
\vspace{0mm}
\caption{\footnotesize \textbf{\color{blue} Decomposing $F$ into its network and component driven risks}.\ 
\textbf{a} Single-source network ($S$, yellow) with one disconnected link ($l = 1$, red cut). Despite the link failure, all nodes remain connected to the source via alternative pathways (yellow dashed paths); thus, $l = 1$ does \textit{not} constitute a cut-set.\
\textbf{b} A simultaneous failure of the link set $X = \{1,2\}$ disconnects a total of $7$ nodes ($\overline{G}_S(X)$, red shaded). The set $X$ is therefore a cut-set of size $2$, with weight $w_X$ equal to the total weight of all seven disconnected nodes in $\overline{G}_S(X)$.\
\textbf{c} Summing over the contributions of all cut-sets with sizes $m = 1,\dots,L$, we arrive at the overall network risk $F$. 
The coefficients $Z_m$ depend on the network structure in \textbf{e} (blue), while the powers $p^m$ and $(1 - p)^{L - m}$ capture the contribution of the component risk, illustrated in \textbf{d} (orange). Hence, to minimize $F$ we can either invest resources in the network layout — namely, add redundancy $R$ — or enhance component durability, \textit{e.g.}, by laying links underground. Here, we take $p$ as given and focus on minimizing $Z_m$ by optimizing the network structure and layout.
\textbf{f} Components of $Z_m$:\ The coefficient $Z_m$ is driven by (i) the number of $m$-sized cut-sets, $|\X_m|$, which sets the number of terms in the summation; (ii) the link lengths within these cut-sets, encapsulated in $\gamma_X$; and (iii) the total weight of their disconnected components, $w_X$. 
\textbf{g - i} For example, in $Z_2$ we count $|\X_m| = 3$ distinct $2$-link cut-sets, $X_1$, $X_2$, and $X_3$. For each cut-set we  mark its disconnected component (red), and explicitly calculate its $\gamma_X$ and $w_X$. For instance, $X_1$ represents the simultaneous failure of links $1$ and $2$ thus $\gamma_{X_1} = \lambda_1 \lambda _2$. This cut-set results in service denial to nodes $a,b,c,d$, leading to a total disconnected weight of $w_{X_1} = w_a + w_b + w_c + w_d$. 
}
\label{FigCutSets}
\end{figure}

\textbf{\color{blue} Evaluating SAIDI}.\ 
As illustrated in Fig.\ \ref{FigCutSets}a, sporadic link failures do not necessarily lead to service denial. A failed link $j \nleftrightarrow v$ can often be bypassed, allowing $v$ to remain connected to the sources through an alternative path $S \leftrightarrow v$. The dominant contributions to $F$ therefore arise from \textit{cut-sets} — sets of links $X \subseteq E$ whose simultaneous failure disconnects at least one node from $S$. Each cut-set partitions $G$ into two regions: the subgraph $G_S(X) \subset G$ that remains connected to $S$, and the complementary region $\overline{G}_S(X) = G / G_S(X)$ containing the disconnected nodes. An example of a cut-set with $|X| = 2$ is shown in Fig.\ \ref{FigCutSets}b. 

We denote the set of all cut-sets by $\X = \{ X \subseteq E \mid \overline{G}_S(X) \ne \emptyset \}$, and the set of $m$-sized cut-sets by $\X_m = \{ X \in \X \mid |X| = m \}$. Therefore, we have $\X = \bigcup_{m = 1}^L \X_m$, a union of all equal sized cut-sets. In Supplementary Section~2.1 we use this decomposition to write~\supercite{rodionov2016practical}

\begin{equation}
F \approx \dfrac{1}{W} \sum_{m = 1}^L Z_m p^m (1 - \av \lambda p)^{L - m}
\label{eq:SAIDIfGX}
\end{equation}

a binomial expansion with coefficients

\begin{equation}
Z_m = \sum_{X \in \X_m} \gamma_X w_X.
\label{Zm}
\end{equation}

The expansion in \eqref{eq:SAIDIfGX} constructs $F$ as a probabilistic decomposition. The powers $p^m(1 - \av \lambda p)^{L - m}$ represent the probability of failure of an $m$-link cut-set $X \in \X_m$. The coefficients $Z_m$ capture the contribution of these $m$-sized cut-sets to $F$:\ $\gamma_X = \prod_{l \in X} \lambda_l$ accounts for the lengths of the links in $X$, reflecting the increase of failure probability with $\lambda_l$, and $w_X = \sum_{v \in \overline{G}_S(X)} w_v$ measures the total weight of the nodes disconnected by $X$.

Equation~\eqref{eq:SAIDIfGX} encapsulates our first key insight, disentangling risk embedded in the \textit{network structure} from that rooted in \textit{component durability} (Fig.\ \ref{FigCutSets}c-f). Structural vulnerability is fully encoded in the coefficients $Z_m$, whereas the reliability of individual components enters solely through the powers $p^m$. In the limit of small $p$, the expansion is dominated by its lowest-order terms, implying that reducing $F$ requires minimizing $Z_m$ for $m = 1,2,\dots$. Ideally, one seeks network architectures for which these coefficients vanish at small $m$, by promoting structures whose minimal cut-sets involve many links, such that $\X_m = \emptyset$ for small $m$.

Next, we use Eq.\ \eqref{eq:SAIDIfGX} to analyze a set of archetypical network structures, often encountered in infrastructure settings.

\begin{figure}
\begin{centering}
\includegraphics[width=0.95\linewidth]{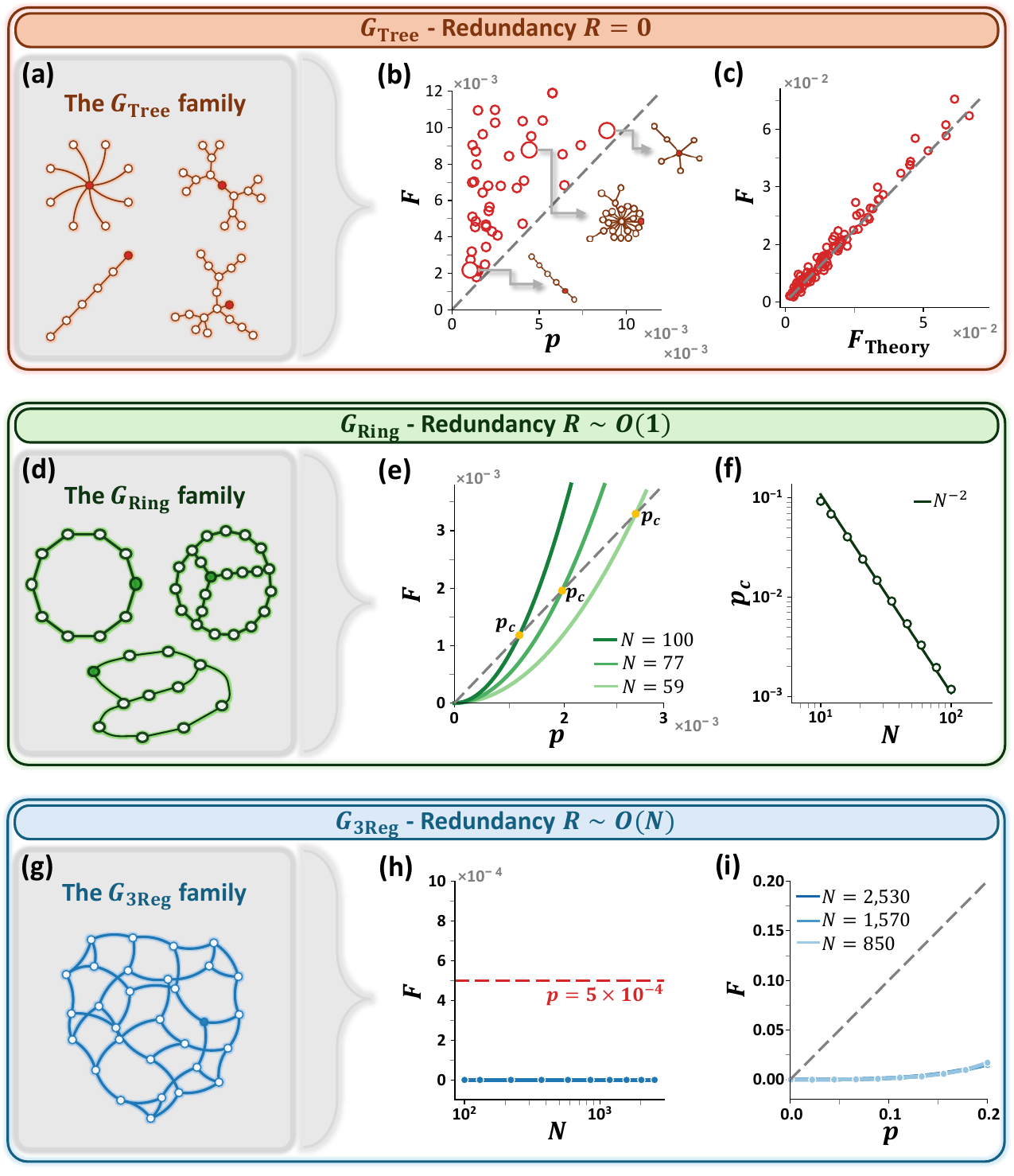}
\par\end{centering}
\vspace{-4mm}
\caption{\footnotesize \textbf{\color{blue} Taxonomy of network archetypes}.\
\textbf{a} Zero-redundancy networks $G_{\text{Tree}}$ have exactly $L = N - 1$ links and $R = 0$. Such networks are tree-like and therefore lack loops.\
\textbf{b} We generated $100$ zero-redundancy networks of varying size and structure, and assigned their component risk uniformly at random in the range $0 \le p \le 10^{-2}$. For each network we measured $F$ numerically and display $F$ versus $p$. For three representative networks we show the network structure explicitly. We find that almost $100\%$ of the time the risk reside above the $F = p$ line (dashed). Hence, the $G_{\text{Tree}}$ family is inherently unstable.\
\textbf{c} Observed SAIDI $F$ versus the theoretically predicted $F_{\text{Theory}}$, as obtained from Eq.\ \eqref{FTree}. The scatter observed in panel \textbf{b} collapses tightly around the $x = y$ line (dashed), corroborating the theoretical prediction.\
\textbf{d} $2$-connected networks are designed to have no cut-sets smaller than $m = 2$. These networks require a marginal redundancy of $R \sim O(1)$, with the most extreme case represented by the ring network $G_{\text{Ring}}$, achieved with $R = 1$.\
\textbf{e} $F$ versus $p$ as obtained for rings of varying size $N$ (green shades). $G_{\text{Ring}}$ exhibits a window of reliability, but at $p = p_c$ (yellow dot) it crosses the $F = p$ line (grey dashed) and enters the unreliable regime. The width of the reliability window narrows as $N$ increases.\
\textbf{f} The critical component risk $p_c$ versus $N$, as obtained from numerical simulations of $G_{\text{Ring}}$ (circles). The theoretical prediction of \eqref{FOpt} is also shown (solid line).\
\textbf{g} We consider the $3$-regular family of networks $G_{\text{3Reg}}$. This network family is practically is risk-free, featuring $F \sim p^3 \to 0$. This robustness, however, comes at a cost of $R \sim O(N)$.\
\textbf{h} We revisit the NYC network, this time generating $3$-regular network layouts. As predicted by \eqref{F3Regular}, $F$ is now independent of $N$ (blue), as hence in $G_{\text{3Reg}}$ network size no longer plays a destabilizing role. The component risk $p$ is shown in red.\
\textbf{i} $F$ versus $p$ for $3$-regular networks of varying size $N$ (blue shades). We observe $F$ consistently below the $F = p$ threshold.
}
\label{FigNetworks}
\end{figure}

\vspace{3mm}
\textbf{\color{blue} Zero-redundancy networks $G_{\rm Tree}$} (Fig.\ \ref{FigNetworks}a–c).
We begin by setting $R = 0$, corresponding to the minimal network deployed in Fig.\ \ref{FigNYC}a. This case captures tree-like structures with no loops, including the star configuration, generic trees, and the linear chain illustrated in Fig.\ \ref{FigNetworks}a. With $R = 0$, each of the $N$ links forms a cut-set, and Eq.\ \eqref{eq:SAIDIfGX} is therefore dominated by the leading $Z_1$ term, which is linear in $p$. In Supplementary Section~2.3.1 we show that for this network family

\begin{equation}
F_{\rm Tree} = \av {w_v \lambda_{v \to S}}\cdot p + O(p^2),
\label{FTree}
\end{equation}

where $\lambda_{v \to S}$ denotes the total length of all links along the shortest path from $v$ to $S$, and $\av{w_v \lambda_{v \to S}} = (1/W) \sum_{v = 1}^N w_v \lambda_{v \to S}$ its node-weighted average.

For $p \to 0$ the linear network may seem robust. However, since $\av {w_v \lambda_{v \to S}}$ is typically greater than unity, such tree-like networks almost always reside in the $F \ge p$ regime — precisely the runaway zone that we seek to avoid.

In Fig.\ \ref{FigNetworks}b we systematically analyze $10^2$ zero-redundancy networks spanning a range of sizes and connectivity patterns. For each network, the component risk is drawn randomly from $p \in (10^{-3},10^{-2})$. As predicted, we find that for nearly all realizations $F$ lies above the $F = p$ diagonal. To test Eq.\ \eqref{FTree}, we next plot $F$ against $\av{w_v \lambda_{v \to S}}p$. Strikingly, the simulated data — previously scattered across the entire $F \ge p$ region — collapse onto the predicted linear function (Fig.\ \ref{FigNetworks}c, dashed line). We thus arrive at two conclusions:\ (i) zero-redundancy networks are intrinsically unreliable; (ii) this reliability risk is accurately captured by our prediction \eqref{FTree}.

\vspace{3mm}
\textbf{\color{blue} $2$-connected networks $G_{\rm Ring}$} (Fig.\ \ref{FigNetworks}d–f).
A natural strategy to mitigate the vulnerability of $G_{\rm Tree}$ is to introduce a small number of redundant links, $R \sim O(1)$. For clarity, we consider the limiting case $R = 1$, corresponding to a single redundant link. In this limit, the objective of eliminating small cut-sets is optimally achieved by the ring configuration $G_{\rm Ring}$ shown in Fig.\ \ref{FigNetworks}d. In $G_{\rm Ring}$ there are no $m = 1$ cut-sets, therefore $\X_1 = \emptyset$ and $Z_1 = 0$. Consequently, $F$ in Eq.\ \eqref{eq:SAIDIfGX} is dominated by the $p^2$ term. The number of $m = 2$ cut-sets is $|\X_2| = \binom{L}{2} \sim L^2$, and their average weight is $w_X = W/3$. Substituting into Eq.\ \eqref{eq:SAIDIfGX} yields

\begin{equation}
\label{FRing}
F_{\rm Ring} \sim L^2 p^2 + O(p^3) \approx p^2 \lambda_{\text{Net}}^2,
\end{equation}

where, on the r.h.s., we replace the number of links $L$ by their total length $\lambda_{\text{Net}}$.

In Eq.\ \eqref{FRing}, the quadratic dependence on $p$ ensures that for $p \ll 1$ we have $F_{\rm Ring} < p$. However, as $L$ increases, the system again reaches $F > p$, entering the runaway regime. For a given $N$, Eq.\ \eqref{FRing} predicts that this transition occurs at $p_c \sim N^{-2}$ (since in $G_{\rm Ring}$, $N \approx L$). Thus, while the ring network provides a finite window of reliability, this window narrows as $N$ increases.

In Fig.\ \ref{FigNetworks}e, we measure $F_{\rm Ring}$ for rings of varying size $N$. For each ring, we identify $p_c$, as the point at which $F_{\rm Ring}$ intersects the $F = p$ line. This defines the window of reliability $p < p_c$. Plotting the extracted $p_c$ versus $N$, we find that, as predicted, that the window size scales as $N^{-2}$ (Fig.\ \ref{FigNetworks}f). Thus, larger networks exhibit a diminished stability window. 

These results, though derived from simplified network structures, reproduce the $F$ patterns observed in our NYC analysis. The network remains robust for small $p$, but once $p > p_c$ reliability degrades rapidly. The qualitative trends in Fig.\ \ref{FigNYC}f,g are thus supported by quantitative, testable predictions. Most importantly, the analysis confirms our initial insight that increasing system size $N$ drives $p_c$ toward zero, thereby narrowing the reliability window.

\vspace{3mm}
\textbf{\color{blue} $3$-regular networks $G_{\rm 3Reg}$} (Fig.\ \ref{FigNetworks}g–i).
We next consider $3$-connected networks in which all nodes have degree $k = 3$. This configuration requires $L = 3N/2$ links, a redundancy of $R = 1 + N/2$ and $\rho \approx 1/2$. In Supplementary Section~2.3.3, we use Eq.\ \eqref{eq:SAIDIfGX} to show that, with an appropriate structural arrangement,

\begin{equation}
F_{3\rm Reg} \sim N^0 p^3 + O(p^4),
\label{F3Regular}
\end{equation}

exhibiting cubic dependence on $p$ and no polynomial dependence on $N$. This represents an ideal configuration:\ the $p^3$ scaling guarantees $F_{3\rm Reg} \ll p$, while the independence from $N$ ensures that the reliability window remains constant with system size. In other words, in the $3$-regular network we have $p_c = 1$, providing network-driven reliability across all scales.

The robustness of $F_{3 \rm Reg}$ follows naturally from Eq.\ \eqref{eq:SAIDIfGX}. In $G_{\rm 3Reg}$, no cut-sets exist with size smaller than $m = 3$, so that $Z_1 = Z_2 = 0$. Unlike tree networks — where each of the $N$ links forms a cut-set, or ring networks — where every one of the $\sim N^2$ link pairs is a cut-set, here - most link triplets are \textit{not} cut-sets. Consequently, of the $\sim N^3$ possible link trios, only a handful pose a service risk. As a result, the total number of \textit{relevant} cut-sets in $\X_3$, that contribute to $Z_3$ remains nearly independent of $N$. The outcome - an ideally robust network, which maintains reliability regardless of system size.

To test this, we apply the $3$-regular wiring scheme to the NYC network. As predicted, unlike $G_{\rm Tree}$ and $G_{\rm Ring}$, $F$ under this construction does not increase appreciably with $N$, consistently remaining well within the $F < p$ domain (Fig.\ \ref{FigNetworks}h,i).

The $3$-regular networks offer ideal robustness. On one hand, $F_{3\rm Reg}$ decreases sharply with $p$; on the other, it remains independent of $N$. This robustness, however, comes at a substantial cost: we must add $R \sim O(N)$ links, increasing the network density from $L = N$ to $L = 3N/2$. This implies a linear increase in the cost \eqref{eq:Cost}, corresponding here to a $50\%$ budget boost, which may be practically prohibitive. In the following, we therefore focus on constructing networks that optimize reliability under a restricted budget.

\vspace{3mm}
\textbf{\Large \color{blue} Optimal networks}

Up to this point, we considered two limiting cases:\ $G_{\rm Ring}$, with $R \sim O(1)$, and $G_{3\rm Reg}$, where $R \sim O(N)$. We now turn to the intermediate regime by considering a finite redundancy

\begin{equation}
R \sim N^\alpha,
\label{Rho}
\end{equation}

with $0 \le \alpha \le 1$. For large $N$ and $\alpha < 1$, this scaling ensures a vanishing relative redundancy $\rho = R/N \to 0$, with smaller $\alpha$ corresponding to lower investment. Our objective is to determine the optimal network structure for a given component risk $p$, system size $N$, and available redundancy $R$ (or $\alpha$).

\begin{figure}
\begin{centering}
\vspace{-3mm}
\includegraphics[width=0.47\linewidth]{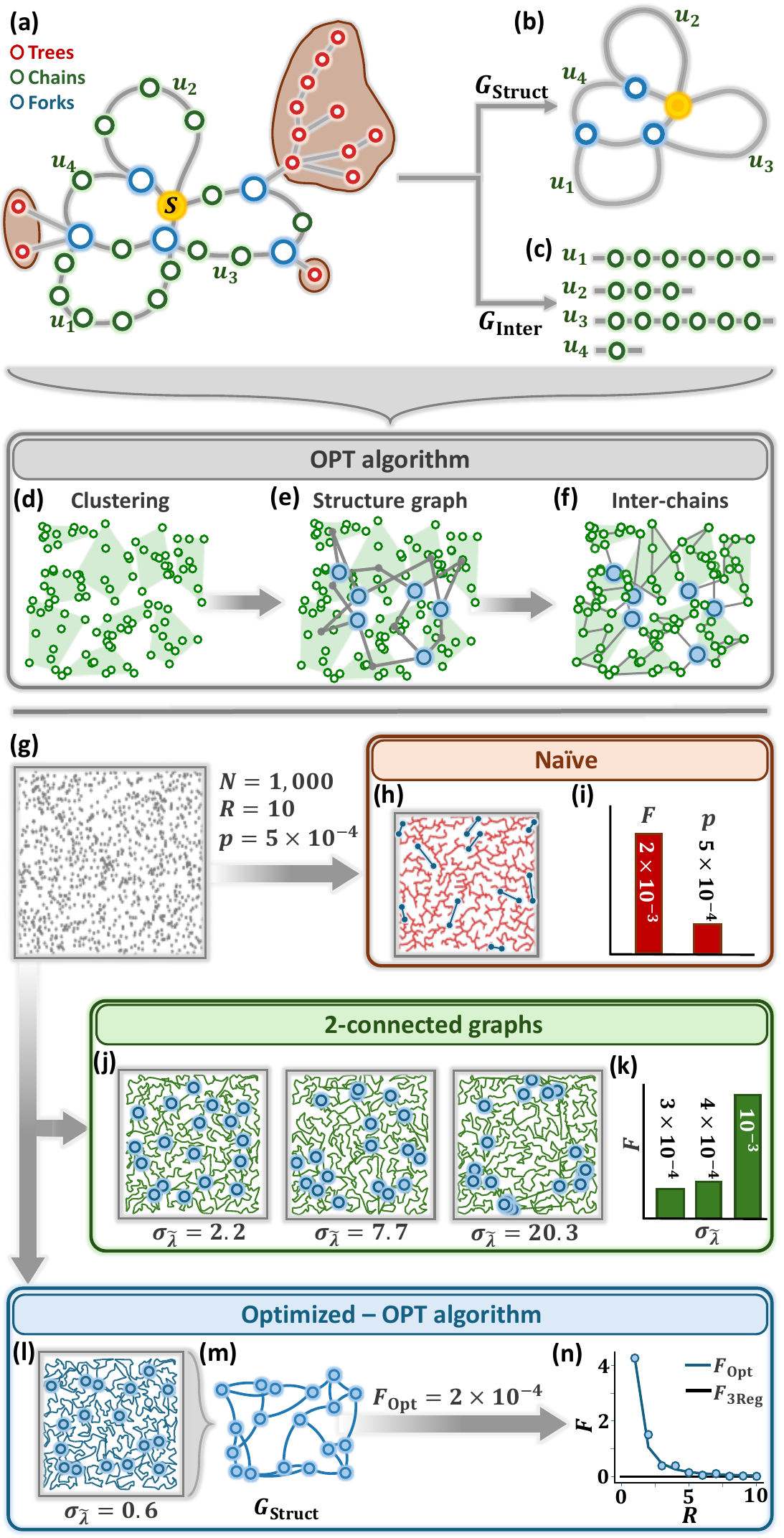}
\par\end{centering}
\vspace{-4mm}
\caption{\footnotesize \textbf{\color{blue} Network optimization – step by step} 
(Full size image appears in page 19).\
\textbf{a} We identify three types of network components:\ Forks – nodes with degree $k \ge 3$ (blue); Chains $U_q$ – sequences of nodes with $k = 2$ that link two fork nodes (green); Trees – radial motifs that lack loops (red). Trees introduce risk due to their single-link cut-sets ($Z_1 > 0$ in \eqref{Zm}). We therefore eliminate all trees in our optimized networks.\
\textbf{b} After eliminating all trees, the resulting $2$-connected graph is collapsed into $G_{\text{Struct}}$, in which all intermediate chains are reduced to single links. The effective length of these links, $\tilde \lambda_q$, captures the internal structure of the corresponding chain $U_q$, as described in Eq.\ \eqref{InterRisk}.\
\textbf{c} The intermediate chains appear in $G_{\text{Inter}}$. The resulting SAIDI $F$ can now be decomposed into $F_{\text{Struct}}$, obtained from $G_{\text{Struct}}$, and $F_{\text{Inter}} \gg F_{\text{Struct}}$, which captures the risks arising from all intermediate chains. Our optimization seeks the best selection of forks and the optimal partition into intermediate chains to minimize $F_{\text{Inter}}$.\
\textbf{d} Step I.\ Clustering.\ Partition the network into clusters $U_q$ (green shaded), that will later form the intermediate chains. We aim for a partition that groups together roughly equal node sets located within confined spatial regions. This helps construct chains with homogeneous $\tilde \lambda_q$.\
\textbf{e} Step II.\ Structure graph.\ Select the optimal set of $2R$ forks (blue nodes). Each fork links three chains, so we choose forks proximal to the circumcenters of chain trios. To construct the $3$-regular $G_{\text{Struct}}$, we link all forks through their adjacent chain clusters (grey links) .\
\textbf{f} Step III.\ Inter-chains.\ Expand the links of $G_{\text{Struct}}$ to traverse all nodes within each intermediate chain cluster at minimal wiring cost. The resulting optimized network includes $2R$ fork nodes with degree $k = 3$ (blue, large), and $3R$ intermediate chains that link parsimoniously between them.\
\textbf{g} We constructed a spatially embedded network with $N = 10^3$ nodes, redundancy $R = 10$ ($\alpha = 1/3$), and component risk $p = 5 \times 10^{-4}$. 
\textbf{h} In the Na"{i}ve construction we generate a minimum spanning tree (red) and assign the $R$ excess links (blue) to close as many loops as possible. 
\textbf{i} The resulting risk is $F \approx  2\times 10^{-3}$, significantly greater than $p$.\
\textbf{j} Using our algorithm, we construct $2$-connected networks of forks (blue) and intermediate chains (green), eliminating all trees. The network risk $F$ is dramatically reduced. However, these graphs still exhibit significant variance $\sigma_{\tilde \lambda}$ and are therefore sub-optimal. 
\textbf{k} As predicted in \eqref{FSigma}, higher $\sigma_{\tilde \lambda}$ leads to larger $F$.\
\textbf{l} Using our three-step algorithm, we construct the optimal network in which $\sigma_{\tilde \lambda}$ is minimized ($0.6$). The resulting SAIDI is $F = 2 \times 10^{-4}$, significantly below the component risk $p$me.
\textbf{m} The structure graph of the optimized network.
\textbf{n} The optimal $F_{\rm Opt}$, as obtained from our algorithm, versus $R$ (circles). The theoretical prediction of Eq.\ \eqref{FOpt} is also shown (blue solid line). Already at $R \gtrsim 5$, a small redundancy of $\rho \lesssim 1\%$, $F$ approaches the ideal $F_{3\text{Reg}}$ (black solid line).
}
\label{FigOptimization}
\end{figure}

\vspace{3mm}
\textbf{\color{blue} Eliminating trees} (Fig.\ \ref{FigOptimization}a).\
The first insight derived from Eqs.\ \eqref{eq:SAIDIfGX}–\eqref{F3Regular} is that tree-like components in $G$ constitute an intrinsic vulnerability, introducing multiple cut-sets of order $m = 1$. Our analysis shows that this vulnerability can be eliminated by introducing a marginal redundancy $R \sim O(1)$, sufficient to remove all tree structures without incurring a significant additional cost (Supplementary Section~3.1).

\vspace{3mm}
\textbf{\color{blue} Components of $2$-connected graphs} (Fig.\ \ref{FigOptimization}a,b).\
Once trees are eliminated, the optimized networks are restricted to $2$-connected configurations, which contain no cut-sets of order $m < 2$ (that is, $Z_1 = |\X_1| = 0$ in \eqref{eq:SAIDIfGX}). In such networks, nodes fall into two categories:\ nodes of degree $k \ge 3$, which act as \textit{forks} (blue), and nodes of degree $k = 2$, which assemble into $Q$ \textit{intermediate chains} $U_q$ ($q = 1,\dots,Q$) connecting these forks (green). Each intermediate chain forms a closed, $2$-connected sequence of $N_q$ nodes and $L_q = N_q + 1$ links, terminating at both ends at fork nodes. 

We use this partition to construct the \textit{structure graph} $G_{\text{Struct}}$, obtained by reducing $G$ to its fork nodes. Each intermediate chain $U_q$ is collapsed in $G_{\text{Struct}}$ into a single weighted link connecting the corresponding forks. The link weight is defined as $w_q = \sum_{v \in U_q} w_v$, equal to the total node weight along the collapsed chain, and its effective length $\lambda_q = \sum_{l \in U_q} \lambda_l$ represents the total chain length (Supplementary Section~3.2).

\vspace{3mm}
\textbf{\color{blue} SAIDI in $2$-connected graphs}.\ 
In Supplementary Section~3.2.1 we show that, for this family of networks,

\begin{equation}
F = F_{\text{Struct}} + F_{{\text{Inter}}},
\label{FBreakdown}
\end{equation}

decomposing SAIDI into contributions from the two structural components of $G$ (Fig.\ \ref{FigOptimization}b,c). The first term on the r.h.s., $F_{\text{Struct}}$, captures the contribution of the structure graph, as computed independently of the collapsed chains. The second term,

\begin{equation}
F_{{\rm Inter}} = \dfrac{1}{W} \sum_{q = 1}^Q w_q F_{{\rm Inter},q},
\label{FInter}
\end{equation}

represents the weighted average contribution of the $Q$ intermediate chains.

Equation \eqref{FBreakdown} constitutes our second key result. It provides a framework for expressing $F$ in terms of our previously analyzed building blocks. The structure graph $G_{\text{Struct}}$ contains only nodes of degree $k \ge 3$, and therefore it can be designed to obey Eq.\ \eqref{F3Regular}. The collapsed intermediate chains are $2$-connected and hence follow Eq.\ \eqref{FRing}. Together, this decomposition enables a systematic derivation of $F$ for a general $2$-connected network by reducing it to its two fundamental structural units:\ forks and intermediate chains.

The crucial point is that the two contributions in \eqref{FBreakdown} scale differently with $p$. The structure graph exhibits $F_{\text{Struct}} \sim p^3$ (Eq.\ \eqref{F3Regular}), whereas Eq.\ \eqref{FRing} predicts $F_{{\rm Inter}} \sim p^2$. Consequently, $F_{{\rm Inter}} \gg F_{\text{Struct}}$, and \eqref{FBreakdown} can be approximated as $F \approx F_{\text{Inter}}$, with the failure risk dominated by the intermediate chains.

To complete the derivation, we use Eq.\ \eqref{FRing} to express the intermediate-chain contribution $F_{{\rm Inter}}$ as (Supplementary Section~3.3)

\begin{equation}
\label{InterRisk}
F \approx F_{\text{Inter}} \sim p^2 \av {\tilde \lambda^2},
\end{equation}

where $\tilde \lambda_q = \lambda_q \sqrt{w_q}$ denotes the effective \textit{link length} associated with the collapsed chain $U_q$, and $\av {\tilde \lambda^2} = (1/Q) \sum_{q = 1}^Q \tilde \lambda_q^2$ is its second moment. Accordingly, the total link length $\lambda_{\rm Net}^2$ appearing in \eqref{FRing} for a single-ring network is replaced in \eqref{InterRisk} by the effective chain length $\tilde \lambda_q^2$, averaged over all intermediate chains.

Next, we express $\av{\tilde \lambda^2}$ in terms of the chain-length variance $\sigma_{\tilde \lambda}^2$, writing $\av{\tilde \lambda^2} = \sigma_{\tilde \lambda}^2 + \av{\tilde \lambda}^2$, and thus bringing $F$ to its final form,

\begin{equation}
\label{FSigma}
F \sim
\left(
1 + \dfrac{\sigma_{\tilde \lambda}^2}{\av {\tilde \lambda}^2}
\right) p^2 \av {\tilde \lambda}^2.
\end{equation}

This expression yields our final insight, identifying the network characteristics that control $F$. Each of the $Q$ intermediate chains is characterized by an effective length $\tilde \lambda_q$, defining a chain-length distribution $P(\tilde \lambda_q)$. From this distribution, network risk is governed by two quantities:\ the mean $\av{\tilde \lambda}$ and the variance $\sigma_{\tilde \lambda}^2$. In what follows, we use these insights to construct minimal-risk networks.

\vspace{3mm}
\textbf{\color{blue} Constructing optimized networks}.
Equation~\eqref{FSigma} shows that minimizing $F$ requires both short chains (small $\av {\tilde \lambda}$) and low chain-length variance (small $\sigma_{\tilde \lambda}^2$). We therefore, first maximize $G_{\rm Struct}$ to increase the number of forks, leaving fewer nodes per intermediate chain. We then assign the remaining non-fork nodes to chains, aiming to produce a narrow distribution $P(\tilde \lambda)$, thereby minimizing $\sigma_{\tilde \lambda}^2$ and reducing the network risk.

Under a given redundancy $R$, the maximal $G_{\rm Struct}$ contains exactly $2(R - 1) \approx 2R$ fork nodes, each with degree $k = 3$. Nodes with higher degree ($k = 4,5,\dots$) reduce the size of $G_{\rm Struct}$, since the total number of redundant links is fixed. Thus, the most effective use of the $R$ redundant links is to create exclusively degree-three forks. Once the $2R$ fork nodes are selected, the remaining $N - 2R$ nodes are distributed among the $3(R - 1) \approx 3R$ resulting intermediate chains, averaging $(N - 2R)/3R \approx N/3R$ nodes per chain. The challenge is then to find the optimal partition of $G$ into $2R$ forks and $3R$ chains that minimizes both $\av{\tilde \lambda}$ and $\sigma_{\tilde \lambda}^2$.

In Supplementary Section~6, we introduce the Optimal Path Tearing (OPT) algorithm to implement this strategy. Illustrated in Fig.~\ref{FigOptimization}d-f, the algorithm takes as input the spatial configuration of all nodes and the allowed redundancy $R$, and performs a controlled tearing of the network into $3R$ intermediate chains and $2R$ fork nodes, producing the optimal partition that minimizes both average chain length and variance 
$\bullet$
\textbf{Step I.\ Intermediate chain-clustering}.\
First, all nodes are partitioned into $3R$ spatially confined clusters $q$, each of which will form an intermediate chain $U_q$. The clustering is optimized to produce chains with approximately homogeneous $\tilde \lambda_q$ 
$\bullet$ 
\textbf{Step II. Structure graph construction}.
Next, we identify the optimal set of $2R$ fork nodes. Each fork connects three chains, so we select forks located near the circumcenters of neighboring cluster trios. Each fork is then linked to three other forks via its adjacent intermediate chains, forming the network’s structure graph 
$\bullet$
\textbf{Step III. Chain construction}.
After constructing $G_{\rm Struct}$, each link is expanded into a linear chain connecting all nodes within the corresponding cluster. Links in each cluster are arranged to traverse all the cluster nodes with minimal total link length.

The resulting network consists of a highly resilient $2R$-node structure graph, connected by $3R$ chains of roughly uniform length. The resulting network SAIDI closely matches that of the average chain (Eq.~\eqref{InterRisk}), and therefore the chains’ properties effectively determine the overall risk. Each chain, a $2$-connected subgraph of $\sim N/3R$ nodes, follows Eq.~\eqref{FRing}, hence its average risk scales as

\begin{equation}
\label{FOpt}
F \sim \frac{p^2 N^2}{R^2}.
\end{equation}

Thus, under this optimal construction, network risk decays quadratically with link redundancy $R$, reflecting the maximal benefit attainable from each added link.

\vspace{3mm}
\textbf{\color{blue}Testing the optimized networks}.\
We evaluate our network optimization for $N = 10^3$ consumer nodes and $R = 10$ excess links, corresponding to a redundancy of $\alpha = 1/3$ in \eqref{Rho}. We begin with a na\"{i}ve deployment, connecting all $N$ nodes via a random spanning tree and then adding the $R$ redundant links to close as many loops as possible (Fig.~\ref{FigOptimization}h,i). This network still contains tree-like subgraphs, which results in a relatively large $F$ due to the abundance of $|X| = 1$ cut-sets.

To mitigate this vulnerability, we eliminate trees by constructing a strictly $2$-connected network (Fig.~\ref{FigOptimization}j). We then introduce $2R$ fork nodes (blue) to deploy the redundancy. At this stage, the chains are not yet optimized for length homogeneity, producing a substantial variance, $\sigma_{\tilde \lambda} = 2.2,7.7$ and $20.3$. While these networks perform better than the na\"{i}ve configuration, they remain sub-optimal due to the chain length heterogeneity. As predicted by our analysis, $F$ increases in networks with greater heterogeneity (Fig.~\ref{FigOptimization}k). 

Finally, in Fig.~\ref{FigOptimization}l,m, we implement the OPT algorithm, following steps (i)–(iii), as illustrated in Fig.\ \ref{FigOptimization}d-f. We designate $2(R - 1) = 18$ fork nodes with degree $k = 3$ (blue) and assign the remaining $982$ nodes to $3(R - 1) = 27$ intermediate chains of (roughly) equal length. This construction achieves a minimized $\sigma_l^2$ of $0.6$ and yields a SAIDI index of $F \approx 2 \times 10^{-4}$,  lower than any of the other equal-redundancy network considered. These results demonstrate that our protocol effectively minimizes the network’s failure risk.

To further evaluate our optimization, we varied $R$ and measured the resulting $F$ as obtained from the OPT algorithm (Fig.~\ref{FigOptimization}n, circles). The observed values closely follow the theoretical prediction of Eq.~\eqref{FOpt} (blue solid line). Remarkably, already at $R \gtrsim 5$, the optimized network achieves a SAIDI comparable to that of the ideal $G_{3\text{Reg}}$ (black solid line). This is despite the fact that $G_{3\text{Reg}}$ employs two orders of magnitude more redundancy.

Hence, our formulation provides a clear recipe to balance cost and risk: given the network size $N$, component durability $p$, and allowed redundancy $R$, it generates the optimal network $G$ that ensures the lowest achievable SAIDI $F$.     

\vspace{3mm}
\textbf{\color{blue} Optimization boundaries}.
Our analysis of the NYC network indicates that as $N$ increases, the network crosses the critical point into the $F > p$ regime. Taking $F$ from \eqref{FOpt} and $R$ from \eqref{Rho}, the optimal construction gives $F \sim p^2 N^{2(1 - \alpha)}$. At criticality ($F = p$), this yields

\begin{equation}
p_c \sim N^{-\beta},
\label{PcOpt}
\end{equation}

where $\beta = 2(1 - \alpha)$. In the limit $\alpha = 0$ ($R \sim O(1)$), we recover $\beta = 2$, as observed for the $G_{\text{Ring}}$ configuration (Fig.~\ref{FigNetworks}f). In the opposite limit, $\alpha = 1$ ($R \sim O(N)$), we obtain $\beta = 0$ and $p_c \to 1$ - indeed, consistent with $3$-connected networks. Between these extremes ($0 \le \alpha \le 1$), Eq.~\eqref{PcOpt} delineates how the bounds of the reliability window are set by $N$ and $R$ (or $\alpha$).

\begin{figure}
\begin{centering}
\includegraphics[width=0.9\linewidth]{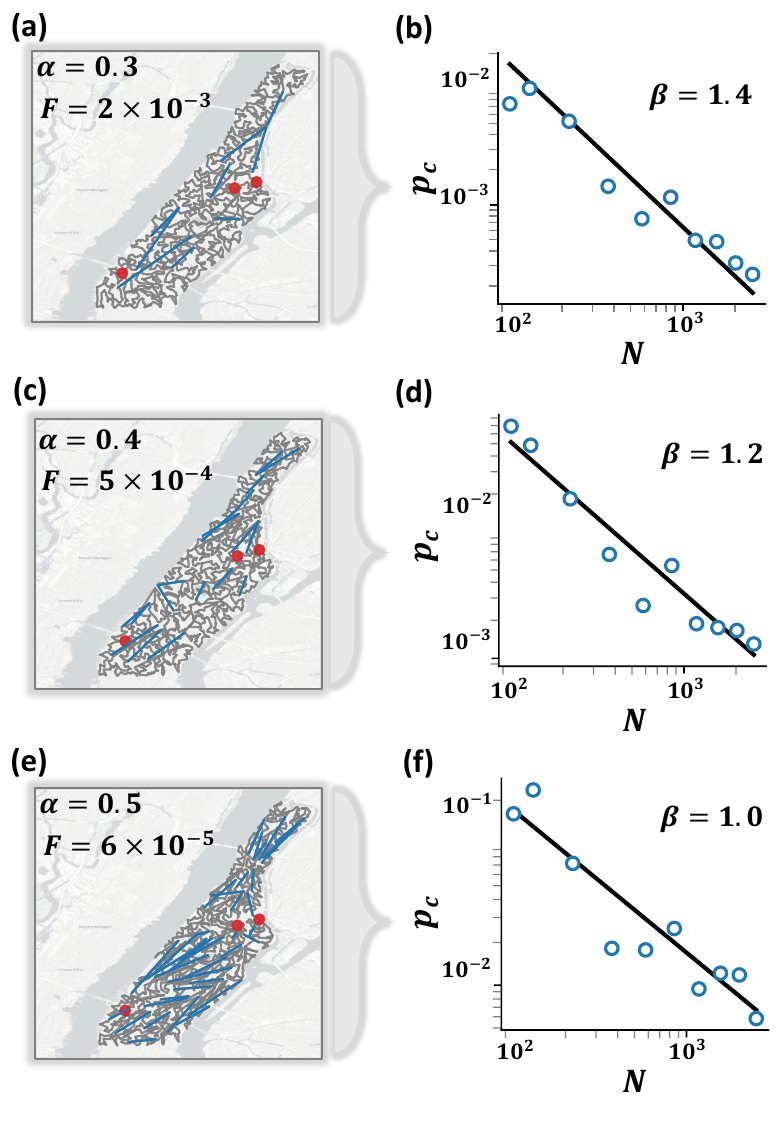}
\par\end{centering}
\vspace{-5mm}
\caption{\footnotesize \textbf{\color{blue} Cost-effective optimization of the NYC network}.\
\textbf{a}-\textbf{c} We apply our algorithm to construct optimized power networks for the NYC spatial node layout under three levels of redundancy:\ $\alpha = 0.3, 0.4$ and $0.5$ in \eqref{Rho}. For each network we show the resulting SAIDI $F$, ranging from $2 \times 10^{-3}$ for the smallest redundancy, to $6 \times 10^{-5}$ for the most high-cost construction.\
\textbf{d} The critical component risk $p_c$ vs.\ $N$ as obtained from the NYC network with $\alpha = 0.3$ (circles). The numerical results closely follow the theoretical scaling of Eq.\ \eqref{PcOpt} with $\beta = 1.4$, as predicted (solid line).\
\textbf{e}-\textbf{f} Corresponding results for $\alpha = 0.4$ ($\beta = 1.2$) and $0.5$ ($\beta = 1$).
}
\label{FigNYCOpt}
\end{figure}

This represents our final result, capturing the impact of $N$ on network reliability. Larger networks exhibit a narrower stability window, as $p_c$ decreases with $N$, in effect, quantitatively corroborating the qualitative trends observed in Fig.~\ref{FigNYC}i. Importantly, this window can be widened by controlling the scaling exponent $\beta$ through the optimal deployment of the $N + R$ available links. To examine this, we revisit the NYC network of Fig.~\ref{FigNYC}, constructing its optimal $G$ under redundancy levels $\alpha = 0.3, 0.4,$ and $0.5$. In Fig.~\ref{FigNYCOpt}b,d,f we plot $p_c$ versus $N$ (circles) for all three cases. The data align closely along three distinct slopes, $\beta = 1.4, 1.2,$ and $1$ (solid lines), in excellent agreement with the predictions of Eq.~\eqref{PcOpt}.

These results provide direct guidance for network planning, highlighting the trade-off between component durability $p$ and network reliability $F$. For a fixed $p$, the network must remain within the stability window, $p < p_c$, imposing an upper bound on $\beta$ in \eqref{PcOpt}, which translates to a minimum redundancy, $\alpha = 1 + \ln p / 2\ln N$, in \eqref{Rho}. Conversely, if $R$ is fixed, reflecting a constrained budget, $p$ must satisfy $p < N^{-\beta}$, necessitating investment in component durability, \textit{e.g}., laying wires underground.

\vspace{3mm}
\textbf{\color{blue} \Large Empirical network optimization}
 
\begin{figure}
\begin{centering}
\includegraphics[width=0.92\linewidth]{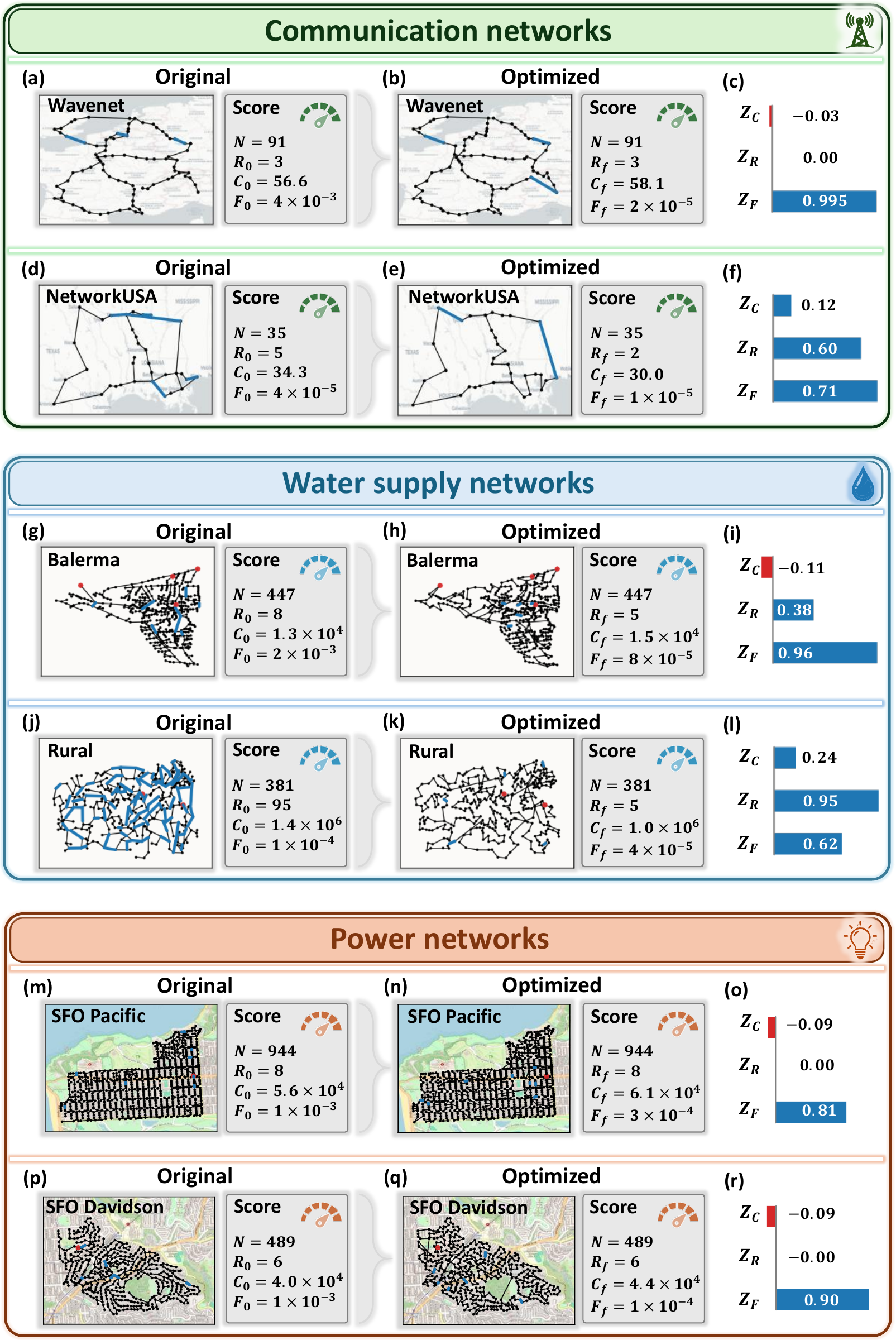}
\par\end{centering}
\vspace{-2mm}
\caption{\footnotesize \textbf{\color{blue} Optimizing real-world infrastructure networks}.\
\textbf{a} The Wavenet communication network, with $N = 91$ nodes and $R = 3$ redundant links (blue).
\textbf{b} The network after optimization using our algorithm.
\textbf{c} Optimization scores $Z_C$, $Z_R$, and $Z_F$ from Eq.\ \eqref{ZCZRZF}. Reliability improves significantly ($Z_F = 0.995$) while maintaining the same redundancy ($Z_R = 0$), and a marginal inclrease in cost ($Z_C = -0.03$).
\textbf{d}-\textbf{f} NetworkUSA:\ optimization reduces network risk ($Z_F = 0.71$) while requiring fewer redundant links and reduced cost ($Z_R = 0.50. Z_C = 0.12$).
\textbf{g}-\textbf{l} Similar results for the Balerma and Rural water networks.
\textbf{m}-\textbf{r} Optimization of the SFO Pacific and SFO Davidson power networks.
}
\label{FigRealNets}
\end{figure}

To evaluate our optimization in real-world settings, we collected benchmark infrastructure networks across multiple domains $\bullet$ \textit{Communication}.\ Wavenet, NetworkUSA~\supercite{knight2011internet} $\bullet$ \textit{Water distribution}.\ Balerma, Rural~\supercite{wdsrd,marchi2014,bietal2015,reca2008,zheng2011,bietal2015} $\bullet$ \textit{Power}.\ SFO Pacific, SFO Davidson~\supercite{smartds,elliott2020smartds,smartds_readme} (Fig.~\ref{FigRealNets}). These networks span diverse spatially embedded infrastructures with varying sizes $N$ and redundancy levels $R$.

For each of the six networks, we first measured the initial redundancy $R_0$, cost $C_0$, and SAIDI $F_0$. We then applied our optimization protocol to propose an alternative wiring that enhances reliability at comparable or lower cost. To quantify the improvements, we define

\begin{equation}
\begin{array}{ccc}
Z_C = \dfrac{C_0 - C_f}{C_0}, \,\,\, & Z_R = \dfrac{R_0 - R_f}{R_0}, \,\,\, & Z_F = \dfrac{F_0 - F_f}{F_0},
\end{array}
\label{ZCZRZF}
\end{equation}

where $C_f,R_f$ and $F_f$ denote the optimized network parameters. Hence $Z_X \to 1$ represents a $100\%$ improvement in $X$, whereas $Z_X < 0$ indicates degraded performance. 

Our optimization consistently improves network reliability — often dramatically — while preserving or even reducing overall cost. In Wavenet, for example, without adding any redundancy ($Z_R = 0$), the reliability score reaches $Z_F = 0.995$, reflecting a two–orders-of-magnitude boost from $F_0 = 4 \times 10^{-3}$ to $F_f = 2 \times 10^{-5}$. This improvement was achieved solely through better use of the available links. In the Rural water distribution network, we achieve a more modest $Z_F = 0.62$, but this is accompanied by a substantial reduction in redundancy and cost ($Z_R = 0.95$, $Z_C = 0.24$). In the SFO–Davidson power network, a slight increase in cost ($Z_C = -0.09$) delivers a major reliability gain, $Z_F = 0.9$, highlighting the value of optimized wiring even under constrained resources. These and all remaining performance metrics are summarized in Fig.~\ref{FigRealNets}. 

Together, these results demonstrate that our algorithm can substantially enhance infrastructure planning, ensuring robust service continuity under limited resources and practical constraints. A detailed analysis of all networks in Fig.~\ref{FigRealNets} is provided in Supplementary Section~7.3.

\vspace{3mm}
\textbf{\color{blue} \Large Discussion and outlook}

Infrastructure robustness presents a complex multivariate challenge negotiating limited resources, spatial constraints, and reliability benchmarks. Our framework isolates the key parameters governing this trade-off:\ network size $N$; budget, captured via $R$, $\rho$, or $\alpha$; and component durability, quantified by $p$. For any given set of these parameters, our method enables the construction of the optimal network graph and the evaluation of its reliability through Eqs.\ \eqref{FSigma} - \eqref{PcOpt}. This allows us to determine whether the system lies within the desired reliability window, $p \le p_c$, or outside it, in which case no feasible solution exists under the given constraints.

The proposed algorithm is designed to find the exact optimum by minimizing both $\av{\tilde \lambda}$ and $\sigma_{\tilde \lambda}^2$ in \eqref{FSigma}. However, because the space of all possible $P(\tilde \lambda)$ distributions is vast, computing the exact solution is often infeasible. Accordingly, our implementation — whose results appear in Figs.\ \ref{FigOptimization}–\ref{FigRealNets} — relies on approximations. In particular, we employed different proxies for $\tilde \lambda$, such as optimizing for the number of nodes per chain or for the standard chain length $\lambda$ instead of the effective $\tilde \lambda$. A detailed description of our implementation and its limitations is provided in Supplementary Section~6. Importantly, our results show that these approximations have minimal practical impact, achieving near-optimal network configurations. 

Interestingly, the optimal networks we identify are characterized by remarkably homogeneous topologies:\ the degrees of all nodes are confined to $k = 2$ or $3$, and the lengths $\tilde \lambda$ of all chains are designed to be as uniform as possible. This sharply contrasts with canonical results suggesting that network heterogeneity — particularly scale-free degree distributions — yields the most robust topologies against random failures. 

The origin of this apparent discrepancy lies in the value of $p$. Heterogeneous networks are advantageous under large-scale failures, where a substantial fraction of nodes or links fail simultaneously and the network risks fragmentation. In our analysis, by contrast, we focus on the $p \to 0$ limit, where failures are sporadic and scattered, and the primary challenge is minimizing individual node downtime. Thus, the optimal strategy, maximizing heterogeneity or promoting uniformity, depends critically on the type of risk considered:\ either a major, large-scale breakdown, or the inevitable routine failure of individual components (see extended discussion in Supplementary Section 4).

While analytical solutions for complex network optimization are rarely tractable, algorithmic approaches remain challenging due to the combinatorial and highly constrained nature of the problem. Yet, as cities expand and our dependence on power, communication, and other technological infrastructures deepens, the design of scalable and reliable networks is emerging as a central bottleneck to growth and development~\cite{lei2025privacy}. The approach presented here — algorithmic optimization guided by network-physics principles — offers a general and practical framework for improving the robustness and efficiency of such systems.  

\clearpage

\clearpage

\clearpage

\clearpage

\clearpage

\clearpage

\clearpage

\begin{figure}
\begin{centering}
\vspace{-3mm}
\includegraphics[width=0.77\linewidth]{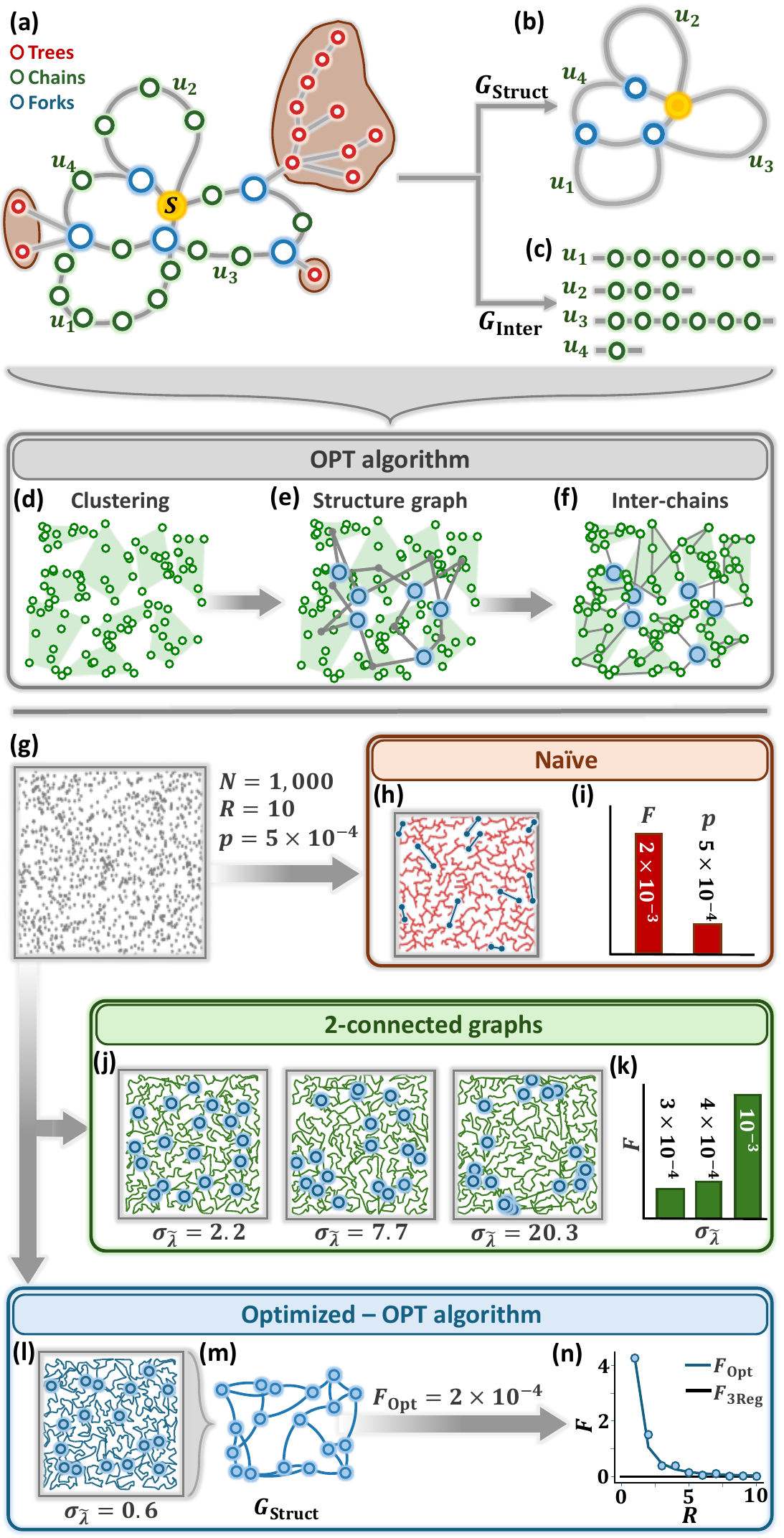}
\par\end{centering}
\begin{center}
\footnotesize \textbf{\color{blue} Figure 4.\ Network optimization – step by step - Full size image.}
\end{center}
\end{figure}

\clearpage

\printbibliography[heading=subbibliography,title={References}]
\end{refsection}

\clearpage

\setcounter{equation}{0}
\setcounter{section}{0}
\setcounter{subsection}{0}
\setcounter{subsubsection}{0}
\setcounter{figure}{0}
\setcounter{table}{0}
\renewcommand{\theHsection}{S.\arabic{section}}
\renewcommand{\theHsubsection}{S.\arabic{section}.\arabic{subsection}}
\renewcommand{\theHsubsubsection}{S.\arabic{section}.\arabic{subsection}.\arabic{subsubsection}}
\renewcommand{\theHequation}{S.\arabic{equation}}
\renewcommand{\theHfigure}{S.\arabic{figure}}
\renewcommand{\theHtable}{S.\arabic{table}}

\begin{center}
{\color{blue} \LARGE \textbf{Cost-effective network robustness \\[10pt] Supplementary information}}
\end{center}

\vspace{2mm}

\hspace{-2.27mm}

\begin{refsection}

\section{Modeling framework}
\label{supp:model}
\subsection{Network model}

To model infrastructure networks we consider an undirected graph $G(S,V,E)$, where $V = \{1,\dots,N\}$ represents the $N$ nodes and $E = \{1,\dots,L\}$ denotes the $L$ physical links between them. A subset $S \subseteq V$ represents the sources, $S = \{1,\dots,K\}$. Each consumer node $v \in V$ is assigned a weight $w_v$, representing its importance. A critical consumer, for example a hospital, is therefore assigned a higher weight than a non-critical node, such as a domestic unit. 

We distinguish between two classes~\supercite{colbourn1987combinatorics,shier1991network} of  networks:\
\textit{Centralized networks} describe systems such as power~\supercite{allan2013reliability} or water distribution~\supercite{trifunovic2006water}, in which few sources supply all consumers, and hence $K \ll N$, typically of order $K \sim O(1)$. The vast majority of nodes in $V$ are, therefore, consumers, and their service relies on their connection to one of the few sources. Alongside these systems, we also consider \textit{distributed networks}, such as communication~\supercite{neumayer2011assessing} or transportation systems~ \supercite{SANCHEZSILVA200547}, wherein every node acts as both a source and a consumer, such that $S = V$ and $K = N$.

In the main text, our analysis focuses exclusively on centralized networks. Our results, however, as shown, \textit{e.g}., in Fig.\ 6 of the main text cover both types. Here, we show the generalized analysis, designed to treat both network classes.    

Each link $l$ is associated with a construction cost $c_l$. For simplicity, we assume a constant cost per unit length $c$, such that a link of length $\lambda_l$ incurs a cost $c\lambda_l$, yielding the total network construction cost as

\begin{equation}
C = \sum_{l\in E} c\,\lambda_l = c\,\lambda_{\mathrm{Net}}.
\end{equation}

Here $\lambda_{\mathrm{Net}} = \sum_{l\in E} \lambda_l$ is the total network length. This can be generalized to account for link-dependent costs, where $c = c_l$, for example, distinguishing underground from overground wiring or accounting for terrain-driven construction expenses.

In centralized networks, a node $v$ is considered operational if at least one operational path connects $v$ to a source $s \in S$. In distributed networks, as all nodes are also sources, and therefore we consider pairwise operability:\ a pair $\{u,v\} \subseteq V$ is considered operational if at least one operational path connects between them. The combined weight of this node pair is given by $w_{uv} = w_u + w_v$. 

\vspace{2mm}
Each link $l \in E$ may fail independently with probability $p_l$, and is, therefore, operational with probability $1 - p_l$. To evaluate $p_l$, we assume a uniform failure probability per unit length $p$~\supercite{allan2013reliability}, capturing the links' \textit{intrinsic failure probability}. Each infinitesimal line segment of length $\dif \lambda$ fails independently with probability $p\,\dif \lambda$. In the limit where $p \ll 1$, this leads to a Poisson failure probability

\begin{equation}
p_l = 1 - e^{-p\lambda_l} = p\lambda_l + O(p^2 \lambda_l^2),
\label{eq:pl}
\end{equation}

where $\lambda_l$ is the total $l$-link length. Networks with underground wiring or well-maintained equipment typically exhibit smaller values of $p$, whereas aging, exposed, or poorly maintained infrastructure tends to have larger $p$. The model readily generalizes to heterogeneous failure probabilities, allowing different values of $p$ for different link types or environments.

\vspace{2mm}

\textbf{Redundancy}.\ To quantify the network redundancy, let $L = |E|$ denote the total number of links. Because a connected graph containing $N$ nodes requires at least $N - 1$ links, we define the redundancy as

\begin{equation}
R = L - (N - 1),
\end{equation}

\textit{i.e}.\ the number of excess links above the sheer minimum. The relative redundancy is thus

\begin{equation}
\rho = \frac{R}{N},
\end{equation}

the fraction of excess links. This formulation shows Eq.~(3) in the main text. The total link count is thus

\begin{equation}
L = (1+\rho)N - 1.
\end{equation}

\subsection{Reliability indices}
\label{sec:reliability_indices}

We quantify the reliability of a network by the expected loss of connected weight over time due to link failures. Distinguishing between centralized and distributed networks this is expressed via

\begin{equation}
F =
\begin{cases}
\displaystyle 
\frac{1}{W_{\text{C}}} \sum_{v \in V} w_v \, 
P(S \nleftrightarrow v \mid G,p)
& \text{Centralized networks} 
\\[20pt]
\displaystyle 
\frac{1}{W_{\text{D}}} \sum_{s < t} w_{st} \, 
P(s \nleftrightarrow t \mid G,p)
& \text{Distributed networks},
\end{cases}
\label{supp_reliability_indices}
\end{equation}

where $P(x \nleftrightarrow y \mid G,p)$ is the probability that $y$ is disconnected from $x$, given the graph structure $G$ and the intrinsic link-failure probability $p$. 

For centralized networks, Eq.\ \eqref{supp_reliability_indices} captures the standard \emph{SAIDI} (System Average Interruption Duration Index), which quantifies the expected fraction of customer weight that becomes disconnected from all sources $S$~\supercite{allan2013reliability,mousareliability} over time. For distributed networks, as all consumers are also sources ($S = V$), $F$ provides the \emph{pairwise reliability index}, defined as the expected fraction of weighted node pairs that become disconnected from one another~\supercite{rodionov2016practical,neumayer2011assessing}. The pair weight, as defines above, is $w_{st} = w_s + w_t$. To ensure $0 \le F \le 1$ we normalize by the total weight $W_{\text{C}} = \sum_{v \in V} w_v$ for centralized networks, and $W_{\text{D}} = \sum_{s<t} w_{st}$ for distributed.

To simplify the analysis, we note that any centralized network with multiple sources can be reduced to an equivalent single-source network by contracting all vertices in $S$ into a single super-source. Such mapping preserves all failure events and results in an effective multigraph with only one source. Therefore, without loss of generality, in our analysis, we may assume that all centralized networks contain a single source.

\subsection{Evaluating $F$}

We consider two numerical processes by which to extract $F$ in \eqref{supp_reliability_indices} - a dynamic and a static process.

\subsubsection{Dynamic state-space model}
\label{supp:state_space_model}

The dynamic state-space model represents a frequently used framework for modeling the stochastic behavior of network components over time~\supercite{monte_carlo,bolch2006queueing,allan2013reliability}. Each link $l$ in the network constitutes a two-state continuous-time Markov process, alternating between an \emph{Up} state $Q_l(t) = 1$ (operational) and a \emph{Down} state $Q_l(t) = 0$ (failed). Transitions between these states are governed by a failure rate $\alpha_l$ and a repair rate $\beta_l$, which are taken to be exponentially distributed.

For a single link $l$, let $u_l$ denote the mean time to failure (MTTF), and let $r_l$ denote the mean time to repair (MTTR). 
The total cycle time is therefore

\begin{equation}
t_l = u_l + r_l,
\end{equation} 	

for which the corresponding transition rates are

\begin{equation}
\alpha_l = \frac{1}{u_l}, \qquad \beta_l = \frac{1}{r_l}.
\end{equation}

Each system state is defined by the subset of links that are operational at a given time $t$. We then track the total weight of disconnected consumers (or consumer pairs) to obtain $F(t)$. This provides the reliability of the network over the time window $t \in (0,T)$ as  

\begin{equation}
F = \dfrac{1}{T} \int_0^T F(t) \dif t. 
\end{equation}

This construction provides a complete description of the stochastic dynamics of the network under random failures and repairs. The challenge is that the number of system states grows exponentially with the number of links, rendering exact analytical treatment infeasible for large-scale networks. This prompts us to introduce a more scalable approach below.

\subsubsection{Static reliability model}

The dynamic model described in Sec.\ \ref{supp:state_space_model} helps track the time evolution of all link states. In the limit $T \to \infty$ it helps capture the \emph{steady-state unavailability} of a link, defined as the link's expected down time over the course $0 \le t \le T$. This expectation provides the probability $p_l$ that $Q_l(t) = 0$ at any given time. For a specific link $l$ this becomes

\begin{equation}
p_l = \frac{\alpha_l}{\alpha_l + \beta_l} = \frac{r_l}{u_l + r_l}.
\end{equation}

reducing the stochastic up--down dynamics of all links into their effective failure probabilities. Hence, the link states represent Bernoulli variables with independent failure probabilities $p_l$, whose value can be directly linked to the failure/repair cycles. 

In simple terms, for $T \to \infty$, the dynamic state-space model reduces to a \emph{static reliability model} in which each link fails independently with probability $p_l$. This static model is equivalent to the steady-state behavior of Sec.\ \ref{supp:state_space_model}'s dynamic model~\supercite{colbourn1987combinatorics}.

To summarize, the static model presented here offers computational simplicity and scalability, while overlooking potential stochastic and corelation driven nuances. The dynamic formulation, on the other hand, which tracks the explicit stochastic dynamics of all links, provides a more realistic description of network behavior over finite time horizons, allowing to capture temporal correlations, rather than only the system's steady-state statistics.

In our work, we used the dynamic model to perform our detailed numerical simulations, and the static reliability model allows for our analytical advances. 

\vspace{20mm}
\section{Analysis}
\label{supp:general_model}

We now outline our analytical derivation of the reliability index $F$, using our cut-set decomposition formulation.

\subsection{Cut-set decomposition}

A cut-set $X \subseteq E$ is a set of links whose simultaneous failure disconnects at least one node from $S$. The failure of a cut-set partitions the network into two regions:\ the subgraph $G_S(X) \subset G$, consisting of all nodes that remain connected to the sources after the failure of $X$, and its complement $\overline{G}_S(X)$, which contains all nodes disconnected from the sources as a result of the $X$-failure. For distributed networks, which lack designated sources, the distinction between the two subgraphs is arbitrary, hence we select an arbitrary node $s$ and treat it as a source for the purpose of defining $G_S(X)$ vs.\ $\overline{G}_S(X)$. 

Next we use

\begin{equation}
\X = \{X \subseteq E \mid \overline{G}_S(X) \neq \emptyset\},
\end{equation}

to denote the set of all cut-sets in $G$, and $\X_m = \{X \in \X \mid |X| = m\}$ for the set of cut-sets of size $m$. With this, we can evaluate $F$ by accumulating the contribution of all cut-sets. The contribution of a cut-set $X \in \X_m$ is captured by

\begin{equation}
F_X = p_X\,w_X.
\end{equation}

Here $p_X$ is $X$'s failure probability, namely the probability that all links in $X$ simultaneously fail, while all remaining links in $\overline X$ remain operational. For a given $p$ and $\lambda_l$ we have
 
\begin{equation}
p_X = \prod_{l \in X} (\lambda_l\, p) \prod_{l \notin X} (1 - \lambda_l\, p),
\label{eq:pX}
\end{equation}

where $\lambda_l p = p_l$ is the individual failure of link $l$, which we take to be independent of all other link failures. The cut-set cumulative weight $w_X$ is the total weight of all disconnected nodes as a result of $X$, which we express by

\begin{equation}
w_X = 
\begin{cases} 
\displaystyle\sum_{v \in \overline{G}_S(X)} w_v & 
\text{Centralized networks} 
\\[20pt]
\displaystyle\sum_{u \nleftrightarrow v} w_{uv} & 
\text{Distributed networks},
\end{cases}
\end{equation}

where $u \nleftrightarrow v$ denotes pairs of nodes disconnected after the failure of $X$. Since different cut-sets capture non-overlapping events, we can partition the probability space by cut-sets, and apply the law of total probability, to obtain

\begin{equation}
\label{supp:contribution_sum}
F = \frac{1}{W} \sum_{X \in \X} F_X.
\end{equation}

For simplicity, we use $W$ to denote the sum of all weights, taken to be $W_{\text{C}}$ or $W_{\text{D}}$ for centralized and distributed networks, respectively (see Eq.\ \eqref{supp_reliability_indices} in Sec.\ \ref{sec:reliability_indices}).

To simplify $p_X$ in \eqref{eq:pX} we assume that the link lengths are locally uniform. This allows us to approximate the terms in the second product as $1 - \lambda_l p \approx (1 - \av{\lambda} p)$, where $\av{\lambda}$ is the average link length. Next, we define

\begin{equation}
\gamma_X = \prod_{l \in X} \lambda_l,
\label{eq:GammaX}
\end{equation}

allowing us to express the first product on the l.h.s.\ of \eqref{eq:pX} as $\gamma_X \prod_{l \in X} p = \gamma_X p^m$, where $m = |X|$. Together, these two simplifications allow us to rewrite Eq.\ \eqref{eq:pX} as 

\begin{equation}
p_X \approx \gamma_X \, p^m\, (1 - \av{\lambda}p)^{L - m},
\label{eq:pMGamma}
\end{equation}

from which we obtain 

\begin{equation}
\label{supp:contribution_approx}
F_X \approx \gamma_X\, w_X\, p^m (1-\av{\lambda}p)^{L-m}.
\end{equation}

Substituting Eq.~\eqref{supp:contribution_approx} into Eq.~\eqref{supp:contribution_sum} yields

\begin{equation}
\label{supp:Zm_decomposition}
F \approx \frac{1}{W} \sum_{m=1}^{L} Z_m\, p^m\, (1-\av{\lambda}p)^{L-m},
\end{equation}

where

\begin{equation}
\label{supp:Zm}
Z_m = \sum_{X \in \X_m} \gamma_X\, w_X.
\end{equation}

Equations~\eqref{supp:Zm_decomposition} and~\eqref{supp:Zm} reproduce Eqs.~(4) and~(5) of the main text, respectively. They help expresses $F$ as a polynomial in the failure probability $p$. The coefficients $Z_m$ aggregate the contribution of all cut-sets of size $m$, \textit{i.e}.\ all $X \in \X_m$.

Equation~\eqref{supp:Zm_decomposition} separates between the two distinct contributions to the network reliability. The factor $p^m (1-\av{\lambda}p)^{L-m}$ encodes the durability of individual components, while the coefficient $Z_m$ captures the topological structure of the network through the number ($|\X_m|$), size ($m$), and impact ($\gamma_X\,w_X$) of all cut-sets. 

Our result in \eqref{supp:Zm_decomposition} is closely related to the highly studied network \emph{reliability polynomial}~\supercite{colbourn1987combinatorics}. The reliability polynomial measures the \emph{all-terminal reliability}, namely the probability that the network remains fully connected under independent link failures. In its classic formulation, the reliability polynomial is derived under uniform failure probability ($p_l=p$), uniform link lengths ($\gamma_X=1$), and equal weights for all cut-sets ($w_X = 1$), effectively assigning the same importance to all disconnection events. As a result, the coefficient $Z_m$ reduces to the number of cut-sets of size $m$, without accounting for the magnitude, location, or impact of their subsequent disconnected components. This coarse treatment makes the reliability polynomial substantially less informative than the weighted reliability index $F$ considered here. However, due to its relative simplicity, the reliability polynomial has been extensively studied in classical graph-theoretic and reliability textbooks.

\subsection{Minimal cut-sets and risk decomposition}

In general, the total number of cut-sets in a typical network is enormous. Therefore, to reduce the complexity of our analysis we consider \emph{minimal cut-sets}, namely cut-sets that do not contain any other smaller cut-set. Such minimal cut-sets serve as the fundamental building blocks of all cut-sets, and their number is typically much smaller. We denote by $\mathcal{M} \subset \X$ the set of all minimal cut-sets, and by $\mathcal{M}_{sv} \subseteq \mathcal{M}$ the set of all minimal cut-sets that separate the source $s$ and a node $v$.

Each event where $s \nleftrightarrow v$ can be described by a collection of failing minimal cut-sets $\mathcal{A} \subseteq \mathcal{M}_{sv}$. The source component $G_S(\mathcal{A})$ is defined as the set of nodes that remain connected to $s$ after the failure of all cut-sets in $\mathcal{A}$, and the disconnected component $\overline{G}_S(\mathcal{A})$ is its complementary set. These components must remain consistent with the disconnections induced by each individual cut-set $M \in \mathcal{A}$; hence we express them as

\begin{equation}
    G_S(\mathcal{A}) = \bigcap_{M\in \mathcal{A}} G_S(M), \quad \overline{G}_S(\mathcal{A}) = \bigcup_{M\in \mathcal{A}} \overline{G}_S(M).
\end{equation}

Since each component in ${G_S(M)}_{M \in \mathcal{A}}$ is a connected component containing $s$, and each component in ${\overline{G}S(M)}{M\in \mathcal{A}}$ is a connected component containing $v$, both $G_S(\mathcal{A})$ and $\overline{G}_S(\mathcal{A})$ are non-empty connected components. Let $M = \partial G_S(\mathcal{A})$ denote the set of links separating $G_S(\mathcal{A})$ from the rest of the network. The set $M$ is a minimal cut-set and is uniquely determined by the collection of failing minimal cut-sets $\mathcal{A}$. As a result, the probability space of all events where $s \nleftrightarrow v$ can be partitioned according to the events $\partial G_S = M$.

We define the \emph{risk} associated with each minimal cut-set $M$ as

\begin{equation}
F_M = \sum_{v \in V} w_v\, P(\partial G_S = M),
\end{equation}

that is, the expected disconnected weight conditioned on $M$ being the separating cut. Using this conditioning, the risk can be decomposed as

\begin{equation}
\label{supp:risk_def}
F_M = p_M\, w_M\, P(S \leftrightarrow M),
\end{equation}

where $p_M = p^{|M|} \prod_{l\in M} \lambda_l$ is the probability that all links in $M$ fail, $w_M$ is the expected disconnected weight resulting from the separation induced by $M$, and $P(S \leftrightarrow M)$ is the \emph{reaching probability}, namely, the probability that for each link in $M$, at least one of its end-points is reachable from $S$. This factor ensures that $M$ is indeed the first separating cut encountered when traversing the network from the sources.

We can now use \eqref{eq:GammaX} to express $p_M$ as

\begin{equation}
p_M = \gamma_M p^m,
\end{equation}

where $m = |M|$, and consider only the leading powers in the limit where $p \ll 1$, \textit{i.e}.\ assuming reliable components. In this limit the reaching probability satisfies $P(S \leftrightarrow M) \sim O(1)$. Together, this allows us to approximate $F_M \approx p_M \, w_M = \gamma_M w_M p^m$, up to minor corrections of order unity.

Summing over the individual contributions of all minimal cut-sets we obtain the \emph{risk decomposition} of $F$ as

\begin{equation}
F = \frac{1}{W} \sum_{M \in \mathcal{M}} F_M \approx 
\frac{1}{W} \sum_{M \in \mathcal{M}} \gamma_M \, w_M p^m,
\label{eq:FMinimalCutSets}
\end{equation}

a power series expansion in terms of the link intrinsic failure probability $p$. Compared to the full cut-set decomposition of Eq.\ \eqref{supp:Zm_decomposition}, this formulation is slightly simpler to compute and is especially useful for analyzing complex network topologies. It is particularly tailored to the $p \ll 1$ regime.

\subsection{Simple network topologies}

We now apply the above framework to three fundamental network typologies.

\subsubsection{Tree networks}
\label{supp:subsec_tree_networks}

A tree network contains no cycles and therefore has a redundancy of $R = 0$. Under these conditions, the failure of any single link disconnects
parts of the network, namely all minimal cut-sets are of size $m = 1$. Between every pair of nodes $v,s$ exists a unique path $\mathrm{Path}(s,v)$, containing the sequence of links connecting them. Thus the probability of $v$ becoming disconnected from $s$ is

\begin{equation}
P(s \nleftrightarrow v)
=
1 - \prod_{l \in \mathrm{path}(s,v)} q_l,
\end{equation}

where $q_l = 1 - p \lambda_l$ is the probability that link $l$ is operational. Consequently, 

\begin{equation}
\label{supp:F_tree}
F_{\mathrm{Tree}} = \frac{1}{W}\sum_{v\in V} \left(1-\prod_{l\in \mathrm{path}(s,v)} q_l\right) w_v.
\end{equation}

Using the fact that $p$ is small, we write

\begin{equation}
1 - \prod_{l \in A} q_l =
1 - \prod_{l \in A} (1 - p \lambda_l) =
p\sum_{l\in A}\lambda_l + O(p^2),
\end{equation}

allowing us to extract, by first-order approximation, the risk as

\begin{equation}
\label{supp:F_tree_approx}
F_{\mathrm{Tree}} = \av{w_v \lambda_{v\to s}}\,p + O(p^2),
\end{equation}

where 

\begin{equation}
\lambda_{v\to s} = \sum_{l\in \mathrm{path}(s,v)} \lambda_l
\end{equation}

is the total length of all links along the $s \leftrightarrow v$ path. The mixed moment $\av{w_v \lambda_{v\to s}}$, therefore, captures the weighted distance from all nodes to the source. Thus, to leading order in $p$, $F_{\mathrm{Tree}}$ depends solely on the average weighted path lengths in $G$. Equation \eqref{supp:F_tree_approx} above retrieves Eq.~(6) of the main text.

Note that both the product $\prod_{l\in \mathrm{path}(s,v)} q_l$ and the path length $\lambda_{v\to s}=\sum_{l\in \mathrm{path}(s,v)}\lambda_l$ can be computed rather efficiently using dynamic cumulative products and sums along the tree, resulting in linear-time
evaluation of $F_{\mathrm{Tree}}$.  

\vspace{3mm}\textbf{Star graph}.\
A star graph is a particular realization of the tree network family, in which all nodes are directly connected to the source by a single link. In this case, al nodes $v$ have $\lambda_{v \to s} = \lambda_l$. Under uniform node weights we obtain

\begin{equation}
F_{\mathrm{Star}} = \frac{1}{N}\,\lambda_{\mathrm{Net}}\,p + O(p^2),
\label{eq:FStar}
\end{equation}

capturing the maximally achievable reliability under the tree topology, minimizing the distance between all consumers and the source.

\vspace{3mm}\textbf{Linear chain}.\
In the opposite limit, we consider a linear chain, in which nodes are connected sequentially, forming a single path emanating from the source. This topology, capturing the most vulnerable graph within the tree family, is particularly important, as it serves as one of the building blocks in our analysis of general network structures.

For simplicity, let us assume uniform weights $w_v = 1$ and link lengths $\lambda_l = \lambda$. Hence, the total network length is $\lambda_{\mathrm{Net}} = L \lambda$, and its average is $\av{\lambda_{v \to s}} = \lambda_{\mathrm{Net}}/2$. With this, we arrive at

\begin{equation}
\label{supp:F_linear}
F_{\mathrm{Linear}} = \frac{1}{2}\lambda_{\mathrm{Net}}p + O(p^2),
\end{equation}

replacing the $1/N$ factor of $F_{\mathrm{Star}}$ in \eqref{eq:FStar} with $1/2$. Indeed, the linear chain, in the limit of large $N$ represents an extremely unreliable topology.

\vspace{3mm}\textbf{Minimal cut-set analysis}.\
We now revisit our derivation and express it using our minimal cut-set formulation of Eq.\ \eqref{eq:FMinimalCutSets}. Here, since all individual links are minimal cut-sets, we have $\M = \{ l \}_{l \in E}$. Therefore, in \eqref{eq:FMinimalCutSets} we have $\gamma_M = \lambda_l$ and $m = 1$. To evaluate $w_M$ we seek the total weight of nodes downstream from $l$. Hence we define $V_l = \{v \in V \mid l \notin \textrm{Path}(s,v) \}$, allowing us to write $w_{\{l\}} = \sum_{v \in V_l} w_v$. Collecting all terms we arrive at

\begin{equation}
F = \frac{1}{W} \sum_{l\in E} \sum_{v \in V_l} \lambda_l w_v p + O(p^2),
\end{equation}

which, by rearranging the order of summation provides 

\begin{equation}
F = \frac{1}{W} \left( \sum_{v \in V_l} w_v 
\sum_{l\in \textrm{Path}(s,v)} \lambda_l \right) p
+ O(p^2).
\end{equation}

Expressing the sum in the form of an average, we obtain

\begin{equation}
F = \av{w_v \lambda_{v\to s}}\,p + O(p^2),
\end{equation}

precisely the result of \eqref{eq:FMinimalCutSets} above, or equivalently, Eq.\ (6) of the main text.

Finally, for distributed networks, we consider all nodes as sources, and track the disconnection of all pairs $u$ and $v$. This provides

\begin{equation}
F_{\mathrm{Tree}} = \frac{1}{W} \sum_{u<v} \left(1-\prod_{l\in \mathrm{path}(u,v)} q_l\right) w_{uv},
\end{equation}

which, following a similar derivation, yields

\begin{equation}
F_{\mathrm{Tree}} = \av{w_{uv} \lambda_{uv}}\,p + O(p^2),
\end{equation}

an analogous formula to Eq.\ \eqref{eq:FMinimalCutSets}, in which the path distances are averaged over \textit{all} node pairs.

\subsubsection{Ring networks}
\label{sec:RingNetworks}

A ring network consists of a single cycle connecting all nodes, so that we have a redundancy of $R = 1$, and every pair of nodes is connected by exactly two distinct paths. We enumerate all nodes from $0$ to $N -1$, setting node $0$ as the source, and all links from $0$ to $L = N$, such that link $l$ is situated between nodes $l$ and $l + 1$ (link $L$ connects nodes $N$ and $0$). For a specific node pair spanning a segment of the ring, a disconnection occurs if and only if at least one link fails on each of the two disjoint paths connecting them. 

For centralized networks, the exact reliability index is therefore

\begin{equation}
\label{supp:F_ring_exact}
F_{\mathrm{Ring}} = \frac{1}{W} \sum_{v=1}^{N} w_v \left(1-\prod_{l=1}^{v}q_l\right) \left(1-\prod_{l=v+1}^{L}q_l\right).
\end{equation}

Equation \eqref{supp:F_ring_exact} is exact for any $p$, large or small. It can be computed in linear time using a dynamic programming algorithm to extract the cumulative products $\prod_{l=1}^{v}q_l$. This allows us to efficiently calculate the precise value of $F_{\mathrm{Ring}}$ using numerical tools.

\vspace{3mm}\textbf{Multiple sources}.\
If the ring network contains multiple sources, they can be contracted into a single super-source as described in Section~\ref{supp:model}. This contraction naturally decomposes the network into a collection of rings attached to the super-source. The reliability of the multi-source ring network can therefore be analyzed by applying the above results independently to each resulting ring component.

\vspace{3mm}\textbf{Cut-set analysis}.\ 
In the ring configuration all $\binom{N}{2}$ link pairs of the form $\{l_i,j_j\}$ represent minimal cut-sets of size $m = 2$. Hence the set of minimal cut-sets $\M$, in this case, is simply $\M = \X_2 = \{(l_i,l_j) \mid i,j = 0,\dots,N, i < j \}$. As a result, $F$ in \eqref{eq:FMinimalCutSets} follows

\begin{equation}
F_{\textrm{Ring}} = \sum_{i = 0}^{N} \sum_{j = i}^{N}
\gamma_{\{l_i,l_j\}} w_{\{l_i,l_j\}} p^2.
\label{eq:FRing1}
\end{equation}

In case of uniform weights $w_v = w = W/N$ and lengths $\lambda_l = \lambda = \lambda_{\textrm{Net}}/N$, we can simplify \eqref{eq:FRing1} and bring it to a closed form. First we note that the summation on the r.h.s.\ of the equation comprises $\binom{N}{2}$ terms. The coefficient $\gamma_{\{l_i,l_j\}} = \lambda^2 = (\lambda_{\textrm{Net}}/N)^2$, identically for all link pairs, and can therefore be extracted from the summation. Finally, in a ring topology, on average, a link pair failure disconnects $1/3$ of the nodes, and thus $\av{w_{\{l_i,l_j\}}} = wN/3 = W/3$. With this, Eq.\ \eqref{eq:FRing1} reduces to

\begin{align}
\label{eq:FRing2}
F_{\mathrm{Ring}} 
&= \frac{1}{3} \binom{N}{2} \left(\frac{\lambda_{\mathrm{Net}}}{N}\right)^2 p^2 + O(p^3)
\\ \nonumber
&= \frac{1}{6} (\lambda_{\mathrm{Net}})^2 
\left(1 - \frac{1}{N} \right) p^2 + O(p^3). 
\end{align}

In the limit of large $N$, this simplifies to

\begin{equation}
\label{supp:F_ring}
F_{\mathrm{Ring}} \sim \lambda_{\mathrm{Net}}^2 p^2 + O(p^3),
\end{equation}

reproducing Eq.~(7) of the main text.

While the specific expression of $F_{\mathrm{Ring}}$ in \eqref{eq:FRing2} relies on the simplifying assumptions of uniform weights and link lengths, the asymptotic scaling of \eqref{supp:F_ring} is more general, and can, in fact be extracted also from the exact expression in~\eqref{supp:F_ring_exact}. Indeed, while the specific pre-factors and coefficients may depend on the particular form of the weight and link-length statistics, the leading power of $\sim p^2$ is intrinsic to the ring topology, whose minimal cut-sets are all of order $m = 2$. 

For distributed networks, we construct $F_{\textrm{Ring}}$ in a similar fashion to the above derivation by taking consecutively every node as the source. This provides $N$ distinct reliability indices $F_{\textrm{Ring},v}$ - one per node $v = 0,\dots,N$. Averaging over all $v$ we obtain $F_{\textrm{Ring}}$ for the distributed network. Thanks to the rotational symmetry of the ring configuration we have $F_{\textrm{Ring},v} = F_{\textrm{Ring},u}$, independently of the specific source, and hence the outcome is identical to that obtained for centralized networks, namely  

\begin{equation}
F_{\mathrm{Ring}}^{\mathrm{Distributed}} = F_{\mathrm{Ring}}^{\mathrm{Centralized}}.
\end{equation}

\subsubsection{$3$-regular networks}
\label{supp:subsec_3reg}

A $3$-regular network $G_{\textrm{3Reg}}$ is a network in which each node has exactly three
incident links. By the handshaking lemma, the total number of links is
$L=\frac{3}{2}N$, leading to a linear redundancy
$R=\frac{1}{2}N \sim O(N)$, corresponding to a relative redundancy of $\rho \sim \frac{1}{2}$. 

By appropriately assigning the $3N/2$ links one can ensure that the $3$-regular network is also $3$-edge-connected. Indeed, according to Harary~\supercite{harary1962maximum}, $3$-regular graphs that are also $3$-connected graphs exist for all even $N \ge 2$. One explicit construction starts from a cycle of $N$ nodes and adds a chord connecting each node to its opposite node, yielding the desired graph characteristics.

Under such constructions we guarantee that all minimal cut-sets in $G_{\textrm{3Reg}}$ are of size $m \ge 3$, lacking any cut-sets with only $1$ or $2$ links. As a result, the first two coefficients in \eqref{supp:Zm_decomposition} vanish, namely $Z_1 = Z_2 =0$, leaving us with the leading order $\sim p^3$. The crucial point is that in $G_{\textrm{3Reg}}$ the vast majority of link trios are \textit{not} cut-sets. In fact, it can be shown that the only cut-sets in $\X_3$ are \emph{trivial cut-sets}, consisting of the three links incident to a single node. This results in exactly $N$ minimal cut-sets. Among these cut-sets, the vast majority ($N-1$) disconnect a bounded region of the network, and only a few selected cut-sets, corresponding to the nodes adjacent to the source, disconnect and order $O(N)$ of nodes.

Taken together, substituting these statistics into \eqref{eq:FMinimalCutSets} we arrive at an asymptotic scaling of

\begin{equation}
F_{\mathrm{3Reg}} = N^0\, p^3 + O(p^4),
\label{eq:F3Reg}
\end{equation}

retrieving Eq.~(8) of the main text.

In contrast to trees and rings, whose reliability degrades with system size, a $3$-regular, $3$-connected network satisfies $F_{\mathrm{3Reg}} < p$ for all $p < 1$, and, under sufficiently small $p$, it guarantees  $F_{\mathrm{3Reg}} \ll p$. Most crucially, in $G_{\mathrm{3Reg}}$ the reliability is independent of $N$. Recent work~ \supercite{brand2026most} further suggests that to optimize the coefficients of order $p^4$, $p^5$, and higher, one should use $3$-regular graphs of high girth, thereby avoiding short cycles that generate additional low-order cut-sets.

\vspace{20mm}
\section{General networks}
\label{supp:subsec_mesh}

We now turn to derive $F$ for general network structures. Our approach builds on the expansion of $F$ in terms of powers $p^m$, as appears in Eqs.\ \eqref{supp:Zm_decomposition} and \eqref{eq:FMinimalCutSets}. We, therefore, decompose the network into its fundamental structural components:\ trees, which contribute a linear $p$ dependence, rings and chains, whose leading contribution is quadratic $\sim p^2$, and the remaining $3$-connected structure, which only contributes to higher powers of $p$.  Ultimately, this decomposition leads to Eqs.~(10)–(13) of the main text, laying the analytical foundation for the topology optimization algorithm described in main text Fig.\ 4.

\begin{figure}
\centering
\includegraphics[width=0.68\linewidth]{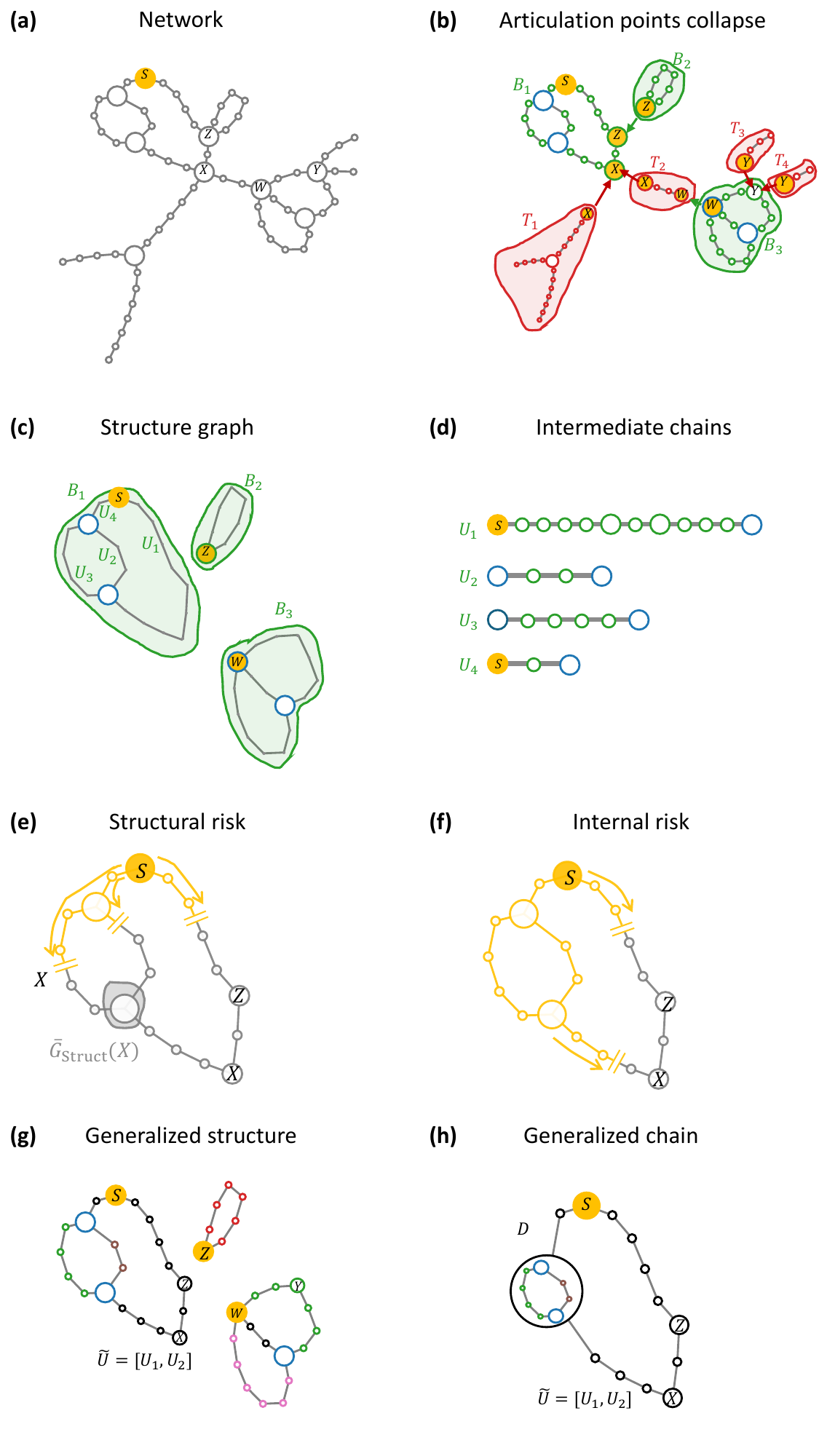}
\vspace{-5mm}
\caption{\footnotesize \textbf{\color{blue} Reliability analysis – step by step.}
\textbf{a} A network with a single source $s$. 
\textbf{b} The network contains four articulation points $X$ and $Y$, whose removal disconnects the sub-trees $T_1,T_2$ and $T_3,T_4$, and $Z$ and $W$, whose removal disconnects the ring components $B_2,B_3$. After contracting all sub-trees, the network reduces to three $2$-edge-connected components, $B_1$, $B_2$, and $B_3$. The articulation are then treated as unreliable sources (yellow).
\textbf{c} The structure graph. All $2$-edge-connected component decompose into intermediate chains (grey), linked through fork nodes (blue), whose degree is $k \ge 3$. 
\textbf{d} For example, component $B_1$ contains four chains of lengths $11$, $5$, $3$, and $2$.
\textbf{e} The structural risk $F_{\text{Struct}}$ is rooted in the structure graph, where the minimal cut-sets containing at independent failures from at least three separate chains. Since $G_{\text{Struct}}$ is in the $G_{\text{3Reg}}$ family, it poses a negligible risk.
\textbf{f} The internal risk $F_{\text{Inter}}$ arises from the intermediate chains, as failure of two links within the same chain disconnects part of the network. This represents the main component of the graph's risk $F$. The disconnected weight decomposes into the contribution from $\overline{G}_{\mathrm{Struct}}(X)$ and the additional disconnection occurring within each chain belonging to $X$. 
\textbf{g,h} Generalized structure. Chains $U_{q_1}$, and $U_{q_2}$ are transformed into a generalize chain, in which the component $D$ is transformed into a node within the chain.
}
\label{supp:strc_fig}    
\end{figure}

\subsection{Articulation point collapse}
\label{supp:subsubsec_eliminate_trees}

We begin our discussion by focusing on cut-sets of size $m = 1$. Such cut-sets arise from \emph{bridges} or \emph{articulation points} - namely, links or nodes whose removal disconnects parts of the network. A bridge necessarily induces articulation points at its two endpoints, however articulation points may exist even in the absence of bridges. Therefore, to isolate all order one cut-sets we focus on the network's set of articulation points $\A$.

Let $a \in \A$ be an articulation point, whose removal separates the network into the source-connected component $G_S(a)$ and its complement $\overline{G}_S(a)$. For any node $v \in \overline{G}_S(a)$, all paths from $S$ to $v$ must traverse through $a$, and therefore

\begin{equation}
P(S \leftrightarrow v) = 
P(S \leftrightarrow a)\, P(a \leftrightarrow v).
\label{eq:PsvFactorization}
\end{equation}

Note that this factorization follows solely from the topology of the graph and does not rely on independence assumptions beyond those already implicit in the failure model.

Equation \eqref{eq:PsvFactorization} allows us to treat the articulation point $a$ as an effective source of the subgraph $\overline{G}_S(a)$. This source has an effective failure probability of $P(S \nleftrightarrow a)$. When source $a$ fails the entire subgraph $\overline{G}_S(a)$ becomes disconnected. This allows us to contract $\overline{G}_S(a)$ into the single node $a$, by updating $a$'s weight according to

\begin{equation}
w^\prime_a = w_a + W_{\overline{G}_S(a)} - F_{\overline{G}_S(a)}.
\label{eq:waprime}
\end{equation}

Here $W_{\overline{G}_S(a)} = \sum_{v \in \overline{G}_S(a)} w_v$ is the total node weight of $\overline{G}_S(a)$ and $F_{\overline{G}_S(a)}$ is $\overline{G}_S(a)$'s reliability index with $a$ treated as its source. Hence, $w^\prime_a$ is the expected connected weight when $a$ becomes detached from the source, \textit{i.e}.\ the total weight of $\overline{G}_S(a)$, subtract by the expected \textit{a priori} disconnected weight due to internal failures within $\overline{G}_S(a)$. If $\overline{G}_S(a)$ consists of multiple connected components, this procedure is applied to each component independently, and their contributions are summed.

Applying this contraction recursively to all articulation points collapses all subgraphs $\{\overline{G}_S(a)\}_{a \in \A}$ in $G$ into effective single nodes, resulting in a reduced graph that contains no $m = 1$ cut-sets. This pre-processing, therefore, allows us to analyze the structure graph and its intermediate chains as appears in the main text and derived below.

In Supplementary Figure~\ref{supp:strc_fig}b we illustrates this process on a graph with four articulation points $\A = \{ X,Y,Z,W \}$. These points lead to the potential isolation of several graph components, as indicated by the shaded regions. Following the articulation point collapse we arrive at the effective $2$-connected graph in Supplementary Figure~\ref{supp:strc_fig}c, in which the four nodes in $\A$ are assigned their effective weights as appear in \eqref{eq:waprime}. 

\subsection{Structure graph}
\label{supp:subsubsec_structure}

Once all articulation points are collapsed, we arrive at an effective $2$-connected network, with all cut-sets of order $m \ge 2$ (Supplementary Figure \ref{supp:strc_fig}c). The next step is to apply a similar collapse to all the network's $2$-connected blocks by contracting nodes of degree $k = 2$. In this process, we detect the network's \emph{intermediate chains} $U_q$, which comprise all maximal sequences of consecutive degree $2$ non-source nodes. The sub-network $B_1$ in Supplementary Figure \ref{supp:strc_fig}c, for example, features four intermediate chains, $U_1,U_2,U_3$ and $U_4$. All intermediate chains are terminated by nodes of degree $k \ge 3$, \textit{i.e}.\ \textit{fork nodes}. 

To treat these intermediate chains we employ \emph{chain contraction}, in which we collapse $U_q$ into effective links, all while accounting for the weights of the collapsed nodes and the failure probabilities of the collapsed links. This collapses a sequence fork $\{ v \leftrightarrow$ chain $U_q \leftrightarrow$ fork $u \}$ into two directly linked forks:\ $\{v \leftrightarrow$ link $q \leftrightarrow$ $u\}$. 

To account for the characteristics of the collapsed chain $U_q$, we characterized it by three effective quantities

\begin{align}
\begin{aligned}
\lambda_q &= \sum_{l \in E(U_q)} \lambda_l, 
\\[7pt]
w_q       &= \sum_{v \in V(U_q)} w_v, 
\\[7pt]
p_q       &= 1 - \prod_{l \in E(U_q)} q_l.
\end{aligned}
\label{supp:p_q}
\end{align}

The first two, $\lambda_q$ and $w_q$, represent the total chain length and weight, respectively. The parameter $p_q$ measures the intermediate chain's failure probability, defines as the probability that at least one of its links fails.

Upon collapsing the intermediate chain $U_q$ into the effective link $q$, we reassign the parameters in Eq.\ \eqref{supp:p_q} as follows:\ the link length is taken to be $\lambda_q$, and its failure probability $p_q$. The chain weight is split evenly between the two terminal fork nodes $v$ and $u$ as

\begin{equation}
\begin{array}{ccc}
w_v^\prime = w_v + \dfrac{1}{2} w_q
&
\,\,\,
&
w_u^\prime = w_u + \dfrac{1}{2} w_q.
\end{array}
\label{eq:ForkWeightsSingleChain}
\end{equation}

This weight updating process is repeated for all chains. Therefore, as the structure graph is formed the final weights of all fork nodes follows

\begin{equation}
w_v^\prime = w_v + \dfrac{1}{2} \sum_{U_q \in K_v} w_q,
\label{eq:ForkWeights}
\end{equation}

where $K(v)$ is the set of all intermediate chains incident on fork node $v$. This way, if all chains connecting to $v$ are severed, the disconnected fork accounts also for the weight of the intermediate nodes within the severed chains. We assume, on average, that a severed chain leads to a disconnection of half of its overall weight, and hence the chain weights $w_q$ are split evenly between their two terminal fork nodes.

After collapsing all intermediate chains we arrive at the structure graph $G_{\mathrm{Struct}}$ in which all nodes are forks with degree $k \ge 3$. The resulting network, thus, belongs to the $G_{\mathrm{3Reg}}$ family of Sec.\ \ref{supp:subsec_3reg}. The links in this graph are weighted according to $\lambda_q$ and their failure probability is taken to be $p_q$, as appear in Eq.\ \eqref{supp:p_q}. The weights of all nodes are updated via \eqref{eq:ForkWeights}.

The main advantage of the structure graph is the substantial reduction in size. While the original network comprises $N \gg 1$ nodes, the number of forks in $G_{\mathrm{Struct}}$ is bounded by the redundancy $R$. To observe this consider the case where the $R$ redundant links are used in the most effective way possible, by spreading them among as many nodes as possible. This will generate the greatest number of forks, all of which will have the minimal fork-degree of exactly $k = 3$. The total number of forks obtained under this construction is $2(R-1)$, providing a strict upper bound for the size of $G_{\mathrm{Struct}}$. Consequently, the topological complexity of the network, after collapsing articulation points and intermediate chains, is compressed a graph of size $O(R)$, typically orders of magnitude smaller than $N$.

In Supplementary Figure~\ref{supp:strc_fig}d we show the structure graph corresponding to each $2$-connected block. For example, block $B_1$ contains four chains with lengths $L_q = 11, 5, 3$, and
$2$.

\subsubsection{Risk decomposition}
\label{supp:subsubsec_internal_structural}

Our analysis prompts us to distinguish between three types of minimal cut-sets:\

\textbf{Tree cut-sets} $\M_{\mathrm{Tree}}$.\ 
Cut-sets of size $m = 1$ within tree-like subgraphs. In our analysis all trees are collapsed into single articulation points, and hence their overall risk is already encapsulated within $G_{\mathrm{Struct}}$. What remains in $\M_{\mathrm{Tree}}$ is the risk of single-link cut-sets within each tree, that can potentially disconnect parts of the tree.

\textbf{Inter-chain cut-sets} $\M_{\mathrm{Inter}}$.\ 
Failures within an intermediate chain may disconnect part of the chain. Also here, the risk of a chain $U_q$'s failure is already encapsulated within its collapsed link $q$. What $\M_{\mathrm{Inter}}$ captures are internal failures of links inside $U_q$. Since all intermediate chains are $2$-connected we have all cut-sets in $\M_{\mathrm{Inter}}$ of size $m = 2$ (Supplementary Figure~\ref{supp:strc_fig}d).

\textbf{Structural cut-sets} $\M_{\mathrm{Struct}}$.\ 
Failures that disconnect fork nodes in $G_{\mathrm{Inter}}$. All cut-sets in $\M_{\mathrm{Inter}}$ have $m \ge 3$, and must involve failed links from within different intermediate chains Supplementary Figure~\ref{supp:strc_fig}e).

Together the set of all minimal cut-sets can be constructed from the unity of the three sets, as

\begin{equation}
\mathcal{M} = \mathcal{M}_{\mathrm{Tree}} \;\dot{\cup}\;
\mathcal{M}_{\mathrm{Inter}} \;\dot{\cup}\; \mathcal{M}_{\mathrm{Struct}},
\end{equation}

partitioning the risk into three disjoint sets, capturing tree- ($m = 1$), chain- ($m = 2$) and fork-related ($m \ge 3$) cut-sets.

We can now use Eq.\ \eqref{eq:FMinimalCutSets} to break down $F$ into the cumulative risks associated with all cut-sets. We write

\begin{align}
F_{\mathrm{Tree}} &= \dfrac{1}{W}
\sum_{M \in \mathcal{M}_{\mathrm{Tree}}} F_M 
\\[7pt]
F_{\mathrm{Inter}} &= \dfrac{1}{W}
\sum_{M \in \mathcal{M}_{\mathrm{Inter}}} F_M 
\label{eq:FInterSum}
\\[7pt]
F_{\mathrm{Struct}} &= \dfrac{1}{W}
\sum_{M \in \mathcal{M}_{\mathrm{Struct}}} F_M,
\label{eq:FStructSum}
\end{align}

accounting separately for each of the three cut-set types. Together this provides a decomposition of $F$ as

\begin{equation}
\label{supp:total_risk}
F = F_{\mathrm{Tree}} + F_{\mathrm{Inter}} + F_{\mathrm{Struct}}.
\end{equation}

Equation \eqref{supp:total_risk} breaks down the network risk into the order-$p$ contribution of trees, the order-$p^2$ contribution of intermediate chains, and the higher order ($p^3,p^4,\dots$) contribution arising from the structure graph itself. 
Note that, in principle, $G_{\rm Struct}$, despite having $k \ge 3$, may have cut-sets of order two, which can potentially introduce $O(p^2)$ contributions into $F_{\rm Struct}$. To bypass this, we show in Supplementary Section~\ref{sec:OptimalNetworkCharacteristics} how to design $3$-regular graph that are guaranteed to also be $3$-connected. With this construction algorithm, $G_{\rm Struct}$ can, indeed, be built to have no order $2$ cut-sets.

\textbf{Tree elimination}.\ 
The decomposition of \eqref{supp:total_risk} indicates that the leading source of risk in $G$ is embedded in its tree components, \textit{i.e}.\ $F_{\mathrm{Tree}}$, which contribute a linear $p$ dependence in \eqref{eq:FMinimalCutSets}. The crucial point is that tree-like components are easily avoidable at relatively low cost. Indeed, a tree of $N$ nodes, comprising $N - 1$ links, can be rewired into a ring configuration by adding just a single link. Therefore, as we optimize the network we seek to avoid any tree components, replacing them with rings or intermediate chains, thus setting $F_{\mathrm{Tree}} = 0$. Under these conditions we have

\begin{equation}
\label{supp:FInterStruct}
F = F_{\mathrm{Inter}} + F_{\mathrm{Struct}},
\end{equation}

precisely recovering Eq.\ (10) of the main text.

In \eqref{supp:FInterStruct} the internal risks is controlled by the geometry and weight distribution along all intermediate chains, which is largely independent of the topology of the structure graph. The structural risk, in contrast, is governed by the connectivity of $G_{\mathrm{Struct}}$, in which the impact of the chains is fully encapsulated within the effective node weights and the link effective lengths and failure probabilities. In what follows we analyze these two contributions - internal and structural - separately.

\subsection{Intermediate chain internal risk}
\label{supp:subsubsec_inter_risk}

Consider an intermediate chain $U_q$ whose two endpoint fork nodes $v$ and $u$ are both reachable from $S$. Effectively, $U_q$ behaves as a linear chain with $v$ and $u$ serving as its two effective sources. Contracting these two sources into a single source transforms $U_q$ into a ring motif (Supplementary Figure~\ref{supp:strc_fig}d). Therefore, $U_q$'s risk index can be evaluated within the framework of $G_{\mathrm{Ring}}$ in Sec.\ \ref{sec:RingNetworks}.

We express the internal risk associated with the chain $U_q$ as

\begin{equation}
F_{\mathrm{Inter},q} = w_q\, F_{\mathrm{Ring},q}\, P(S \leftrightarrow U_q),
\label{eq:FInterq}
\end{equation}

where $F_{\mathrm{Ring},q}$ is the ring reliability of $U_q$ as given by \eqref{supp:F_ring}, and $P(S \leftrightarrow U_q)$ is the \emph{reaching probability}, namely the probability that both endpoints of $U_q$ are reachable from $S$. The reaching probability ensures that failures inside $U_q$ indeed generate a minimal cut-set. In the limit of small $p$, since the end points of $U_q$ are forks in the structure graph, their probability of being disconnected is negligible, and hence $P(S \leftrightarrow U_q) \sim O(1)$. We therefore approximate \eqref{eq:FInterq} as $F_{\mathrm{Inter},q} \approx w_q \, F_{\mathrm{Ring},q}$. 

Introducing the individual ring risk of \eqref{eq:FInterq} into the total internal risk in \eqref{eq:FInterSum} provides, by summing over all $q = 1,\dots,Q$ rings

\begin{equation}
F_{\mathrm{Inter}} \approx \frac{1}{W} 
\sum_{q=1}^{Q} w_q\,F_{\mathrm{Inter},q},
\end{equation}

thus retrieving Eq.~(11) of the main text.

To estimate $F_{\mathrm{Inter}}$ we consider the case of homogeneous link lengths and node weights. The allows to use the ring approximation of Eq.~\eqref{supp:F_ring} to obtain

\begin{equation}
\label{supp:F_inter}
F_{\mathrm{Inter}} \sim p^2 \sum_{q=1}^{Q} \lambda_q^2\, w_q.
\end{equation}

Defining the \emph{effective chain length} $\tilde{\lambda}_q = \lambda_q \sqrt{w_q}$, we rewrite the sum in \eqref{supp:F_inter} as

\begin{equation}
\label{supp:effective_length}
F_{\mathrm{Inter}} \sim p^2\, \av{\tilde{\lambda}^2},
\end{equation}

bringing us to Eq.\ (12) of the main text. Next, we express the second moment on the r.h.s.\ of \eqref{supp:effective_length} via

\begin{equation}
\av{\tilde{\lambda}^2} = \sigma_{\tilde{\lambda}}^2 + \av{\tilde{\lambda}}^2,
\end{equation}

where $\sigma_{\tilde{\lambda}}^2$ is the variance of $\tilde{\lambda}$ across the $Q$ chains. Substituting this decomposition into Eq.~\eqref{supp:effective_length} yields

\begin{equation}
\label{supp:effective_length_variance}
F_{\mathrm{Inter}} \sim \left( 1 + \frac{\sigma_{\tilde{\lambda}}^2}{\av{\tilde{\lambda}}^2} \right) p^2\, \av{\tilde{\lambda}}^2,
\end{equation}

which corresponds to Eq.~(13) of the main text. 

Equation~\eqref{supp:effective_length_variance} shows that minimizing the internal risk $F_{\mathrm{Inter}}$ can be achieved by (i) lowering the mean effective chain length $\av{\tilde{\lambda}}$ and (ii) reducing its variance $\sigma_{\tilde{\lambda}}^2$. The first objective is rather intuitive, simply indicating that the shorter are the chains the smaller is the risk. It is of limited insight however, as the average chain length is governed by the number of redundant links, and therefore, if $R$ is set, we have little control over $\av{\tilde{\lambda}}$. The second objective, however, is key to our analysis, indicating that to optimize network reliability, we must aim, as much as possible, to construct uniform length chains, pushing $\sigma_{\tilde{\lambda}}^2 \to 0$. This observation provides a direct design principle for cost-effective network reliability, which we use below in our optimization algorithm.

\subsection{Structural risk}
\label{supp:subsec_structural_risk}

Structural risks arise from minimal cut-sets that contain at most one failing link from each intermediate chain. Unlike internal risks, which are localized within a single chain, structural risks reflect the global topology of the structure graph $G_{\mathrm{Struct}}$, in which the intermediate chains serve as single links. Therefore, at the level of the structure graph, a structural minimal cut-set corresponds to a set of chains

\begin{equation}
M = \{U_{q_1}, \dots, U_{q_m}\}
\end{equation}

whose simultaneous failure separates a subset of fork nodes from $S$. An example of a structural cut-set is shown in Supplementary Figure~\ref{supp:strc_fig}e.

Because each chain can contribute at most one failing link, a cut-set $M$ in $G_{\mathrm{Struct}}$ generally induces multiple minimal cut-sets in the original network, obtained by selecting one failing link along each chain. In particular, if each of the chains $U_{q_i}$ contains $L_{q_i}$ links, then $M$ induces $\prod_{i = 1}^{m} L_{q_i}$ distinct minimal cut-sets in the original graph.

To analyze the contribution of such cut-sets, we label the links of chain $U_{q_i}$ as

\begin{equation}
U_{q_i} = (l_{i1}, l_{i2}, \dots, l_{iL_{q_i}}).
\end{equation}

The link labels are sorted such that $l_{i1}$ is the link closest to the source-connected component and $l_{iL_{q_i}}$ is the one most distant from $S$. Assume the failing link in $U_{q_i}$ is $l_{ij}$, then exactly $L_{q_i} - j$ nodes along that chain become disconnected. Therefore, under structural cut-sets, the failing intermediate chains behave like the single source linear chain, analyzed in Eq.~\eqref{supp:F_linear}.

Hence, the structural risk associated with a cut-set $M$ combines three ingredients:\ (i) the probability that the chains inside $M$ fail, (ii) the probability that $M$ is reachable from the sources, and (iii) the expected disconnected weight. 

For item (i), the failure probability of $M$ is given by

\begin{equation}
p_M = \prod_{i = 1}^{m} p_{q_i},
\label{eq:pMStructureProd}
\end{equation}

where $p_{q_i}$ is $U_{q_i}$'s probability of failure as defined in Eq.~\eqref{supp:p_q}. We express $p_{q_i}$ using the intrinsic link-failure probability $p$ as $p_{q_i} = \lambda_{q_i} p + O(p^2)$, neglecting the nonlinear contribution in the limit where $p \ll 1$. This allows us to write $p_M$ as

\begin{equation}
p_M = \gamma_M p^m + O(p^{m + 1}),
\label{eq:pMStructure}
\end{equation}

where $\gamma_M = \prod_{i = 1}^{m} \lambda_{q_i}$, thus bringing it to the same form as in \eqref{eq:pMGamma}.

To capture item (ii), we use the reaching probability $P(S \leftrightarrow M)$, that ensures $M$ is the effective separating cut-set, rather than being shielded by an upstream failure.

Finally, to extract the expected disconnected weight in (iii), we denote the source-disconnected subgraph of $G_{\mathrm{Struct}}$ by  $\overline{G}_{\mathrm{Struct},M}$, and use

\begin{equation}
w_M = \sum_{v \in \overline{G}_{\mathrm{Struct},M}} w_v^\prime,
\end{equation}

to express the cumulative effective weight of all disconnected fork nodes. In this summation, we take $w_v^\prime$ from \eqref{eq:ForkWeights}, thus accounting for both the fork weight itself and the internal weights disconnected from its neighboring intermediate chains.  

Collecting all terms we arrive at the $M$-driven structural risk

\begin{equation}
F_M = P(S \leftrightarrow M)\, p_M w_M.
\end{equation}

which taking $p_M$ from \eqref{eq:pMStructure} provides us with

\begin{equation}
F_M = \gamma_M w_M p^m + O(p^{m+1}),
\label{eq:FMStruct}
\end{equation}

where $\gamma_M$ and $w_M$ are both given in terms of the effective node weights and link lengths of the $G_{\mathrm{Struct}}$, as outlined in Sec.\ \ref{supp:subsubsec_structure}. Summing over all structural minimal cut-sets, as appears in \eqref{eq:FStructSum}, we obtain

\begin{equation}
F_{\mathrm{Struct}} = \frac{1}{W} 
\sum_{M \in \mathcal{M}_{\mathrm{Struct}}} F_M,
\label{eq:FStruct}
\end{equation}

where $F_M$ is taken from \eqref{eq:FMStruct}

Thanks to our mapping in Secs.\ \ref{supp:subsubsec_eliminate_trees} - \ref{supp:subsubsec_structure}, which accounts for the internal weights and link properties of all trees and intermediate chains, we can treat $G_{\mathrm{Struct}}$ as a standalone network, whose structural risk can be evaluated independently of its tree or chain components. This is observed by the fact that Eqs.\ \eqref{eq:FMStruct} and \eqref{eq:FStruct} follow the form of our general network risk analysis in \eqref{eq:FMinimalCutSets}, enabled thanks to $G_{\mathrm{Struct}}$'s effective node and link parametrization.

The crucial point is that by its construction, $G_{\mathrm{Struct}}$ belongs to the $3$-connected family of networks (Sec.\ \ref{supp:subsec_3reg}), and hence has no minimal cut-sets of size $m < 3$. Consequently its risk $F_{\mathrm{Struct}}$ follows Eq.\ \eqref{eq:F3Reg}, which ensures that

\begin{equation}
F_{\mathrm{Struct}} \sim p^3,
\label{eq:FStructScaling}
\end{equation}

a cubic scaling with $p$ and no scaling dependence on the network size. Recalling Eq.\ \eqref{supp:effective_length_variance}, in which the inter-chain risk $F_{\mathrm{Inter}}$ was shown to scale quadratically with $p$ we conclude that $F_{\mathrm{Struct}} \ll F_{\mathrm{Inter}}$. Using the $F$-decomposition of \eqref{supp:FInterStruct} this provides

\begin{equation}
F \approx F_{\mathrm{Inter}} 
\sim \left( 1 + \frac{\sigma_{\tilde{\lambda}}^2}{\av{\tilde{\lambda}}^2} \right) p^2\, \av{\tilde{\lambda}}^2,
\label{eq:FInter}
\end{equation}

indicating that the network risk arises primarily from the structure of $G$'s intermediate chains. This insight underlies our optimization algorithm provided in Sec.\ \ref{sec:OPTAlgorithm} below.

\subsection{Generalized chains}
\label{GenChains}
One caveat of the reliability decomposition in Eq.~\eqref{supp:total_risk}

\begin{equation}
F = F_{\rm Tree} + F_{\rm Inter} + F_{\rm Struct}    
\end{equation}

is that the three terms do not perfectly separate by order in $p$.
While $F_{\rm Tree}$ contains the $O(p)$ contributions and $F_{\rm Inter}$ contains the $O(p^2)$ contributions arising from failures within a single chain, the structural term $F_{\rm Struct}$ may also contain $O(p^2)$ terms. This occurs when the structure graph contains minimal cut-sets of order two. For example, suppose we add an internal link from a chain to itself. The new link splits the chain into three smaller chains and adds an order-two cut-set. Therefore, the new link changes the graph structure, although it only shortens the modified chain length and should not fundamentally change the graph structure.

To solve this, we show that structural cut-sets of order two can be transformed into generalized rings.
Let $X=\{U_{q_1},U_{q_2}\}$ be such a cut-set, where $U_{q_1}$ and $U_{q_2}$ denote two chains of the structure graph.
Any pair of links chosen within $U_{q_1}$ forms a cut-set, and the same holds for pairs within $U_{q_2}$.
In addition, any pair consisting of one link from $U_{q_1}$ and one link from $U_{q_2}$ also disconnects the network. Therefore, the two chains act together as a single generalized chain whose endpoints are $G_S(X)$ and $\overline{G}_S(X)$.

Motivated by this observation, we introduce the notion of a \emph{generalized chain}.
A generalized chain is defined as a maximal collection of chains

\begin{equation}
\tilde{U} = [U_{q_1},\dots,U_{q_k}]    
\end{equation}

such that every pair of chains in the collection forms a structural cut-set. To construct the generalized chain, we remove all links belonging to the chains in $\tilde{U}$. The graph then decomposes into a set of connected components $\{A_i\}$. Each component $A_i$ is added as a node of the generalized chain, with weight $W(A_i)$. An example of such a generalized chain is shown in Supplementary Figure~\ref{supp:strc_fig}f where chains $U_{q_1}$ and $U_{q_4}$ are combined together into a generalized chain $U=[U_{q_1},U_{q_4}]$ with the component $D$ as a node.

Using generalized chains, the reliability decomposition can be written as

\begin{equation}
F = F_{\rm Tree} + \tilde{F}_{\rm Inter} + \tilde{F}_{\rm Struct}.
\end{equation}

Here $\tilde{F}_{\rm Inter}$ accounts for the internal failure risk within the generalized chains, while $\tilde{F}_{\rm Struct}$ accounts for structural cut-sets of order $k \ge 3$. In this representation, the structural term $\tilde{F}_{\rm Struct}$ contains only $O(p^3)$ contributions. Thus, the generalized-chain construction restores the clean asymptotic separation between the different orders of the $F$-expansion.

To formally justify that chains can, indeed, be partitioned into generalized chains, we define an equivalence relation on the set of chains by $U_{q_i} \sim U_{q_j}$ if together the two chains form a structural cut-set. This relation is transitive because two links in a network form an order-two cut-set if and only if they belong to exactly the same simple cycles. Consequently, if $U_{q_i} \sim U_{q_j}$ and $U_{q_j} \sim U_{q_k}$, then $U_{q_i} \sim U_{q_k}$. We therefore define the generalized chains as the equivalence classes of this relation.

\subsection{Distributed networks}

We next extend our analysis to treat distributed networks, in which $F$ captures the pairwise reliability index. As stated, the pairwise index can be written as an average of SAIDI, running over all potential selections of source nodes. The consequence is that the classification of minimal cut-sets and their failure probabilities remains unchanged, and only the subsequent disconnected weight must be reevaluated.

\textbf{Internal risk}.\
We begin with the internal contribution arising from double-link failures within a single intermediate chain $U_q$. As in the SAIDI case, such failures disconnect a contiguous segment of the chain from the rest of the network.

The internal risk associated with the chain $U_q$ can, thus, be written as

\begin{equation}
F_{\mathrm{Inter},q} = w_q\,F_{\mathrm{Ring},q}\, P(S\leftrightarrow U_q),
\end{equation}

where $F_{\mathrm{Ring},q}$ is the pairwise reliability of the ring obtained by contracting the endpoints of $U_q$ and assigning its weight to the ring-endpoints. Since the reaching probability is $O(1)$, we approximate $F_{\mathrm{Inter},q} \approx w_q\,F_{\mathrm{Ring},q}$.

Assuming uniform link lengths and uniform node weights along the chain, and using the pairwise weight definition $w_{uv} = w_u + w_v$, we can extract $F_{\mathrm{Ring},q}$ explicitly:\ for a ring of length $L_q$, there are $L_q - k$ pairs of failing links that disconnect a segment containing $k$ nodes. Each such pair occurs with probability $(\lambda_q/L_q)^2 p^2$ and disconnects a segment of total weight $k\,\overline{w}_q$, where $\overline{w}_q = w_q/N_q$. The corresponding pairwise disconnected weight is therefore $k\,\overline{w}_q\,(W-k\,\overline{w}_q)$.

Summing over all possible segment sizes $k$ yields

\begin{equation}
F_{\mathrm{Inter},q} \approx p^2\left(\frac{\lambda_q}{L_q}\right)^2 \frac{w_q}{N_q} \sum_{k=1}^{N_q} k\,(L_q-k)\, \left(W-k\frac{w_q}{N_q}\right) + O(p^3),
\end{equation}

which, after evaluating the summation, provides

\begin{equation}
\label{supp:pairwise_interq}
F_{\mathrm{Inter},q} \approx p^2\, \left(\frac{\lambda_q}{L_q}\right)^2 \binom{L_q+1}{3} \overline{w}_q \left(W-\frac{1}{2}\,\overline{w}_q\, L_q\right) + O(p^3).
\end{equation}

Equivalently, one may view the internal risk as follows: each of the $\binom{L_q}{2}$ possible pairs of failing links disconnects, on average, a segment whose pairwise weight is $\frac{1}{3}\,\overline{w}_q\,(L_q+1) \left(W-\frac{1}{2}\,\overline{w}_q\,L_q\right)$.

To estimate the total internal pairwise contribution, we observe that $\frac{1}{L_q^2}\binom{L_q+1}{3}\overline{w}_q \sim w_q$. Moreover, under the assumption that $w_q \ll W$, the $(W-\tfrac{1}{2}\overline{w}_q L_q)$ is of the order of $W$, and hence after normalization by $1/W$ its actual contribution is of order unity. Thus together, summing Eq.~\eqref{supp:pairwise_interq} over all intermediate chains, we obtain

\begin{equation}
\label{supp:A_inter_asym}
F_{\mathrm{Inter}} \sim p^2 \sum_{q=1}^Q w_q\,\lambda_q^2 = p^2 \sum_{q=1}^Q \tilde{\lambda}_q^{\,2},
\end{equation}

where $\tilde{\lambda}_q=\lambda_q\sqrt{w_q}$ is the effective chain length. The resulting scaling is, thus, identical to that obtained for centralized networks in \eqref{supp:F_inter}.

\vspace{2mm}
\textbf{Structural risk}.\
We now turn to evaluate the structural contribution to $F$ under pairwise disconnection. Let $M = (U_{q_1},\dots,U_{q_m})$ be a structural minimal cut-set of the structure graph, with failure probability

\begin{equation}
p_M = \prod_{i=1}^m p_{q_i} \approx 
\gamma_M p^m, \qquad \gamma_M = \prod_{i=1}^m \lambda_{q_i},
\end{equation}

as appears in \eqref{eq:pMStructureProd} and \eqref{eq:pMStructure}.

Conditioned on the failure of $M$, the disconnected set consists of two parts: (i) the component $\overline{G}_{\mathrm{Struct},M}$ comprising all detached fork node pairs, at total weight $w_M$; (ii) the partially disconnected segments along each chain $U_{q_i}$. Using the effective weight $w^\prime_v$ of \eqref{eq:ForkWeights}, the resulting pairwise disconnected weight becomes $w_M'\, (W-w_M')$.

Combining these contributions yields the structural risk

\begin{equation}
F_{\mathrm{Struct},M} = \gamma_M\, w_M'\, (W - w_M') p^m + O(p^{m+1}),
\end{equation}

which, again, preserves the $p^3$ scaling observed earlier in \eqref{eq:FStructScaling} for centralized networks. Therefore, our analysis and its conclusions remain the same also for decentralized networks:\ the risk is primarily embedded in the structure of the intermediate chains. Minimizing this risk entails reducing, as much as possible, the effective lengths of the chains $\av{\tilde \lambda}$ and their length variance $\sigma^2_{\tilde \lambda}$.

\vspace{20mm}

\section{Optimal network characteristics}
\label{sec:OptimalNetworkCharacteristics}

We now collect all the insights provided in Sec.\ \ref{supp:subsec_mesh} to summarize the structural characteristics of minimum risk networks, under limited redundancy $R$. These insights will later guide us in designing the OPT algorithm, which was outlined in Fig.\ 5 of the main text and will be described in detail in Sec.\ \ref{sec:OPTAlgorithm} below. 

Our analysis distinguishes between two asymptotic regimes:\

\textbf{Reliable components}.\ 
This represents the limit $p_l \to 0$ where the majority of  components are operational at any given time.

\textbf{Fragile components}.\
Here $p_l \approx 1$ and failures are highly frequent, typically covering a macroscopic fraction of the network at any given time.

Our main focus in this work is on the first regime ($p_l \to 0$), which is most relevant in real-world infrastructure networks. We include the fragile regime to complete our analysis, and observe its fundamentally different emergent solution. The distinction between these two limits is discussed briefly, in qualitative terms, in the main text's Discussion and outlook section. And now supported quantitatively in the analysis below.

\textbf{Overview}.\ 
The limit $p_l \to 0$ has been studied previously in the context of optimizing network connectivity~\supercite{harary1962maximum,wang1997structure}.
In this regime, it has been found, that the most reliable networks exhibit homogeneous degree distributions, with the gap between maximum and minimum degrees bounded by one. Our analysis show that this structural property remains relevant also in the context of centralized and distributed networks, and even in the presence of heterogeneous link lengths and node weights. Our results further indicate that optimality also relies on maximizing the size of the structure graph, thus minimizing the average length of the intermediate chains, and pushing the chain length distribution to be as homogeneous as possible.

In contrast, in the limit $p_l \to 1$ optimality is achieved under the opposite conditions, favoring extreme heterogeneity, \textit{e.g}., scale-free networks~\supercite{cohen2000resilience,Albert2000ErrorAA,Janson2008OnPI,Bollobs2004RobustnessAV}.
Our results recover also this behavior and further show that a star network with multiple parallel links is optimal in this regime.

\subsection{Reliable components $p \ll 1$}

Our analysis shows that the optimal mesh network under a given redundancy $R \ll N$ satisfies four conditions:\ 

\begin{enumerate}
\item The network has a cubic, $3$-edge-connected structure graph $G_{\mathrm{Struct}}$.

\item All fork nodes are linked through intermediate chains of degree $2$ nodes. Their effective lengths $\tilde \lambda_q$ should be minimal.

\item The length distribution $P(\tilde \lambda)$ should be as homogeneous as possible, \textit{i.e}.\ have a minimal $\sigma^2_{\tilde \lambda}$.

\item The network should have no tree components, and therefore all nodes must have $k > 1$.
\end{enumerate}

To satisfy condition 2, we seek to include as many nodes as possible in the structure graph, thus maximizing the number of forks, and reducing the number of nodes along the intermediate chains. To achieve this, we spread the $R$ redundant links across $R$ different nodes, avoiding \textit{wasting} more than one link per node, thus resulting in exactly $2(R - 1)$ degree $3$ nodes comprising $G_{\mathrm{Struct}}$. 

Next, we must link the selected fork nodes such that the resultant $3$-regular $G_{\mathrm{Struct}}$ is also $3$-connected. The links of this structure graph will then form the basis for the intermediate chains. Each link is expanded into a chain $U_q$, a total of $3(R-1)$ chains, to which we must assign all remaining $N - 2(R - 1)$ nodes. 

The optimal chain construction selects the nodes in a way that renders all chain lengths as homogeneous as possible, as per condition 3. In case all weights are uniform, and the nodes are evenly distributed in space, the natural construction assigns a roughly equal number of nodes per chain.

This construction naturally yields $2$-connected networks, with all nodes having degree $k \in \{2,3\}$, in line with previous analyses that bounded the span of degrees by one. It also avoids $k = 1$ nodes, thus satisfying condition 4 above.

\subsection{Fragile components $p \sim 1$}

For simplicity, in this discussion we assume uniform link lengths set to $\lambda_l = 1$, and hence we have $p_l = p \approx 1$. Under these conditions, the binomial expansion in \eqref{supp:Zm_decomposition} takes the form

\begin{equation}
F = \frac{1}{W} \sum_{m = 1}^{L} Z_m \, p^m(1 - p)^{L - m},
\label{eq:FBinom}
\end{equation}

where

\begin{equation}
Z_m = \sum_{X \in \mathcal{X}_m} w_X
\end{equation}

is the total disconnected weight associated with minimal cut-sets of size $m$.

Here, since $p \to 1$, we rewrite \eqref{eq:FBinom} in terms of $q = 1 - p \to 0$, obtaining

\begin{equation}
F = \frac{1}{W} \sum_{m = 1}^{L} Z_m \, q^{L - m}(1 - q)^m,
\label{eq:FBinomq}
\end{equation}

whose leading powers are driven by $Z_L,Z_{L-1},Z_{L-2},\dots$ in descending order. Therefore, to lower the risk $F$ we seek to minimize these high order coefficients. Since $Z_L = W$ is fixed for all networks, we focus on the leading nontrivial coefficient $Z_{L-1}$.

The coefficient $Z_{L-1}$ corresponds to cut-sets of size $m = L = 1$, capturing events in which exactly one link remains operational. If that link is incident to the source, exactly one downstream node remains connected; otherwise, no node remains connected. Therefore,

\begin{equation}
Z_{L-1}
=
W-\sum_{v \sim s} w_v ,
\end{equation}

in which the sum runs over nodes directly adjacent to the source. This quantity is minimized when the source is connected directly to all nodes, \textit{i.e}., when the network is a star. Hence, in the limit $p_l \to 1$, the star topology is optimally reliable. 

Note that when $R > 0$, the optimal star topology is realized by a
\emph{multigraph}, where redundancy is implemented by multiple parallel links between the source and selected nodes.

Within this class of star-like networks we can further enhance reliability by minimizing the next-to-leading coefficient $Z_{L-2}$, this time corresponding to events in which exactly two links remain operational.

Let $E_{sv}$ denote the set of links connecting the source $s$ to node $v$, and let us denote the size of this set by $L_{sv}=|E_{sv}|$. For a given node $v$, there are two different types of working link pairs that allow $v$ to operate:\ (i) $L_{sv}(L-L_{sv})$ pairs in which exactly one of the two links is incident
on $v$ and the other is incident on a different node; (ii) $\binom{L_{sv}}{2}$ pairs in which both surviving links are incident on $v$.

Summing the contributions of node $v$ over all such pairs, and then over all nodes, yields

\begin{align}
W-Z_{L-2}
&=
\sum_{v} w_v
\left(
L_{sv}(L-L_{sv}) + \binom{L_{sv}}{2}
\right) 
\\[7pt] \nonumber
&=
\left(L-\frac{1}{2}\right)\sum_v w_v L_{sv}
-
\frac{1}{2}\sum_v w_v L_{sv}^2 .
\end{align}

Minimizing $Z_{L-2}$ under the constraint $\sum_v L_{sv}=L$ leads, via standard
Lagrange-multiplier arguments, to the balance condition

\begin{equation}
\frac{w_v}{w_u}
\approx
\frac{L-L_u}{L-L_v},
\end{equation}

for all pairs of nodes $u,v$. Thus, the number of links not incident to a node decays in proportion to its weight. In the special case of uniform node weights, the optimal configuration assigns, as much as possible, equal link multiplicities to all nodes.

\vspace{3mm}\textbf{Interpretation.}
In the fragile regime wherr $p_l \approx 1$, the optimal structures are strongly heterogeneous, characterized by the emergence of a dominant hub connected by many parallel or short links. Such configurations concentrate connectivity around a small number of
high-degree nodes, thereby maximizing the probability that at least one incident link remains operational when failures are frequent.

This behavior is consistent with classical results on the robustness of networks with heterogeneous degree distributions under random link or node failures. In particular, scale-free and hub-dominated networks are known to exhibit enhanced resilience in the frequent failure regime, as the presence of high-degree nodes substantially increases global connectivity despite extensive network damage. This phenomenon has been analyzed extensively in the context of percolation and network robustness~\supercite{cohen2000resilience,Albert2000ErrorAA,Janson2008OnPI,Bollobs2004RobustnessAV},
where it was shown that heterogeneous networks maintain a giant connected component even under the elimination of a majority of the links.

This stands in sharp contrast with our current results, obtained for $p_l \approx 0$, showing that reliability is dominated by low-order cut-sets, for which homogeneous, highly symmetric structures become favorable. The qualitative transition observed here thus mirrors the classical dichotomy between robustness against rare failures, optimized by uniform connectivity, and robustness against frequent failures, optimized by degree heterogeneity.

\vspace{20mm}

\section{Asymptotic behavior}

The network reliability index $F$ depend on four main parameters:\ the network size $N$, the redundancy $R$, the total link length $\lambda_{\mathrm{Net}}$, and the component sensitivity to failure, quantified by the intrinsic failure rate $p$. In our optimal network setting, the network has a skeletal structure $G_{\mathrm{Struct}}$, whose contribution to $F$ is negligible, and $Q$ intermediate chains $U_q$, optimized to be as homogeneous as possible. The intermediate chains are the primary contributor to the risk, hence $F \approx F_{\mathrm{Inter}}$, where

\begin{equation}
\label{eq:F_inter2}
F_{\mathrm{Inter}} \sim \dfrac{1}{W} p^2 \sum_{q=1}^{Q} \lambda_q^2\, w_q.
\end{equation}

a reiteration of Eq.\ \eqref{supp:F_inter}, originally derived in Sec.\ \ref{supp:subsubsec_inter_risk}; note that we reintroduced the normalization factor $1/W$, which we omitted in the original equation.

On average, each chain carries a weight of

\begin{equation}
w_q \approx \dfrac{W}{Q},
\end{equation}

and has a length of

\begin{equation}
\lambda_q \approx \dfrac{\lambda_{\mathrm{Net}}}{Q}.
\end{equation}

With this, Eq.\ \eqref{eq:F_inter2} provides 

\begin{equation}
F \sim \dfrac{1}{W} p^2 \sum_{q = 1}^Q \left( \dfrac{\lambda_{\mathrm{Net}}}{Q} \right)^2 \dfrac{W}{Q} = 
\dfrac{p^2 \lambda_{\mathrm{Net}}^2}{Q^2}
\label{eq:Scaling1}
\end{equation}

Next we take the total number of chains to be $Q \sim R$ and the total length as $\lambda_{\mathrm{Net}} = (N + R) \av{\lambda} \approx N \av{\lambda}$, exact in the limit where $R \ll N$. Introducing this into \eqref{eq:Scaling1} we obtain

\begin{equation}
F \sim \frac{p^2\,N^{2}}{R^{2}},
\label{eq:Scaling2}
\end{equation}

retrieving Eq.\ (14) of the main text. Hence $F$ scales quadratically with network size $N$ and with the component risk $p$. The main advantage is that, thanks to our optimization, we also achieve a quadratic reduction in $R$. This enhanced risk reduction, where $F \sim R^{-2}$ is a direct benefit from our optimal inter-chain construction.

\vspace{3mm}\textbf{Bounded redundancy $R \sim N^{\alpha}$}.\
We parameterize the number of redundant links as

\begin{equation}
R \sim N^{\alpha}, \qquad 0<\alpha<1.    
\end{equation}
    
The case $\alpha = 0$ corresponds to $R \sim O(1)$, capturing a ring-like network, where indeed $F \sim N^2$. In the opposite limit where $\alpha = 1$ we have $R \sim O(N)$. This describes linear redundancy, as discussed for the $3$-regular network family, where $F \sim N^0$, and system size adds no risk factor. Intermediate values $0 < \alpha < 1$ describe cost-effective designs in which the number of redundant links grows sublinearly with system size, and hence for large $N$ we have $R \ll N$. Substituting $R \sim N^{\alpha}$ into the asymptotic scaling in \eqref{eq:Scaling2} provides

\begin{equation}
F \sim p^2 N^{\beta}, \qquad \beta = 2(1-\alpha).
\label{eq:Scaling3}
\end{equation}

The critical component risk $p_c$ is obtained when $F$ crosses the $F = p$ line. To extract $p_c$ we write $F = p_c$, which using \eqref{eq:Scaling3} leads to

\begin{equation}
p_c \sim N^{-\beta},
\end{equation}

precisely Eq.~(15) of the main text.

\vspace{20mm}

\section{The OPT algorithm}
\label{sec:OPTAlgorithm}

\textbf{Problem description}.\ 
In this section, we present a spatial algorithm for constructing cost-effective and highly reliable networks from geometric data. We begin with a set of $N$ spatial points $P = (p_1,\dots,p_N) \subset \mathbb{R}^2$, representing consumer nodes, and a prescribed redundancy level $R$. With that our goal is to construct a spatial network $G$ that satisfies the characteristics outlined in Sec.~\ref{sec:OptimalNetworkCharacteristics}:

\begin{enumerate}
\item
$G$ has exactly $R$ redundant links.
\item
$G$ has a maximal $3$-edge-connected subgraph $G_{\mathrm{Struct}} \subseteq G$.
\item
Nodes in $G_{\mathrm{Struct}}$ are linked via intermediate chains with minimal and nearly uniform effective lengths $\tilde{\lambda}_q$.
\end{enumerate}

\textbf{Challenge}.\
For a given redundancy $R$, the maximal size of $G_{\mathrm{Struct}}$ is fixed at $2(R - 1)$, achieved by creating strictly degree $3$ fork nodes. Therefore, the main algorithmic challenge is to strategically select the $2(R - 1)$ fork nodes and then construct the $Q = 3(R - 1)$ intermediate chains linking them as uniformly as possible, namely ensuring that $P(\tilde{\lambda})$ has a small $\sigma_{\tilde{\lambda}}$. Exact optimization of $P(\tilde{\lambda})$, however, is computationally infeasible in a general spatial setting, as it involves a diverging solution space constrained by a combination of topological, geometrical, and weight-related objectives.

To bypass these challenges, we design an approximate algorithm. First, we assume that each intermediate chain contains roughly the same number of nodes. This assumption is appropriate when the weight distribution is bounded and the spatial node distribution is clustered, such that links are typically local, allowing nodes within each chain to be connected via links of bounded length. Under these conditions, a homogeneous $\tilde{\lambda}_q$ translates to an approximately equal number of nodes per chain. The result is a structurally controlled near-optimal network configuration.

\vspace{3mm}\textbf{Related problems}.\ 
The minimum 2-edge-connected spanning subgraph (2-ECSS) problem is the fundamental survivable-network design task of selecting a minimum-cost edge set that preserves 2-edge-connectivity. Our goal of designing a minimal-cost, 2-edge-connected network with a 3-connected cubic structure and nearly equal chains can be viewed as a structured, spatially constrained variant of this classical problem. The 2-ECSS problem decision version was shown to be NP-hard even in unweighted graphs via reductions from Hamiltonicity \supercite{Frederickson1981ApproximationAF}. Consequently, the literature has focused on approximation algorithms and special geometric or structural cases.

Classical results established constant-factor approximations:\ Khuller and Vishkin presented a 3-approximation based on depth-first-search augmentation \supercite{KhullerVishkin94}, while Frederickson and JáJá \supercite{Frederickson1981ApproximationAF} and later Khuller and Raghavachari \supercite{KhullerRaghavachari92} achieved 2-approximations by augmenting a minimum spanning tree to eliminate bridges. LP-based survivable-network design techniques, pioneered by Jain’s iterative-rounding framework \supercite{Jain1998AF2}, yield a 1.5-approximation for unweighted 2-ECSS and a 2-approximation for the weighted case \supercite{JainRaviSingh03}. In geometric and metric settings, where edge costs obey the triangle inequality, stronger guarantees are known, including a $4/3$-approximation for metric 2-ECSS \supercite{SeboeVygen14} and PTAS results for planar or Euclidean variants \supercite{czumaj2000fast}. Recently, further improvements achieved a $(1.3+\varepsilon)$-approximation~\supercite{kobayashi2023approximation}. 

A variety of algorithmic techniques—tree augmentation, ear decompositions \supercite{monma1989methods}, and degree-bounded constructions—underpin these results and are conceptually aligned with the fork-and-chain structure at the core of our spatial reliability framework. For a comprehensive review of ILP and polyhedral methods in survivable-network design, see the survey of Kerivin and Mahjoub \supercite{kerivin2005design}.

\vspace{3mm}\textbf{Algorithm overview}.
Our algorithm consists of three main stages:\

\begin{enumerate}
\item
\textbf{Constructing chain clusters}.\ 
We first apply a flow-based $k$-means procedure to partition the $N$ spatial points into $3(R-1)$ clusters of nearly equal size. Each cluster will, as the algorithm proceeds, form a single chain. This step therefore enforces approximate uniformity in the number of nodes per chain (Fig.~\ref{supp:optimal_spatial}c).
\item 
\textbf{Constructing $G_{\mathrm{Struct}}$}.\ 
We then select $2(R - 1)$ fork nodes, seeking those that can most efficiently connect through the chain clusters identified in step 1. Once selected, we construct a minimum-cost cubic, $3$-edge-connected structure graph $G_{\mathrm{Struct}}$ over these forks. Each link of the structure graph is routed through the center of exactly one cluster, thereby assigning each cluster to a unique link in $G_{\mathrm{Struct}}$ and fixing the chain–fork connectivity (Fig.~\ref{supp:optimal_spatial}f).
\item 
\textbf{From clusters to intermediate chains}.\ 
For each cluster, we construct a minimal-length spatial path connecting its two associated forks while visiting all points in the cluster. These paths stand in place of $G_{\mathrm{Struct}}$'s links drawn in step 2, thus realizing the geometric embedding of the corresponding intermediate chains (Fig.~\ref{supp:optimal_spatial}h).
\end{enumerate}

\textbf{Optional fine-tuning}.\ 
After these stages, we optionally apply a local-improvement procedure that modifies fork attachment points while keeping the structure graph fixed. This fine-tuning is designed to reduce the total geometric cost at the expense of a controlled increase in $F$ (Fig.~\ref{supp:optimal_spatial}i). This final step permits small deviations from uniform chain lengths while preserving the underlying $G_{\mathrm{Struct}}$ topology, thereby enabling additional cost reductions in practical settings.

Below, we provide a detailed description of each stage of the algorithm.

\subsection{Constructing chain clusters}

To partition the spatial points into $3(R - 1)$ clusters of equal size, we employ a variant of the $k$-means algorithm~\supercite{bradley2000constrained}. The standard $k$-means algorithm begins by randomly selecting $Q = 3(R - 1)$ initial centers (Supplementary Fig.~\ref{supp:optimal_spatial}b), assigning each point to its nearest center, and then updating each center to be the mean of the points assigned to it. These two steps—assignment and centroid update—are repeated until convergence.

In the variant we use here, the assignment step is replaced by an integer linear program (ILP) that enforces equal cluster sizes (Supplementary Fig.~\ref{supp:optimal_spatial}c). For a given set of centers $\mu = (\mu_1,\dots,\mu_Q)$, we formulate an assignment problem on a bipartite graph between centers and points, where the cost $c_{qj}$ is the distance between center $\mu_q$ and point $p_j$. We introduce binary variables $x_{qj}$ indicating whether point $p_j$ is assigned to center $\mu_q$, and impose the constraint that each center $\mu_q$ receives exactly $N_q$ points, such that $\sum_{q=1}^{Q} N_q = N$. Since we aim for homogeneous chains, we bound $N_q$, setting $N_q=\lfloor N/Q \rfloor + c_q$, ensuring all chains reside within a limited range. In our implementation we set $c_q \in \{0,1\}$. This allows a degree of freedom to enable $\sum_{q=1}^{Q} N_q = N$, while preserving cluster homogeneity, as much as possible.

The assignment step is therefore solved by the following ILP:\

\begin{equation}
\label{eq:cluster_ilp}
\begin{aligned}
\min_{x} \quad 
& \sum_{q=1}^{Q} \sum_{j=1}^{N} c_{qj} \, x_{qj} 
\\[6pt]
\text{s.t.} \quad
& \sum_{q=1}^{Q} x_{qj} = 1, && \forall j=1,\dots,N, 
\\[6pt]
& \sum_{j=1}^{N} x_{qj} = N_q, && \forall q=1,\dots,Q, 
\\[6pt]
& N_q = \lfloor\dfrac{N}{Q}\rfloor + c_q 
\\[6pt]
& x_{qj} \in \{0,1\}, && \forall q=1,\dots,Q,\; j=1,\dots,N.
\end{aligned}
\end{equation}

The first condition minimizes the total cost, ensuring that nodes are assigned preferably to the nearest center. The two constraints that follow guarantee that each node is assigned to exactly one center, and that all centers receive their quota of $N_q$ nodes. Finally, the third constraint ensures that all $N_q$ are within a bounded margin of one from the average cluster size, thus yielding (roughly) equal size clusters.  After solving the ILP, the centers are recomputed as the means of their assigned points, and the assignment and update steps are repeated until convergence, \textit{a l\`{a}} the classic $k$-means algorithm.

\subsection{Constructing $G_{\mathrm{Struct}}$}
\label{sec:ConstructingGStruct}

This step of the algorithm involves three tasks:\ 

\begin{enumerate}
    \item 
    \textbf{Fork selection}.\ Selecting the $2(R - 1)$ fork nodes.
    \item 
    \textbf{Link assignment}.\ Assigning the links of $G_{\mathrm{Struct}}$, such that it is $3$-edge-connected.
    \item 
    \textbf{Cluster-chain matching}.\ One-to-one matching of chain clusters to structure links.
\end{enumerate}

\begin{figure}
\centering
\includegraphics[width=1\linewidth]{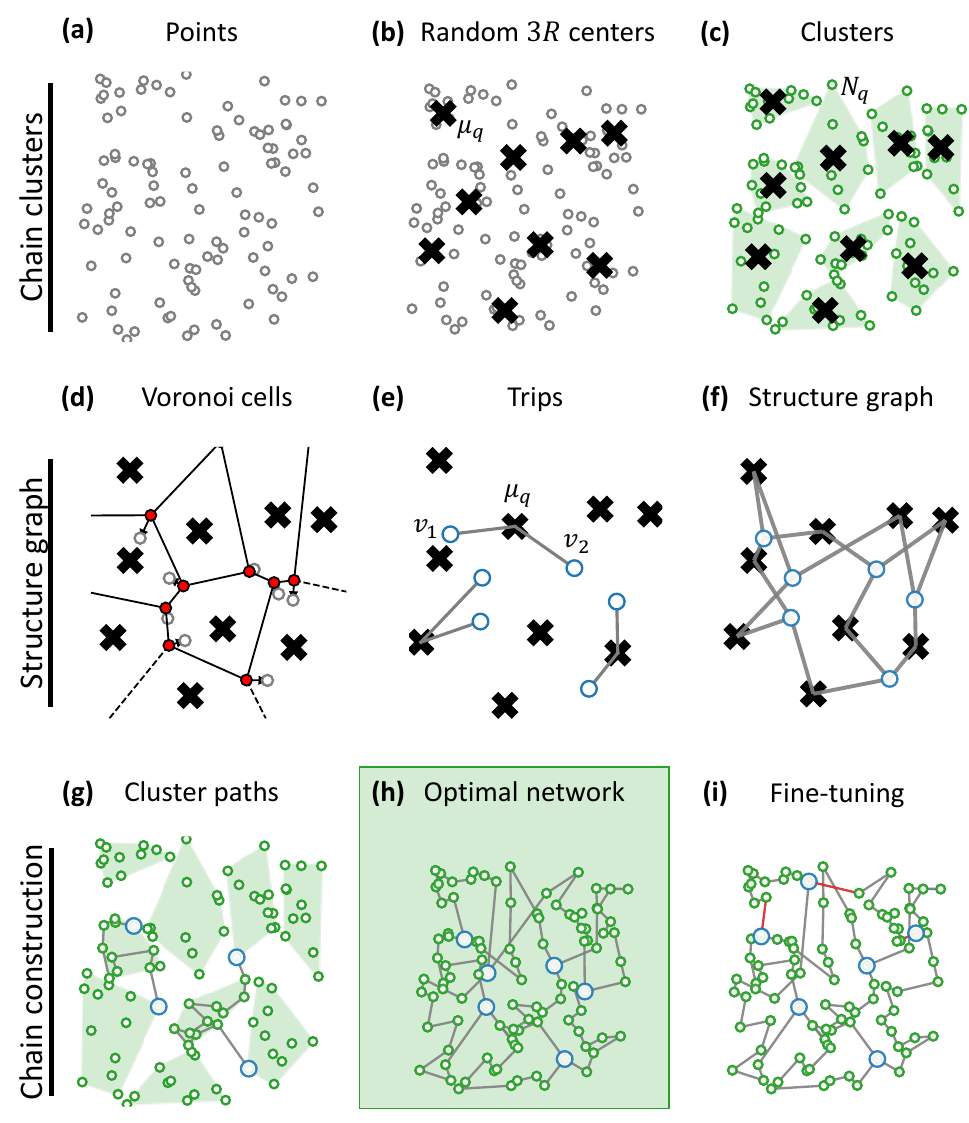}
\caption{\textbf{\color{blue} The OPT algorithm}.\
\textbf{a} We begin with a spatial distribution of points (nodes).
\textbf{b,c} We use $k$-means to divide all nodes into equal size clusters (green shaded). Centers of all clusters appear as $\times$. These clusters will define the network's intermediate chains.
\textbf{d} Using Voronoi cells we designate the location of fork nodes proximal to the cluster-centers.
\textbf{e} We seek to connect all forks (blue) via trips (grey paths) traversing through cluster-centers ($\times$). 
\textbf{f} The lowest cost trips that generate a $3$-connected network form the structure graph.
\textbf{g} We expand the links of $G_{\mathrm{Struct}}$ into chains by seeking minimal paths that traverse all nodes of each cluster.
\textbf{h} The result is the optimized network:\ A set of fork nodes comprising $G_{\text{Struct}}$ (blue), linked via homogeneous intermediate chains that traverse all remaining nodes (green).
\textbf{i} The fine-tuning step suggests local changes to further decrease cost.
}
\label{supp:optimal_spatial}
\end{figure}

\textbf{Fork selection}.\ 
To identify candidate fork locations, we compute the Voronoi diagram of the $Q = 3(R - 1)$ cluster centers $\{\mu_q\}_{q = 1}^{Q}$. For each Voronoi vertex we select the closest point from the input set $\{p_1,\dots,p_N\}$, designated to serve as a potential fork nodes (Supplementary Figure~\ref{supp:optimal_spatial}d). A Voronoi cell of a center is the set of all points in $\mathbb{R}^2$ closer to that center than to any other \supercite{aurenhammer1991voronoi}. Voronoi cells form convex polygons, that, when no three sites are collinear, induce a $3$-regular graph, as every Voronoi vertex is the meeting point of exactly three cell boundaries. This local degree-three structure naturally aligns with the fork-and-chain architecture, making Voronoi vertices ideal candidates for fork placement.

Typically, the diagram contains more than $2(R - 1)$ candidate vertices. If, however, fewer than $2(R-1)$ candidates are available, we augment the set by selecting midpoints of adjacent Voronoi centers to ensure at least $2(R-1)$ feasible fork choices.

\textbf{Link assignment and cluster-chain matching}.\ 
To optimize the links between the designated forks, we construct a multigraph of all feasible \emph{trips}. A trip $v_1 \rightarrow \mu_q \rightarrow v_2$ is a connection between two forks $v_1$ and $v_2$ routed through a chain center $\mu_q$ (Supplementary Figure~\ref{supp:optimal_spatial}e). Each trip corresponds to a potential structure link, hence our objective is to select exactly one trip for each chain center while maintaining a cubic $3$-edge-connected graph over the chosen forks (Supplementary Figure~\ref{supp:optimal_spatial}f). The cost $c_t$ of a trip $t$ is its total euclidean distance, which we seek to minimize. 



Denote by $T$ the set of all feasible trips, by $V$ the set of all potential fork locations, and by $\mathcal{Q}$ the set of chain centers. For any subset of forks $S \subseteq V$, let $\delta(S)$ be the set of all trips having exactly one endpoint in $S$. To ensure that the final structure graph is $3$-edge-connected, we set a binary variable $x_t$ for each trip $t$ and must enforce

\begin{equation}
\sum_{t \in \delta(S)} x_t \geq 3, \qquad \forall\, S \subset V,
\end{equation}

which is exact, but yields an exponential number of constraints. To overcome this we employ a greedy-constraint scheme, whereby we initially omit all connectivity constraints, solve the ILP, compute the $3$-edge-connected components of the resulting solution, and for each component add the corresponding violated cut constraint~\supercite{kerivin2005design}. This process is repeated until no violated cuts remain. The structure selection process is achieved by choosing both the valid trips($x_t$ variables), and the chosen fork($y_v$ variable for each optional fork). Together the construction of $G_{\mathrm{Struct}}$ is formulated through the following ILP:\

\begin{equation}
\begin{aligned}
\min \quad 
& \sum_{t \in T} c_t \, x_t 
\qquad  \\[4pt]
\text{s.t.} \quad 
& \sum_{v \in V} y_v = 2(R-1)
\qquad \text{(number of forks)} \\[4pt]
& \sum_{t \in T} x_t = 3(R-1)
\qquad \text{(number of chains)} \\[4pt]
& \sum_{t \in \delta(v)} x_t = 3\, y_v,
\quad \forall v \in V
\qquad \text{(cubic structure)} \\[4pt]
& \sum_{u,v \in V} x_{uvq} = 1,
\quad \forall q \in \mathcal{Q}
\qquad \text{(each chain center used exactly once)} \\[4pt]
& x_{uvq} \le y_u, \quad x_{uvq} \le y_v,
\quad \forall (u,v,q) \in T
\qquad \text{(activate link only if both forks are selected)} \\[4pt]
\end{aligned}
\nonumber
\end{equation}

\begin{equation}
\label{eq:structure_ilp}
\begin{aligned}
& \sum_{q \in \mathcal{Q}} x_{uvq} \le 1,
\quad \forall u,v \in V
\qquad \text{(at most one link per fork pair)} \\[4pt]
& \sum_{t \in \delta(S)} x_t \ge 3,
\quad \forall S \subset V,\; 1 \le |S| \le |V|-1
\qquad \text{(3-edge-connectivity)} \\[4pt]
& x_t \in \{0,1\},
\quad \forall t \in T
\qquad \text{(trip selection variables)} \\[4pt]
& y_v \in \{0,1\},
\quad \forall v \in V
\qquad \text{(fork selection variables)} .
\end{aligned}
\end{equation}

\vspace{3mm}
Our proposed greedy-constraint scheme can generate a large number of constraints as the iterations progress, potentially leading to slow convergence. To mitigate this, after a pre-defined number of iterations $n$, we construct a simplified ILP for refinement. Let $V_i$ and $T_i$ denote the sets of selected forks and trips at iteration $i$, $i > n$. In iteration $i + 1$ we impose:\ 

\begin{enumerate}
    \item $y_v = 1$ for all $v \in V_i$ and $y_v = 0$ otherwise;
    \item $c_t = 0$ for all $t \in T_i$.
\end{enumerate}

By fixing the fork set and eliminating the cost of previously selected trips, we force the model to \emph{improve} the existing structure rather than jump to a new configuration that merely satisfies earlier cuts, but introduces new violated cuts. Empirically, this significantly reduces runtime for large redundancy levels. In our implementation we set $n = 2$.

An example of a structure graph produced by this model is shown in Supplementary Figure~\ref{supp:optimal_spatial}f.

\subsection{From clusters to intermediate chains}

By the end of steps 1 and 2 of our algorithm we arrive at a structure graph with $2(R - 1)$ nodes and $Q = 3(R - 1)$ structural links. The remaining $N - 2(R - 1)$ nodes are split into clusters $\mu_q$, each of which is associated with a specific structural link. We denote these cluster-links by $v_i \rightarrow \mu_q \rightarrow v_j$, where $v_i,v_j \in G_{\mathrm{Struct}}$ and $\mu_q$ is their intermediate cluster. 

In step 3, we expand the structural links into chains that include all nodes in the associated cluster. For each trio $v_i \rightarrow \mu_q \rightarrow v_j$, we construct a minimal-length path that starts at the first fork $v_i$, terminates at the second fork $v_j$, and visits all points in $\mu_q$ at its path. The result is precisely the desired structure:\ fork nodes that link through intermediate chains.

The only challenge that remains is to determine the optimal ordering of the points along each chain that minimizes the total chain length. This challenge represents a variant of the Traveling Salesman Problem (TSP), specifically a path-TSP, which is a classical NP-hard problem for which numerous exact, approximation, and heuristic methods are available~\supercite{lawler1985tsp}. Here we employ an ILP-based formulation to obtain near-optimal solutions, supported by heuristic assumptions that help accelerate convergence.

Following a similar approach to the construction of $G_{\mathrm{Struct}}$ in Sec.\ \ref{sec:ConstructingGStruct}, we use ILP to solve the TSP via greedy subtour-elimination constraints, which ensure that all sub-path components are $2$-connected. To ensure convergence in reasonable timescales, after a predefined number of initial iterations we modify the objective by assigning zero cost to edges selected in the previous iteration. This encourages the solver to refine the current solution rather than jump to a new configuration that satisfies earlier cuts but introduces new subtours, leading to substantially faster convergence for long chains.

To further accelerate convergence, we also avoid searching the complete solution space. Hence, we restrict the set of candidate links to those belonging to the Delaunay triangulation of the chain’s points~\supercite{aurenhammer1991voronoi}. If no feasible path exists within this subgraph, we fall back to allowing all pairwise links. Using the Delaunay triangulation as the candidate graph is a standard geometric TSP heuristic, justified by the fact that the Euclidean minimum spanning tree, and in practice many edges of the optimal tour, are always contained within the Delaunay triangulation~\supercite{preparata2012computational}.

Since this step corresponds to a standard path-TSP formulation, any alternative TSP solver or heuristic can be substituted here without affecting the essential structure of the algorithm.

\subsection{Fine tuning}
\label{supp:subsubsec_local}

In real network settings, even a small percentage of cost reduction can have a meaningful engineering impact. Therefore, we now present a practical method for locally improving an optimized spatial network by reducing its total construction cost, while allowing a restricted degradation in $F$. Indeed, since our optimal networks already achieve very high reliability, it is often acceptable to trade a slight increase in $F$ for a noticeable reduction in cost. We apply this local improvement strategy to the real network shown in Fig.\ (6) of the main text.

The key idea is to modify chain lengths while keeping the structure graph fixed (Fig.~\ref{supp:optimal_spatial}.i). Since all modifications preserve the cubic $3$-edge-connected nature of the structure graph, the structural risk remains near zero, and the only impact on reliability arises from changes in the internal risk. The only links that can be modified without changing the structure graph are those connecting the end of a chain to a fork. Each fork is incident to three chains, and shifting the attachment point of one chain along another chain modifies their lengths while preserving the underlying structure graph (Fig.~\ref{supp:optimal_spatial}.i).

Consider a fork connecting three chains $U_{q_1}, U_{q_2}$, and $U_{q_3}$ of lengths $L_1$, $L_2$, and $L_3$, respectively. If the endpoint of chain~$3$ is shifted by $k$ links along chain~$1$, the resulting chain lengths become

\begin{equation}
    (L_1, L_2, L_3) \;\longrightarrow\; (L_1 - k,\, L_2 + k,\, L_3).
\end{equation}

The corresponding change in $F$ is given by

\begin{equation}
\Delta F =
\bigl(F_{\mathrm{Ring},q_1'} - F_{\mathrm{Ring},q_1}\bigr)
+
\bigl(F_{\mathrm{Ring},q_2'} - F_{\mathrm{Ring},q_2}\bigr),
\label{eq:DeltaF}
\end{equation}

where $U_{q'_1}$ and $U_{q'_2}$ are the new chains after the link shift.

To evaluate possible modifications, we define a maximum allowed $F$ degradation, measured as the relative change between the modified and original $F$ values. For each admissible move, we compute the score

\begin{equation}
I(l_{\mathrm{old}}, l_{\mathrm{new}}) =
\frac{\Delta C(l_{\mathrm{old}}, l_{\mathrm{new}})}%
{\Delta F(l_{\mathrm{old}}, l_{\mathrm{new}})},
\label{eq:IScore}
\end{equation}

where $\Delta C$ is the reduction in total construction cost associated with replacing the original link $l_{\mathrm{old}}$ by the new link $l_{\mathrm{new}}$, and $\Delta F$ is the corresponding increase in $F$ presented in \eqref{eq:DeltaF}. A larger score indicates a more favorable trade-off between cost reduction and reliability loss.

The local improvement algorithm proceeds iteratively:\

\begin{enumerate}
    \item 
    Identify all links adjacent to forks whose attachment points can be shifted without modifying the structure graph.
    \item 
    For each admissible modification, compute its score $I$ in \eqref{eq:IScore}.
    \item 
    Select the move with the highest score that does not exceed the allowed $F$ degradation.
    \item 
    Apply the move and update the affected chain lengths.
    \item 
    Repeat until no further acceptable improvements remain.
\end{enumerate}

This procedure provides a simple and effective way to fine-tune an already optimal or near-optimal network, achieving potentially non-negligible cost savings alongside minimal impact on reliability.

\vspace{20mm}

\section{Numerical analysis}

We now describe in detail our numerical simulations and data analysis, as appear in the main text.

\subsection{Failure simulation}

\subsubsection{Measuring reliability}
\label{supp:rel_sim}

All reliability results reported in this work are obtained by Monte Carlo simulation, based on the dynamic state space model described in Sec.~\ref{supp:state_space_model}. Each link $l$ is modeled as a two-state continuous-time process that alternates between an \emph{Up} (operational) state and a \emph{Down} (failed) state, with failure rate $\alpha_l$ and repair rate $\beta_l$.

Each simulation is run over a total time horizon $T = 365 \times 5$ days, corresponding to five years of operation. We fix a characteristic cycle time $t_l = u_l + r_l$ for each link $l$. 
For each link $l$, we then simulate an alternating sequence of failure and repair events, starting from the state in which all links are operational. Failure times and repair times are sampled from exponential distributions with means

\begin{equation}
u_l = \frac{1}{\alpha_l},
\qquad
r_l = \frac{1}{\beta_l},
\end{equation}

respectively. Merging the event sequences of all links yields a global list of event times

\begin{equation}
0 = t_0 < t_1 < \cdots < t_k \le T,
\end{equation}

allowing us to extract the network state at each interval $[t_{i-1}, t_i)$.

At each event time $t_i$, we update the set of failed links and compute the disconnected weight $w_i$, defined as the total weight of nodes disconnected from the source (for centralized networks), or the corresponding pairwise disconnected weight (for decentralized networks). The time-averaged reliability index is then obtained by integrating this piecewise-constant process via

\begin{equation}
    F
    =
    \frac{1}{W T}
    \sum_{i=1}^{k} w_i \,(t_i - t_{i-1}),
\end{equation}

where $W$ is the total node weight.

The computational bottleneck of the simulation is the repeated evaluation of the network connectivity after each event. To accelerate this step, we use caching:\ for any event state involving at most two simultaneously failed links, we store the corresponding disconnected weight $w_i$ and reuse it whenever the same failure configuration reappears.

The statistical accuracy of the simulations is governed by the mean number of cycles $T/t_l$ and by the link unavailability probability $p_l$. As the ratio $T/t_l$ increases, the simulation averages over a larger number of independent failure-repair cycles, leading to improved accuracy of the estimated reliability indices. In addition, as $p_l \to 0$, failures become rare events, and thus a larger number of cycles is required to achieve the same level of statistical precision.

Formally, let $\hat{p}_l$ denote the empirical fraction of time that link $l$ is unavailable during the simulation. For sufficiently large $T$, ergodicity of the two-state Markov process implies that $\hat{p}_l$ is approximately normally distributed as

\begin{equation}
\hat{p}_l \sim \mathcal{N}\!\left(
p_l,\;
\frac{2\,p_l(1-p_l)\,t_l}{T}
\right).
\end{equation}

This expression shows that the variance decreases with an increasing number of expected cycles $T/t_l$. Moreover, the relative variance scales as

\begin{equation}
\mathrm{Var}\!\left(\frac{\hat{p}_l}{p_l}\right)
=
2\,\frac{1-p_l}{p_l}\,\frac{t_l}{T}
\end{equation}

indicating that as $p_l$ decreases, estimating rare unavailability events requires proportionally longer simulation times.

\subsubsection{Setting the simulation parameters}
\label{supp:subsec_p_bar}

In our analysis, we characterize the component risk by the intrinsic failure probability $p$, capturing failure probability per unit length. Each link is assigned a length $\lambda_l$. We now show how to link these parameters to the simulation parameters $\alpha_l,\beta_l$ which describe the Monte-Carlo transition probabilities between Up and Down.

First, we assign the link failure probability $p_l$. Assuming that failures occur independently along the link length, we set

\begin{equation}
    p_l = 1 - e^{-\lambda_l\, p},
\end{equation}

providing the specific failure rate of each link. To realize this target failure probability, we set the Monte Carlo transition rates to

\begin{equation}
    \alpha_l = \frac{p_l}{t_l},
    \qquad
    \beta_l  = \frac{1-p_l}{t_l},
    \label{eq:AlphaBeta}
\end{equation}

ensuring that, on average, link $l$ is down $p_l$ percent of the time.

In practice, $p_l$ depends on the geometric scale and on the distribution of link lengths. Therefore a fixed value of $p$ would generally produce different failure statistics across networks with different geometries, procluding us from achieving a meaningful comparison between networks and across simulations. To overcome this, we fix the \emph{mean link failure probability}

\begin{equation}
\bar{p}_l = \langle p_l \rangle,
\end{equation}

and determine the corresponding global failure rate $\bar{p}$ such that

\begin{equation}
\label{supp:p_bar}
    \bar{p}_l = \left\langle 1 - e^{-\lambda_l\, \bar{p}} \right\rangle ,
\end{equation}

where the average is taken over all links in the network.

The parameter $\bar{p}$ is obtained numerically by solving Eq.~\eqref{supp:p_bar}. 
As an initial value, we use

\begin{equation}
\bar{p} \approx \bar{p}_l\, \frac{\lambda_{\mathrm{Net}}}{L},
\end{equation}

which follows from the linear approximation $p_l \approx \lambda_l\, p$, valid in the limit of small $p$. This initial guess offers a reliable starting point for the numerical solver to extract a more fine-tuned solution.

In summary, each simulation proceeds by first fixing the desired mean link failure probability $\bar{p}_l$, then computing the corresponding failure rate per unit length $\bar{p}$, assigning individual link failure probabilities according to

\begin{equation}
p_l = 1 - e^{-\lambda_l\, \bar{p}},
\label{eq:plvspbar}
\end{equation}

and finally, setting the simulation parameters to satify \eqref{eq:AlphaBeta}.

\subsection{Figures description}

We now turn to discuss each figure of the main text and its specific results and implementation details.

\subsubsection{Main text Figure~1}

The figure illustrates the dependence of the system's reliability on the network size $N$, the redundancy $R$, and the intrinsic link-failure probability $p$. The figure relies on the geometry of Manhattan, New York, as its underlying spatial embedding. We simulate ensembles of networks with varying parameters and evaluate their SAIDI values using the Monte Carlo procedure described in Section~\ref{supp:rel_sim}.

To study the joint effect of redundancy $R$ and failure probability $p$ (panels~a–h), we first sample $600$ points uniformly at random inside the Manhattan polygon. In each realization, we construct networks with redundancy levels ranging from $R = 0$ to $R = 20$. We then perform reliability simulations over a range of $p$ values, assigning individual link-failure probabilities $p_l$ according to their geometric lengths, as described in Eq.\ \eqref{eq:plvspbar} of Sec.\ \ref{supp:subsec_p_bar}. 

To construct the ring case network in panel~d ($R = 1$), we constructed Hamiltonian cycles using the Lin-Kernighan heuristic~\supercite{lin1973effective}, implemented via the \texttt{python-tsp} library~\supercite{python_tsp}. Lin-Kernighan is a widely used local-search algorithm for the Traveling Salesman Problem. The method starts from an initial tour and iteratively applies variable-depth edge exchanges to reduce the total tour length, yielding near-optimal solutions. 

For the higher redundancy networks of panels~g–h, we start from the $R = 1$ cycle and add random additional links to reach the desired redundancy level. In all cases, each generated network is assigned a single source node chosen uniformly at random. We emphasize that in this specific analysis we did not add the $R$ redundant links optimally, just a random, as this figure was not intended to examine our optimization algorithm, rather to illustrate the challenges that motivate it. To lower simulation variance, especially at low $ p$-values, we average the results of 20 simulations for each data point.

To study the effect of network size $N$ on SAIDI (panel~i), we simulate a sequence of $10$ networks constructed over progressively expanding north-to-south slices of Manhattan. Each network contains an increasing number of nodes and redundant links:\ the $i$-th network has $N = 100 + 30 i^2$ nodes and $i + 3$ redundant links. In addition, for every $600$ nodes we introduce an additional source, so that the number of sources scales as $K = \lceil N/600 \rceil$. To reduce statistical fluctuations, the entire procedure is repeated five times and averaged, yielding a total of $50$ simulated networks. All networks in this experiment are simulated using a fixed $p = 5\times10^{-4}$.

\subsubsection{Main text Figure~3}

The figure demonstrates the reliability analysis of our three fundamental network topologies:\ trees, rings, and $3$-regular networks.

\textbf{Trees}.\ 
Panels~a-c present the reliability of an ensemble of $100$ randomly generated tre\"{u}s with heterogeneous sizes and topologies. Each tree is generated as follows:\ first, the number of vertices $N$ is drawn uniformly from the range $5 \le N \le 30$. The tree topology is then constructed from a Pr\"{u}fer sequence~\supercite{prufer1918neuer} of length $N-2$, with the sequence statistics chosen to span different structural regimes.

The resulting ensemble consists of three classes of trees:\ 
(i) Near-star trees, generated by Pr\"{u}fer sequences dominated by a single label, producing highly centralized structures with one dominant hub.
(ii) Near-chain trees, generated from random permutations with small perturbations, yielding elongated, path-like structures; and 
(iii) Fully random trees, generated from uniformly random Pr\"{u}fer sequences, corresponding to the classical Cayley model. 

For each tree, the intrinsic failure probability $p$ is drawn uniformly from the interval $[10^{-3},10^{-2}]$, and the corresponding reliability index is computed using the simulation procedure described in Section~\ref{supp:rel_sim}. We assume uniform links length $\lambda_l = 1$.

\textbf{Rings}.\
Panels~d-f present the corresponding reliability analysis for ring networks ($R = 1$). To generate these networks, we construct simple cycles with an increasing number of nodes, ranging from $N=10$ to $N=100$. For each network size, we generate $10$ independent realizations. For every ring, reliability is simulated over $15$ values of the mean link-failure probability $p$, logarithmically spaced in the interval $[N^{-2},\,2N^{-1}]$. This range spans the regime in which the reliability transition is expected to occur.

Panel~e shows the dependence of the reliability index $F$ on the failure probability $p$ for three representative ring networks of different sizes. For each curve, $F(p)$ is obtained from Monte Carlo simulations as described in Section~\ref{supp:rel_sim}.

In panel~f, we estimate the critical failure probability $p_c$ for each network size. For a given ring, we identify $p_c$ as the solution of the fixed-point condition $F(p_c) = p_c$. Numerically, this is done by fitting a second-order polynomial to the simulated $(p,F)$ data and computing the intersection of the fitted curve with the line $F = p$. The resulting values of $p_c$ are then plotted as a function of the network size $N$. To compare the numerical results with the theoretical prediction $p_c\sim N^{-2}$, we estimate a prefactor $A$ by averaging $p_c N^{2}$ over all simulated sizes. We then plot the theoretical scaling $A N^{-2}$ alongside the numerical estimates, demonstrating good agreement between simulations and theory.

\textbf{$3$-Regular networks}.\ 
Panels~g-h present the reliability of $3$-regular networks constructed from the Manhattan spatial embedding shown in main text Figure~1. Starting from the network realization used in Figure~1, we iteratively add random links between pairs of nodes until all vertices have degree three. The resulting networks are therefore cubic but retain the underlying spatial geometry. For each such network, the reliability index is simulated as a function of $p$.

\subsubsection{Main text Figure~4}
The figure illustrates the OPT spatial algorithm for constructing reliable networks from a given set of points in the plane. 
In panels~g-n, we uniformly sample $1,000$ points in the unit square and compare three different spatial network construction methods with a fixed redundancy of $R = 10$ links:\ a na\"{i}ve approach (panels~h-i), a $2$-connected construction (panels~j-k), and the optimal construction introduced in this work (panels~l-n).

\textbf{Na\"{i}ve approach}.\
The na\"{i}ve construction consists of two stages. First, a randomized spanning tree is generated on the point set. To introduce randomness while preserving spatial locality, we perturb the geometric length of each candidate link $l$ by multiplying its Euclidean length $\lambda_l$ by a random factor $0.1f+0.9$, where $f$ is drawn uniformly from $[0,1]$. The minimum spanning tree is then computed using these perturbed lengths, yielding a random but spatially embedded tree.

In the second stage, redundant links are added greedily in order to improve connectivity. Starting from the tree $T$, we consider candidate links not present in $T$.  Each such link closes a unique cycle, corresponding to the unique path between its endpoints in $T$. 
Adding this link promotes all vertices along that cycle into a $2$-edge-connected component. At each iteration, we select the link that maximizes the number of vertices that undergo this promotion from a $1$-edge-connected to a $2$-edge-connected component. We repeat this procedure until we exhaust our budget of $10$ redundant links.

This na\"{i}ve approach improves connectivity in a local and greedy manner, but does not enforce global structural constraints such as cubic degree, uniform chain lengths, or $3$-edge-connectivity. It therefore serves as a baseline for comparison with the more structured $2$-connected and optimal constructions, as shown in panels~j-k and~l-n.

\textbf{$2$-connected networks}.\
In the $2$-connected construction, we generate networks whose structure graph is cubic and $3$-edge-connected, but whose chain lengths are not optimized and hence exhibit a substatal variance $\sigma_{\tilde{\lambda}}$. We generate such networks using a modified version of the OPT algorithm. Specifically, in the clustering stage of the algorithm, instead of partitioning the points into equal-size clusters, we draw the target chain sizes from a normal distribution with variance $\sigma^{2}$. Increasing $\sigma^{2}$ therefore induces increasing heterogeneity in the resulting chain lengths.

After constructing the network, we compute the effective chain lengths $\{\tilde{\lambda}_q\}$ and evaluate their empirical variance $\sigma_{\tilde{\lambda}}^{2}$. We then simulate the network reliability and plot the resulting SAIDI as a function of the chain-length dispersion $\sigma_{\tilde{\lambda}}$, quantifying the impact of chain-length heterogeneity on reliability.

\textbf{Optimal networks}.\
In panels~l-n, we apply the full OPT algorithm to construct spatial networks with a cubic, $3$-edge-connected structure graph and approximately equal numbers of nodes in each chain. These constructions are designed to minimize the effective chain-length variance and thereby optimize reliability.

In panel~n, we compare the reliability index of the resulting optimal networks, denoted $F_{\mathrm{Opt}}$, with that of idealized $3$-regular graphs. We find that the two reliabilities are very close, indicating that the OPT construction achieves near-optimal performance while respecting the underlying spatial constraints. For comparison, $F_{\rm Opt}$ is achieved with $R = 10$ links, whereas $F_{\rm 3Reg}$ requires $N/2 = 500$ redundant links.

\vspace{3mm}\textbf{SAIDI simulation}.\
To ensure that SAIDI values are comparable across spatially different network realizations, we use a common global failure-per-unit-length parameter $p$ and apply the same value of $p$ to all generated networks. The value of $p$ is calibrated as follows:\ we take the point set, and first construct the minimum spanning tree (MST) using Euclidean link lengths. We then choose $p$ such that the mean link-failure probability of the MST equals a fixed reference value of $5\times10^{-4}$. This procedure normalizes the overall failure scale across different spatial embeddings, allowing a fair comparison of reliability between the na\"{i}ve, $2$-connected, and optimal constructions.

\subsubsection{Main text Figure~5}
Here, we demonstrate the application of the OPT algorithm to the New York City networks discussed previously (Manhattan). Using these optimal spatial constructions, we validate the theoretical prediction for the critical failure probability $p_c$, defined by the fixed-point condition $F(p_c) = p_c$.

To construct the optimal networks, we consider three values of the scaling exponent $\alpha \in \{0.3,0.4,0.5\}$. For each chosen $\alpha$, we take the $10$ independent point sets with varying network size $N$ used in main text Figure~1. For each point set, we apply the OPT algorithm to construct a reliable network with

\begin{equation}
    R = N^{\alpha}
\end{equation}

redundant links.

For each resulting network, we evaluate the SAIDI over $10$ values of the global failure parameter $p$, logarithmically spaced in the interval $[10^{-4},10^{-2}]$. From the simulated dependence $F(p)$, we extract the critical value $p_c$ for each network.

Finally, the numerically obtained values of $p_c$ are compared with the theoretical scaling prediction

\begin{equation}
p_c \sim N^{\beta}, \qquad \beta = 2(1-\alpha),
\end{equation}

demonstrating quantitative agreement between the optimal spatial constructions and our analytical predictions.

\subsubsection{Main text Figure~6}

Main text Figure~6 applies the OPT spatial algorithm to a collection of real-world infrastructure networks, including communication, power-distribution, and water-distribution systems. For each network, we compare the original topology with its optimized counterpart using three performance measures:\ relative improvement in total cost, redundancy, and reliability, as defined in Eq.~(16) of the main text.

To ensure a fair comparison of reliability, we first calibrate the failure model for each original network by fixing the mean link-failure probability to $\langle p_l \rangle = 5\times10^{-4}$. From this value, we determine the corresponding failure-per-unit-length parameter $\bar{p}$. The same value of $\bar{p}$ is then applied to the optimized network, ensuring that both networks are evaluated under identical failure statistics.

\subsection{Empirical networks}

In main text Figure~6 we demonstrate the optimization of $6$ real-world networks. Their description is outlined below.

\subsubsection{Communication networks}
The two communication networks - Wavenet and NetworkUSA - are taken from the Internet Topology Zoo project~\supercite{knight2011internet}. The project is a curated collection of real communication-network topologies obtained directly from network operators around the world. Each topology represents the physical backbone of a national or regional Internet service provider, reconstructed from publicly available documentation, technical reports, or operator-supplied data. The project provides networks of diverse sizes, structures, and geographical scales, making it a standard benchmark for research on resilience, reliability, and network design.

Among the networks included in the Zoo, we select two representative cases:\ The US-based \emph{NetworkUSA} and the Vietnamese \emph{Wavenet} network. Both networks are medium-sized backbone infrastructures that exhibit meaningful structural heterogeneity while remaining small enough to visualize and optimize. The NetworkUSA topology corresponds to a long-haul optical backbone spanning major population centers in the United States, typically organized around a sparse mesh of high-capacity fiber routes. The Vietnam network, in contrast, reflects the backbone layout of a national ISP in Southeast Asia, with a more hierarchical structure connecting a small number of regional hubs. These two networks allow us to demonstrate our methods on realistic communication infrastructures with distinct geometric and structural characteristics.

\subsubsection{Water supply networks}

The water distribution networks used in our study are taken from the Water Distribution System Research Database (WDSRD), developed by the ASCE Task Committee on Research Databases for Water Distribution Systems. The committee was established in 2013 with the goal of assembling a comprehensive collection of benchmark networks for research on water distribution system design, analysis, and optimization. The resulting database contains detailed data files and descriptive narratives for more than forty real-world or realistically adapted systems, each provided in EPANET-compatible format together with pipe, node, and demand information~\supercite{wdsrd}. The networks include both well-known benchmark systems, such as the New York Tunnel System, as well as lesser-known irrigation and municipal networks contributed directly by committee members.

Among these networks, we selected the \emph{Rural Network} and the \emph{Balerma Network}, both of which originate from real irrigation systems and have been widely used as benchmark cases in the water distribution optimization literature. The Rural Network is adapted from an existing irrigation system and was first presented by Marchi \textit{et al}.~(2014). It has an average annual demand of 2.21~MGD and contains $N = 381$ nodes, $L = 475$ pipes, and two reservoirs. The network has therefore been the subject of several pipe-sizing optimization studies using genetic algorithms, particle swarm optimization, and differential evolution~\supercite{marchi2014,bietal2015}.

The Balerma Network originates from the Sol-Poniente irrigation district in Almería, Spain, and was first introduced by Reca and Martínez~(2006). It consists of $N = 447$ nodes, $L = 454$ pipes, and four reservoirs, with an average annual demand of 25.2~MGD. Because of its size and meshed structure, the Balerma Network has become a standard benchmark in the WDS optimization community. As in the Rural case, previous works have treated the topology as fixed and optimized pipe diameters to reduce construction cost while satisfying hydraulic performance constraints, using a wide range of metaheuristics and evolutionary algorithms, including simulated annealing, harmony
search, memetic algorithms, differential evolution, and genetic
algorithms~\supercite{reca2006,banos2010,bietal2015}.

These two networks provide complementary structural characteristics—one sparse and irrigation-like, the other meshed and large-scale, thus allowing us to evaluate our reliability-based design methods on realistic and diverse water distribution infrastructures.

\subsubsection{Power networks}

To evaluate our methods on realistic power systems, we use two distribution networks from the \emph{SMART-DS} (Synthetic Models for Advanced, Realistic Testing: Distribution Systems) dataset, released through the Open Energy Data Initiative (OEDI)~\supercite{smartds,elliott2020smartds}. SMART-DS, developed by the National Renewable Energy Laboratory (NREL), provides large-scale synthetic distribution feeders designed to reproduce the spatial and electrical characteristics of real United States distribution systems. These feeders are generated using high-resolution geographic information, building and population data, realistic land-use patterns, and engineering-based design rules, resulting in synthetic networks that preserve the connectivity structures, typical design practices, and operational conditions of actual distribution grids.

In our analysis we use the San Francisco (SFO) region of the database. The first network corresponds to the \texttt{p2U} area, representing the Pacific Heights region, whose highly regular grid-like street geometry induces a relatively homogeneous and structured distribution layout. The second network is taken from the \texttt{p3U} area (Davidson), which features irregular road geometry, elevation changes, and more spatial constraints, leading to a more complex distribution topology.

Together, these two networks provide complementary test cases — one aligned with a near-orthogonal urban grid and one exhibiting heterogeneous and topographically constrained geometry — allowing us to assess our optimization approach under diverse spatial settings.

For both SMART-DS networks we restrict attention to the medium-voltage (MV) system at 12.47\,kV. The sources of the network are extracted from the \texttt{HVMVSubstation\_N} layer, which represents the high-voltage to medium-voltage substations feeding the MV grid. The customer-side vertices are taken from the \texttt{DistribTransf\_N} layer, whose entries correspond to the MV--LV distribution transformers supplying the low-voltage service area. 

\vspace{3mm}\textbf{Enhancing redundancy}.\
The original distribution power networks, as extracted from SMART-DS were tree-like, with redundancy $R = 0$. Hence, every link was a bridge and any single-link failure disconnected a subtree. To enable a fair comparison with our optimized constructions, we augmented each network by adding $R$ additional links while preserving node locations.

Let $G_{\mathrm{tree}} = (V,E)$ be the original tree-like network. We seek a minimal addition of links with maximal reliability gain. For each tree edge $e \in E$, we computed its downstream weight $w_e$,
defined as the total node weight disconnected from the source upon removal of $e$. We then considered candidate chords, between geographically proximal node pairs, whose addition can offer a substantial reliability boost. Adding a chord $q = (u,v)$ creates a single cycle and protects all tree edges along the unique $u \leftrightarrow v$ path. The benefit of $q$ is, therefore, defines as
$B(q) = \sum_{e \in \mathcal{P}(q)} w_e$, where $\mathcal{P}(q)$ is the set of edges along $q$'s path. The cost of adding $q$ is proportional to its geometric length $\ell_q$.

To enhance the SMART-DS networks we added redundancy greedily:\ at each step we selected the chord maximizing $B(q)/\ell_q$, added it to the network, and set $w_e = 0$ for all $e \in \mathcal{P}(q)$. We repeated this process until $R$ additional links were added.

\printbibliography[
  heading=subbibliography,
  title={References}
]
\end{refsection}

\end{document}